\documentclass[a4paper,12pt,Div=12,notitlepage]{scrartcl}
\usepackage[left=2.25cm,right=2.25cm,top=2.25cm,bottom=2.25cm]{geometry}
\usepackage[manualmark]{scrpage2}
\usepackage{fix-cm}
\usepackage[latin1]{inputenc}
\usepackage[ngerman]{babel}
\usepackage[T1]{fontenc}
\usepackage{ae}
\usepackage{amsmath}
\usepackage{amssymb}
\usepackage{graphicx}
\usepackage{pdfpages}
\usepackage[arrow, matrix, curve]{xy}
\usepackage{multirow}
\usepackage{bigdelim}
\usepackage{mathptmx}

\begin{document}

\title{Raum, Zeit und Wechselwirkung in der Quantentheorie der Ur-Alternativen}

\author{Martin Immanuel Kober\footnote{E-mail: Martin.Immanuel.Kober@t-online.de}\\
{\normalsize Kettenhofweg 121, 60325 Frankfurt am Main, Deutschland}}

\date{}

\maketitle

\begin{abstract}
\footnotesize{Die Quantentheorie der Ur"=Alternativen des Carl Friedrich von Weizsäcker versucht, die allgemeine
Quantentheorie basierend auf dem Begriff logischer Alternativen in der Zeit als fundamentalster möglicher Objektivierung
der Natur im menschlichen Geist zu begründen. Basierend auf dieser Interpretation der Quantentheorie soll dann
die Einheit der Physik beschrieben und die Existenz freier Objekte im Raum, deren Symmetrieeigenschaften und deren
Wechselwirkungen hergeleitet werden. Die Alternativen werden durch eine Kombination binärer Alternativen dargestellt,
welche aufgrund ihres logisch fundamentalen Charakters als Ur"=Alternativen bezeichnet werden. Durch Ur"=Alternativen
als elementaren quantentheoretischen Informationseinheiten wird die der Quantentheorie immanente Kopernikanische
Wende in Bezug auf die Raum-Frage in konsequenter Weise realisiert. Diese besteht darin, dass sich nicht die
Objekte der Natur in einem vorgegebenen Raum mit lokalen Kausalitätsrelationen befinden, sondern die
Existenz des Raumes sich umgekehrt nur als eine indirekte Art der Darstellung der Beziehungsstruktur abstrakter
quantentheoretischer Objekte ergibt. Denn die Ur"=Alternativen existieren nicht in einer vorgegebenen feldtheoretisch
verstandenen physikalischen Realität. Vielmehr wird die Existenz des Raumes aus den Ur"=Alternativen überhaupt
erst begründet. Ein solcher Realitätsbegriff steckt implizit hinter der Unbestimmtheitsrelation und drückt
sich in besonderer Weise im berühmten EPR-Paradoxon aus. Es wird in dieser Arbeit mathematisch konsistent gezeigt,
dass ein Zustand im Tensorraum vieler Ur"=Alternativen direkt in einen reellen dreidimensionalen Raum abgebildet
werden kann, sodass mit der Dynamik der Zustände eine Darstellung in einer (3+1)-dimensionalen Raum-Zeit
möglich wird. Über die $G_2$-Lie-Gruppe, die Automorphismengruppe der Oktonionen, kann ein Ansatz für die
Einbindung der inneren Symmetrien der Elementarteilchen vorgeschlagen werden. Desweiteren ermöglichen die Ur-Alternativen
die Konstituierung eines abstrakten Wechselwirkungsbegriffes, der nicht auf punktweisen Produkten von Feldern sondern
auf quantentheoretischen Verschränkungen abstrakter Objekte basiert. Mit Hilfe dessen wird über das Korrespondenzprinzip
versucht, zu einer rein quantentheoretischen Beschreibung des Elektromagnetismus und der Gravitation zu gelangen.
Dem entspricht eine viel prinzipiellere und zudem radikal hintergrundunabhängige Art der Quantisierung.}
\end{abstract}

\fontsize{12.8pt}{14.8pt}\selectfont

\pagestyle{scrheadings}

\chead[\pagemark]{\pagemark}
\ohead[]{}
\cfoot[]{}
\ofoot[]{}

\newpage

\tableofcontents

\section{Einleitung}

\begin{quote}
\textbf{\textit{"`Man kann die theoretische Physik unseres Jahrhunderts noch in die titanische Tradition
deutschen Denkens einordnen. Sie ist zwar international gültig, aber in der Herkunft vor allem deutsch:
Planck, Einstein, der philosophisch deutsch geprägte Bohr, Heisenberg. Sie wäre ohne die ihr immanenten
philosophischen Fragestellungen nie entstanden und sie ist ohne noch entschiedeneres Philosophieren
nicht zu vollenden."'\\
\\
\footnotesize{Carl Friedrich von Weizsäcker, Vortrag "`Der deutsche Titanismus"' abgedruckt in
"`Wahrnehmung der Neuzeit"', Carl Hanser München/Wien 1983, Seite 30/31}}}
\end{quote}

Zunächst muss erwähnt werden, dass diese Arbeit gewissermaßen aus zwei Teilen besteht. Die Abschnitte zwei, drei, vier und
fünf versuchen einerseits zu begründen, warum eine einheitliche Naturtheorie eine radikale Abkehr von klassischen und feldtheoretischen
Begriffen notwendig macht und die Natur daher im Rahmen einer reinen Quantentheorie beschrieben werden muss. Das bedeutet, dass eine
Wende im physikalischen Weltbild vollzogen werden muss, die derjenigen des Kopernikus und derjenigen Einsteins und Heisenbergs
in nichts nachsteht, ja eigentlich die Heisenbergsche Entdeckung überhaupt erst zu ihrer eigentlichen Konsequenz führt. Diese besteht
darin, dass nicht geometrisch definierte Objekte in einem vorgegebenen physikalischen Raum existieren, sondern abstrakte logische Objekte
umgekehrt die Existenz dieses Raumes überhaupt erst begründen. Die genannten Abschnitte versuchen andererseits die Grundidee der
Quantentheorie der Ur"=Alternativen des Carl Friedrich von Weizsäcker darzustellen und zu zeigen, dass sie den konsequentesten
Ansatz darstellt, um genau diese Kopernikanische Wende bezüglich der Raum-Frage zu realisieren. Denn die Ur"=Alternativen sind rein
logischen Objekte, elementare quantentheoretische Informationseinheiten, die keine vorgegebene physikalische Realität voraussetzen,
deren Existenz aber umgekehrt zu begründen gestatten. Die Abschnitte sechs, sieben, acht und neun entwickeln basierend auf der
Grundidee der Quantentheorie der Ur"=Alternativen neue eigene konkrete mathematische Ansätze. Dies sind im Speziellen eine Abbildung
der symmetrischen Zustände im Tensorraum vieler Ur"=Alternativen in die Raum-Zeit, was einer Begründung der Existenz von Quantenobjekten
in der Raum-Zeit entspricht, desweiteren eine Integration der inneren Symmetrien der Elementarteilchen, zudem eine rein quantentheoretische
Fassung des Wechselwirkungsbegriffes in Bezug auf Ur"=Alternativen im Gegensatz zu punktweisen Produkten von Feldern und basierend
auf diesen neuen Konzepten ein Grundansatz zu einer rein quantentheoretischen schlechthin hintergrundunabhängigen Formulierung des
Elektromagnetismus und der Gravitation. In dieser Einleitung muss zunächst einmal die ganze Ausgangsfragestellung deutlich
gemacht werden und in der Zusammenfassung und Diskussion wird das Ergebnis dieser Arbeit in kondensierter Weise dargestellt.

Die zeitgenössische fundamentale theoretische Physik basiert auf zwei grundlegenden Theorien, nämlich der Quantentheorie
und der allgemeinen Relativitätstheorie. Die Quantentheorie repräsentiert in ihrer allgemeinen Formulierung als Theorie
des Hilbert"=Raumes ein abstraktes mathematisches Schema zur Beschreibung der Dynamik beliebiger physikalischer Objekte.
Aber sie sagt in dieser allgemeinen Form zunächst noch nichts darüber aus, welche speziellen Objekte mit welchen
Eigenschaften es in in der Natur überhaupt gibt und welchen Wechselwirkungen sie unterliegen. Insbesondere besteht sie
in dieser abstrakten Formulierung vollkommen unabhängig von der Existenz eines physikalischen Ortsraumes. Lediglich ein
Zeitparameter wird in ihr vorausgesetzt, was aufgrund des noch fundamentaleren Charakters der Zeit auch unumgänglich ist.
Die allgemeine Relativitätstheorie hingegen repräsentiert eine Beschreibung von Raum und Zeit, innerhalb derer zugleich
die Gravitation als einer der fundamentalen Wechselwirkungen integriert ist. Die anderen fundamentalen Wechselwirkungen,
also der Elektromagnetismus und die schwache sowie starke Wechselwirkung, sind im Rahmen relativistischer Quantenfeldtheorien
formuliert. Relativistische Quantenfeldtheorien sind eine Verbindung aus der Quantentheorie und der speziellen Relativitätstheorie.
Das große Problem besteht nun einerseits darin, eine quantentheoretische Beschreibung der allgemeinen Relativitätstheorie
und damit der Gravitation zu finden und andererseits darin, die Gravitation mit den anderen Wechselwirkungen in einer
einzigen Theorie zu vereinheitlichen. Eine solche einheitliche Theorie müsste zudem erklären, warum es all die
konkreten Objekte gibt, die es tatsächlich gibt.

Nun stellt sich allerdings die Frage, in welchem begrifflichen Rahmen sich eine solche Vereinheitlichung vollziehen soll.
Denn die Unterschiedlichkeit der Begriffe, auf welchen die Quantentheorie, die allgemeine Relativitätstheorie und die
verschiedenen Wechselwirkungen basieren, sind ja gerade der Kern der eigentlichen Problematik. Die Quantentheorie in
ihrer abstrakten Formulierung als allgemeiner Theorie des Hilbert"=Raumes aber ist deutlich abstrakter als die
allgemeine Relativitätstheorie und überhaupt jede Feldtheorie. Denn sie basiert auf den Begriffen des Zustandes
als abstraktem Vektor in einem Hilbert"=Raum und dem Begriff der Observable als einem Operator, der auf Zustände
im Hilbert"=Raum wirkt und enthält außer der Zeit, dem fundamentalsten aller physikalischen Begriffe,
noch keinerlei konkrete Annahmen über die Beschaffenheit der Natur. Vor allem enthält sie wie erwähnt
nicht den Begriff eines physikalischen Ortsraumes und damit auch keinerlei feldtheoretische Begriffe.
Der Ursprung der Quantentheorie bei Max Planck lag darin, dass ein Kontinuum möglicher Zustände durch ein
diskretes Spektrum möglicher Zustände ersetzt wurde, um damit das Auftreten von Divergenzen zu vermeiden.
Durch das Raum-Zeit-Kontinuum im Rahmen von Feldtheorien gerät man aber erneut in exakt diese Problematik hinein.
Dies geschieht vor allem basierend auf der Beschreibung der Wechselwirkungen durch kontinuierliche punktweise
Produkte von Feldern. Daher kann man in relativistischen Quantenfeldtheorien nur zu endlichen Ergebnissen kommen,
indem man das Verfahren der Renormierung anwendet, durch welches die unendlichen Werte nachträglich sozusagen künstlich
beseitigt werden. Bei dem Versuch einer quantentheoretischen Beschreibung der allgemeinen Relativitätstheorie
in der Näherung einer relativistischen Quantenfeldtheorie ist eine Renormierung aber bekanntlich unmöglich.
Ein Versuch, die Unendlichkeiten zu beseitigen, besteht darin, dass man in bestimmten Ansätzen
entweder die Raum-Zeit diskretisiert oder eine kleinste Länge einführt. Aber es ist zu erwarten, dass dieser
Versuch noch nicht an die eigentliche Wurzel der Problematik herangeht, weil er den physikalischen Raum
mit seinen lokalen Kausalitätsbeziehungen noch als etwas Fundamentales ansieht, wohingegen die Phänomene
der Quantentheorie deutlich zeigen, dass diese Vorstellung der Realität auf der fundamentalen Ebene
überwunden werden muss.

Der wirkliche begriffliche Kern der ganzen Problematik besteht also darin, dass die Quantentheorie keinerlei
feldtheoretische Begriffe enthält. Die grundsätzliche nicht-Lokalität der Quantentheorie, also die Tatsache,
dass sie unabhängig vom Begriff eines physikalischen Ortsraumes ist, zeigt sich in aller Deutlichkeit sowohl im
mathematischen Apparat der Quantentheorie alsauch in den konkreten Phänomenen wie dem Doppelspaltexperiment und dem
EPR-Paradoxon. Die Grenzen der Gültigkeit einer lokalen Beschreibungsweise der Natur werden exakt durch die
Heisenbergsche Unbestimmtheitsrelation definiert, welche in Wirklichkeit die Grenze der Anwendbarkeit der Begriffe
Ort und Impuls ausdrückt. Wenn man aber diese fundamentale nicht-Lokalität der Quantentheorie mit dem Problem der
Unendlichkeiten im Rahmen einer Kontinuumsbeschreibung zusammen betrachtet, so drängt sich die Vermutung geradezu auf,
dass der physikalische Raum gar keine fundamentale Realität der Natur ist, sondern sich aus einer fundamentaleren
Beschreibungsweise der Natur erst nachträglich ergibt. Diese Beschreibung müsste dann in rein quantentheoretischen
Begriffen erfolgen. Relativistische Quantenfeldtheorien repräsentieren eine Art Hybrid aus einer klassischen
feldtheoretischen und einer quantentheoretischen Beschreibungsweise der Natur. Einstein zeigte aber in seiner
Arbeit von 1935 mit dem EPR-Paradoxon, dass eine quantentheoretische Beschreibungsweise der Natur mit einer
feldtheoretischen Beschreibungsweise nicht vereinbar ist. Entweder das Prinzip der lokalen Kausalität, auf dem alle
Feldtheorien basieren, ist wahr, oder die Quantentheorie ist wahr. Das in dieser Arbeit erdachte Gedankenexperiment
Einsteins wurde allerdings seither in immer neuen Weisen realisiert und immer entschieden die Experimente eindeutig
zu Gunsten der Quantentheorie. Daher besteht die Notwendigkeit, zu einer rein quantentheoretische Beschreibungsweise
der Natur zu gelangen, aus der sich die Existenz des physikalischen Raumes, der dann über die Dynamik mit der Zeit
zumindest formal zu einer (3+1)-dimensionalen Raum-Zeit verbunden werden kann, die entsprechende Darstellung der
Objekte und die allgemeine Relativitätstheorie sowie die Wechselwirkungen erst nachträglich als indirekte
Konsequenz ergeben.

Carl Friedrich von Weizsäcker versuchte seit den 1950er Jahren in seinem Programm der Rekonstruktion der Quantentheorie,
die allgemeine Quantentheorie aus grundsätzlichen erkenntnistheoretischen Postulaten zu begründen. Die hieraus
hervorgehende Quantentheorie der Ur"=Alternativen versteht die Quantentheorie als eine Theorie abstrakter Information
in der Zeit und versucht, die Existenz der konkret existierenden physikalischen Realitäten mit ihrer
spezifischen Struktur herzuleiten. Das Entscheidende hierbei ist, dass Ur"=Alternativen keine Objeke in einem bereits
existierenden Raum oder überhaupt irgendeiner unabhängig existierenden physikalische Realität sind. Vielmehr konstituiert
sich die physikalische Realität einschließlich des Ortsraumes überhaupt erst aus den Ur"=Alternativen. Zudem bilden die
Ur"=Alternativen diskrete Zustandsräume und bieten daher die Aussicht, die Unendlichkeiten von Beginn an zu vermeiden.
Das auf dieser Grundbasis aufsetzende Programm dieser Arbeit wurde oben bereits erwähnt. Zunächst soll auf verschiedenen
Argumentationslinien die These begründet werden, dass auf fundamentaler Ebene eine Beschreibunsgweise der Natur
notwendig ist, bei welcher nicht Objekte in einem vorgegebenen Raum sind, sondern abstrakte rein
quantentheoretische Objekte hinter der Existenz des Raumes stehen. Dies liefert die entscheidende
Rechtfertigung für die Annahme, dass die Quantentheorie der Ur"=Alternativen den bisher vielversprechendsten
und begrifflich konsequentesten Ansatz zu einer einheitlichen Naturtheorie darstellt. Es soll dann weiter
die grundlegende Idee der Rekonstruktion der Quantentheorie und der Quantentheorie der Ur"=Alternativen
entwickelt werden. Anschließend werden meine eigenen neuen Beiträge dargestellt, also
insbesondere ein Ansatz zur Abbildung des Zustandsraumes vieler Ur"=Alternativen
in die Raum-Zeit und damit zur Begründung der Existenz raum-zeitlicher Quantenobjekte,
ein Ansatz zur Einbindung der inneren Symmetrien der Elementarteilchen und eine Möglichkeit,
den Wechselwirkungsbegriff in einer abstrakten rein quantentheoretischen Weise zu fassen. Schließlich wird
basierend auf diesen Konzepten und dem Korrespondenzprinzip ein Vorschlag zur Formulierung eines Modells
des Elektromagnetismus und vor allem der Gravitation im Rahmen der Quantentheorie der Ur"=Alternativen entwickelt.
Dies ist der aus meiner Sicht bisher begrifflich konsequenteste Ansatz zu einer quantentheoretischen Beschreibung
der Gravitation, denn er ist in einem radikalen Sinne hintergrundunabhängig, indem er keinerlei raum-zeitliche
Begriffe voraussetzt, sondern nur auf dem Begriff abstrakter quantentheoretischer Information basiert.
Um die Grundvoraussetzungen für das Argumentationsgebäude zu liefern, müssen zunächst
einige erkenntnistheoretische Grundeinsichten thematisiert werden.

\section{Erkenntnistheoretisches Prolegomena}

\subsection{Normale Wissenschaft und wissenschaftliche Revolutionen}

Der Wissenschaftstheoretiker Thomas Samuel Kuhn unterscheidet zwischen normaler Wissenschaft und wissenschaftlichen
Revolutionen \cite{Kuhn:1962}. Während in der normalen Wissenschaft nach einem festen Paradigma, das auf einem
bestimmten System grundlegender Begriffe und Postulate basiert, spezielle Probleme gelöst werden, wird während
einer wissenschaftlichen Revolution das Paradigma an sich in Frage gestellt und nach einem neuen gesucht, da sich
zunehmend zeigt, dass die Art von konkreten Problemstellungen, mit denen man es zu tun hat, im Rahmen des alten 
Paradigmas nicht mehr behandelt werden kann. In der Geschichte der fundamentalen theoretischen Physik sind die
entscheidenden wissenschaftlichen Revolutionen die Entstehung der klassischen Mechanik, welche sowohl die
Aristotelische Vorstellung der Mechanik alsauch die Trennung der Sphäre des Himmels mit seinen mathematischen
Gesetzen und der der irdischen Sphäre mit ihren mechanischen Gesetzen aufhebt, der Übergang zur Elektrodynamik
mit ihrem Feldbegriff und der Einsicht, dass auch die Kräfte eine innere Dynamik aufweisen, vor allem aber
die beiden Revolutionen zu Beginn des zwanzigsten Jahrhunderts, nämlich die Entstehung der speziellen und
der allgemeinen Relativitätstheorie, in denen sogar die grundlegende Vorstellung der Beschaffenheit von
Raum und Zeit verändert wird, in dem diese mit einer neuen Struktur belegt werden, welche zudem in das
dynamische Geschehen miteinbezogen wird, und die Entstehung der Quantentheorie, in welcher mechanistische
Begriffe und die Vorstellung einer konkret definierten Bewegung von Objekten durch Raum und Zeit
überwunden werden. Während bei normaler Wissenschaft keine begrifflichen und erkenntnistheoretischen
Grundlagenfragen gestellt werden müssen, da der Grundrahmen fest vorgegeben ist, innerhalb dessen die
speziellen Probleme gelöst werden, stehen genau diese Fragen bei einer wissenschaftlichen Revolution
im Zentrum des eigentlichen geistigen Geschehens. Das menschliche Denken basiert auf bestimmen Begriffen,
die sich auf die wirklichen Realitäten in der Natur beziehen. Es ist aber keineswegs selbstverständlich,
dass die Phänomene des neuen Erfahrungsbereiches sich mit den natürlichen Begriffen in unserem Denken adäquat
beschreiben lassen. Deshalb müssen nicht nur die Begriffe in unserem Denken bezüglich ihrer Anwendbarkeit auf
bestimmte Realitäten in der Natur, sondern auch die grundsätzliche Beziehung unseres Denkens zur Natur ganz
grundsätzlich analysiert werden. Genau dies ist der Grund, warum für die meisten ganz großen theoretischen Physiker
des zwanzigsten Jahrhunderts, welche die Relativitätstheorie und die Quantentheorie entdeckten und formulierten,
nämlich Max Planck, Albert Einstein, Niels Bohr, Werner Heisenberg, Wolfgang Pauli und Erwin Schrödinger, die
philosophischen Fragen in Bezug auf diese neuen Theorien von alles entscheidender Bedeutung waren. Und genau
deshalb müssen diese Fragen natürlich auch bei dem Bestreben der Vereinheitlichung der Quantentheorie, der allgemeinen
Relativitätstheorie und aller fundamentalen Wechselwirkungen nicht nur unbedingt miteinbezogen werden, sondern sogar
im Zentrum stehen. Carl Friedrich von Weizsäcker geht sogar soweit, dass er die allgemeine Quantentheorie als
eine fundamentale Naturtheorie vollständig aus rein erkenntnistheoretischen Postulaten begründen möchte. 
Dies alles sollte Grund genug sein, die Bedeutung philosophischer und insbesondere erkenntnistheoretischer
Grundfragen als zentralen Bestandteil der fundamentalen theoretischen Physik in ihrem vollen Umfang
zu würdigen. Deshalb soll sich nun zunächst sehr grundlegenden erkenntnistheoretischen Basisfragen
zugewandt werden, die für die Rechtfertigung und das Verständnis des von Weizsäckerschen Ansatzes der
Quantentheorie der Ur"=Alternativen von zentraler Bedeutung sind.

\subsection{Die grundlegende Idee der Kantischen Erkenntnistheorie}

In der Naturwissenschaft vollzieht sich eine Wechselbeziehung zwischen der Natur als an sich selbst unabhängig vom
Menschen existierender Realität einerseits und dem menschlichen Geist andererseits. Dies aber bedeutet, dass zwar nicht
die Natur selbst, aber die Naturwissenschaft als menschliche Geistestätigkeit immer auch ein in spezifischer Weise
auf den menschlichen Geist bezogener Vorgang ist. Aus eben diesem Grunde ist die Untersuchung des menschlichen
Geistes selbst eine unabdingbare Voraussetzung der Naturwissenschaft, insbesondere der fundamentalen Naturwissenschaft,
die sich mit der grundlegenden Beschaffenheit der Natur in ihrem Inneren beschäftigt. Denn Carl Friedrich von Weizsäcker
brachte die Grundgegebenheit, dass die Naturwissenschaft als Untersuchung einer vom menschlichen Geist unabhängigen
Realität dennoch immer nur basierend auf den grundlegenden Erkenntnisstrukturen des menschlichen Geistes basiert,
auf den folgenden Satz: "`Die Natur ist vor dem Menschen, aber der Mensch ist vor der Naturwissenschaft."'
Die zentrale geistesgeschichtliche Bedeutung des Immanuel Kant besteht darin, dass er jene grundlegende Wende
im menschlichen Denken in letzter Konsequenz vollzog, welche vor allem nach den im menschlichen Geist liegenden
Bedingungen der Möglichkeit von Erkenntnis fragt, welche den grundlegenden Rahmen bilden, innerhalb dessen sich die Natur,
das Ding an sich in der Sprache Kants, das für sich selbst vollkommen unabhängig von diesem Rahmen existiert,
für den Menschen darstellt \cite{Kant:1781},\cite{Kant:1783}. Gemäß der Kantischen Erkenntnistheorie sind es
vor allem zwei Grundgegebenheiten des menschlichen Geistes, welche die grundlegende Art und Weise der Erkenntnis
über die Natur konstituieren. Dies sind zum einen die Grundformen der Anschauung, nämlich Raum und Zeit,
innerhalb derer uns die Objekte der äußeren Realität erscheinen, und dies sind zum anderen die Kategorien,
also grundlegende Begriffe, auf denen das menschliche Denken basiert, und mit denen der menschliche Geist
diese Erscheinungen dann ordnet. Die vielleicht wichtigste unter den Kategorien ist die Kausalität. Raum, Zeit und Kausalität
sind also gemäß Kant gar keine Eigenschaften der Realität an sich, sondern Realitäten, die nur im menschlichen Geist
existieren, aber für die menschliche Erfahrung konstitutiv sind. Das hinter der Erscheinung der Natur
innerhalb des menschlichen Geistes im Rahmen von Raum, Zeit und Kausalität eigentlich existierende
Ding an sich ist seiner eigentlichen Natur nach nicht erkennbar. Kant begründet die Tatsache, dass Raum,
Zeit und Kausalität nicht durch spezielle Erfahrung in den menschlichen Geist gelangt sind, sondern ihm
stattdessen a priori gegeben sind, mit dem Argument, dass man sich überhaupt gar nicht denken und vorstellen könnte,
dass eine Erfahrung sich nicht im Rahmen von Raum, Zeit und Kausalität vollzieht. Wenn man sich zum
Beispiel einen Gegenstand vorstellt und rein gedanklich alle Eigenschaften von ihm wegnimmt,
seine Farbe, die Gravitationswirkung, der er unterliegt, und schließlich sogar das Material, aus dem
er besteht, so wird am Ende doch eine Sache übrig bleiben, die man sich nicht wegdenken kann, und das ist der Raum,
den er ausgefüllt hat. Wenn es aber schlechthin unmöglich ist, von der Räumlichkeit der Gegenstände in der Welt zu
abstrahieren, so kann der Mensch grundsätzlich überhaupt keine Erfahrung über eine Realität in der Natur machen,
ohne dass diese an den Anschauungsraum gebunden ist, und er kann kein Objekt wahrnehmen, ohne dass dies
im Raum erscheint. Daher muss der Raum eine a priori gegebene Realität sein, welche das grundlegende
Wesen unserer Erfahrung über die Welt bestimmt, und gerade keine in der Natur an sich existierende Eigenschaft.
Dies ist der Grund, warum die entscheidenden revolutionären Schritte in der Naturwissenschaft im Denken nur
schwer vollzogen werden können. Unser Denken und vor allem unser Wahrnehmungs- und Vorstellungsvermögen
ist a priori an bestimmte Grundgegebenheiten gebunden, die aber gerade deshalb zumindest auf der fundamentalen
Ebene der Realität gar nicht zwangsläufig Gültigkeit beanspruchen können. Diese Erkenntnis, dass der Raum eine
a priori gegebene Grundform der Anschauung darstellt, die konstitutiv für menschliche Erfahrung ist, ist der
entscheidende Schlüssel zum Verständnis der Paradoxien in der Quantentheorie. Unser Geist
interpretiert die Phänomene in der Natur im Rahmen der Raumanschauung, welche aber auf das
Quantenobjekt selbst, das Ding an sich in der Sprache Kants, überhaupt gar nicht sinnvoll
bezogen werden kann.

\subsection{Der Begriff der Materie bei Platon und Kant}

Bei Platon im Dialog Timaios \cite{Platon:Timaios} wird der Materiebegriff auf eine reine mathematische Struktur zurückgeführt,
die bei ihm allerdings mit den vier regulären Körpern Tetraeder, Würfel, Oktaeder und Ikosaeder identifiziert wird, die sich
ihrerseits aus gleichseitigen Dreiecken zusammensetzen. Die entscheidende Einsicht, die sich hierin ausdrückt, besteht aber
ganz sicher nicht in dieser speziellen geometrischen Vorstellung, die zwar durchaus interessant ist, aber vor dem Hintergrund
der heutigen theoretischen Physik ganz sicher nicht aufrecht erhalten werden kann. Sie besteht jedoch in der grundlegenden
philosophischen Erkenntnis, dass die Natur in ihrem Inneren nicht durch etwas Stoffliches oder Mechanisches charakterisiert ist,
sondern durch reine mathematische Form, durch reine Struktur, also letztendlich durch etwas rein Geistiges. Allerdings ist
Platon diesbezüglich nicht konsequent genug in seinem Denken, indem er die Struktur in einem geometrischen Sinne versteht,
also auf die Raumanschauung bezieht. Damit bleibt diese Anschauung trotz ihres immerhin schon sehr grundlegend mathematischen
Ansatzes des Verständnisses des Wesens der Natur immer noch einem geometrischen, also räumlichen, und damit letztendlich 
doch gegenständlichen Realitätsbegriff verhaftet. Denn gemäß Descartes ist das Materielle durch die Eigenschaft der räumlichen
Ausdehnung charakterisiert. Noch wichtiger aber ist der Einwand gegen eine geometrische Vorstellung bezüglich der fundamentalen
Objekte der Natur, der im Rahmen der zweiten Kantischen Antinomie zum Ausdruck gebracht wird, dass es nämlich schon aus Gründen
der reinen begrifflichen Konsistenz und Widerspruchsfreiheit überhaupt keine kleinsten räumlichen Objekte geben kann.
Denn jedes Volumen lässt sich zumindest begrifflich immer noch weiter in Teilvolumina zerlegen \cite{Kant:1781}.
Dies steht aber in völligem Einklang mit dem grundsätzlich nicht-lokalen Charakter der Quantentheorie. Kant hatte also
durch rein philosophische Argumentation bereits eine Intuition für diese grundsätzliche Problematik, obwohl er von der
Quantentheorie überhaupt noch nichts wissen konnte. Im Dialog Parmenides wird der Begriff des Einen philosophisch erörtert,
das keine Teile und keine räumliche Struktur in sich trägt \cite{Platon:Parmenides}, was seinerseits durch Carl Friedrich
von Weizsäcker in Bezug auf die Objekte der Quantentheorie interpretiert wird \cite{Weizsaecker:1981}. Wir sehen hier also,
dass sich die großen Philosophen bezüglich ihrer begrifflichen Reflexionsebene bereits auf einem Niveau befanden, dass ohne
jede Kenntnis der konkreten empirischen Phänomene der Quantentheorie die Grenzen des feldtheoretischen Denkens deutlich
erkennen lässt. Wenn man aber die Quantentheorie kennt, besitzt man endgültig allen Grund, sich mit diesen begrifflichen
Grundfragen in aller Gründlichkeit auseinanderzusetzen, was schließlich zur vollkommenen Abkehr von einem naiven räumlichen
Realitätsbegriff führen muss.

\subsection{Beziehung zur modernen Naturwissenschaft}

Nun macht es aber die Entwicklung der modernen Naturwissenschaft andererseits auch notwendig, die Kantische Erkenntnistheorie
zumindest in gewissem Sinne zu relativieren und zu modifizieren. Die grundlegende Wahrheit der Kantischen Erkenntnistheorie
kann in ihrem Grundgehalt niemals angetastet werden, denn sie analysiert einfach die im menschlichen Geist nun einmal a priori
gegebenen Grundstrukturen des Denkens, aber sie muss doch unter Einbeziehung naturwissenschaftlicher Erkenntnisse,
die Kant noch nicht zugänglich waren, in einem erweiterten Rahmen interpretiert werden. Eigentlich hätte Kant auch eine solche
Neuinterpretation seiner Erkenntnistheorie unter Einbeziehung naturwissenschaftlicher Erkenntnisse für unmöglich gehalten.
Denn die grundlegenden Strukturen im menschlichen Geist stellen ja gerade Bedingungen der Möglichkeit von Erfahrung da,
welche das grundlegende Wesen aller spezielle Erfahrungsinhalte konstituieren, und sollten deshalb eigentlich durch spezielle
Erfahrung nicht modifiziert werden können. Die dem menschlichen Geist innewohnenden Anschauungsformen und Kategorien,
innerhalb derer wir als Menschen Erfahrung machen, und die Kant ja vollkommen richtig untersucht hat, sind für
die Art und Weise des Zugangs des menschlichen Geistes zur Natur tatsächlich unumgänglich. Deshalb bleibt dieser
entscheidende Kern der Kantischen Erkenntnistheorie nicht nur absolut unangetastet, sondern ist, wie wir sehen werden,
von alles entscheidender Bedeutung für ein wirkliches Verständnis der Quantentheorie. Allerdings ist es durchaus möglich,
und diese Möglichkeit sah Kant noch nicht, dass sich innerhalb des a priori vorgegebenen Grundrahmens menschlicher Erkenntnis
eine Realität indirekt darstellt, die an sich selbst nicht nur allgemeinere Eigenschaften aufweist, die dem teilweise
widersprechen, sondern in diesen Eigenschaften sogar teilweise indirekt erkennbar ist. Denn diese Eigenschaften können
sich innerhalb der dem menschlichen Geist gegeben Grundbeschaffenheit dennoch in indirekter Art und Weise widerspiegeln.
In der speziellen und der allgemeinen Relativitätstheorie Albert Einsteins zeigte sich beispielsweise, dass Raum und Zeit
in einer spezifischen Weise miteinander verknüpft sind und dass der Raum allgemeinere geometrische Eigenschaften aufweist
\cite{Einstein:1905},\cite{Einstein:1914},\cite{Einstein:1915},\cite{Einstein:1916}. Wenn sich aber durch empirische
Untersuchungen herausstellt, wie das bei der Relativitätstheorie der Fall war, dass der reale physikalische Raum
andere Eigenschaften hat, als wir sie ihm gemäß der in unserem menschlichen Geist liegenden Raumanschauung zuschreiben,
so kann dies doch nur bedeuten, dass unsere Raumanschauung doch auch eine in der Realität an sich existierende Entsprechung hat,
es also einen außerhalb unseres Geistes wirklich existierenden Raum gibt. Dies gilt jedoch, und hierin liegt genau
die entscheidende Grunderkenntnis dieser Schrift, nur auf der Oberflächenebene der Natur. Wenn man geistig noch
tiefer ins Innere der Natur vordringt, wie dies in der endgültigen Gestalt der Quantentheorie geschieht
\cite{Heisenberg:1925},\cite{BornJordan:1925},\cite{BornHeisenbergJordan:1925},\cite{Heisenberg:1927}, so verliert
die Kategorie der Räumlichkeit ihre ontologische Bedeutung tatsächlich und eben an jener Stelle erhält die
Kantische Philosophie ihre alles entscheidende Bedeutung für die Interpretation der Quantentheorie und
damit zugleich für die Suche nach dem richtigen begrifflichen Rahmen zur Vereinheitlichung der
fundamentalen Physik. Die Beziehung der Kantischen Philosophie zur Quantentheorie und die
Kopenhagener Deutung der Quantentheorie werden ausführlich behandelt in \cite{Heisenberg:1958},\cite{Heisenberg:1969},\cite{Heisenberg:1979},\cite{Weizsaecker:1971},\cite{Weizsaecker:1985},\cite{Weizsaecker:1992},\cite{Weizsaecker:1999}. Bevor dieser zentrale Gedanke aber im nächsten Abschnitt in aller
Ausführlichkeit dargelegt wird, muss zunächst noch darauf hingewiesen werden, dass auch die
Evolutionstheorie Charles Darwins eine wichtige Ergänzung zur Kantischen Erkenntnistheorie
liefert, wie sie erstmals durch Konrad Lorenz in ihrem vollen Gewicht erkannt wurde und beispielsweise
auch seitens Hoimar von Ditfurth und Gerhard Vollmer vertreten wird. Im Rahmen der evolutionären
Erkenntnistheorie \cite{Lorenz:1973},\cite{Ditfurth:1976},\cite{Vollmer:1975} ergibt sich nämlich eine Erklärung,
warum die dem menschlichen Geist a priori gegebenen Wahrnehmungs- und Denkstrukturen näherungsweise, nämlich auf
einer Oberflächenebene, mit der wirklichen Natur in Übereinstimmung stehen, und dann zu versagen beginnen,
wenn man tiefer in die eigentliche Realität der Natur vordringt. Das menschliche Gehirn, das zwar nicht mit
dem menschlichen Geist identisch ist, der weit darüber hinaus geht, aber doch immerhin seine physische
Basis darstellt, entwickelte sich nämlich wie alle anderen lebenden Strukturen in der Natur im Rahmen der
Phylogenese nach dem Prinzip des Überlebensvorteiles, das es dem Menschen gewährte. Demnach haben sich die
dem Gehirn innewohnenden Strukturen zur Erkenntnis der Natur, die sich dann auch im menschlichen Geist
widerspiegeln, in einer solchen Genauigkeit an die Natur angepasst, dass die dem Menschen dadurch zugängliche
Information über die Natur ihm einen signifikanten Überlebensvorteil einbrachte. Diesbezüglich war also eine
gewisse Übereinstimmung mit der Natur vorteilhaft, aber eine tiefere Erkenntnis des Inneren der Natur
nicht notwendig. Auf die Tatsache, dass die Existenz der menschlichen Seele und menschliche Erkenntnis
nicht ausschließlich in einer naturalistischen Weise zu erklären sind, da sie ja überhaupt erst die
Voraussetzung für die Wahrnehmung eines Objektes in der Natur wie des Gehirnes liefern, kann in diesem Zusammenhang
nicht näher eingegangen werden. Deshalb soll sich nun der Beziehung der Kantischen Philosophie zur Quantentheorie
in Bezug auf die Frage nach der Interpretation der Natur des Raumes näher zugewandt werden, um die es in dieser
Schrift zunächst eigentlich geht, um eine adäquate begriffliche Basis für eine konkrete mathematische Formulierung
einer einheitlichen Naturtheorie zu erhalten. Diese geschieht dann anschließend basierend auf der Rekonstruktion
der Quantentheorie und dem Begriff der Ur"=Alternative, welcher der Kopernikanischen Wende in Bezug auf die
Raumfrage im vollen Sinne Rechnung trägt.

\section{Die der Quantentheorie immanente Kopernikanische Wende}

\subsection{Die Entwicklung der Raum-Frage in der Geistesgeschichte}

Gemäß Hegel vollzieht sich in der Geistesgeschichte eine dialektische Bewegung hin zur Wahrheit,
innerhalb derer einander zunächst widersprechende Gegenpositionen, also These und Antithese, zu einer höheren Synthese
geführt werden, die dann als These der Ausgangspunkt des nächsten dialektischen Schrittes ist. Eine solche
dialektische Bewegung hin zu einer immer tieferen und exakteren Erkenntnis vollzog sich zumindest tendenziell auch
in Bezug auf das Verständnis der eigentlichen Natur des Raumes. Hieran waren Philosophie und Physik in gleicher
Weise beteiligt. Die höchste und für die Suche nach einer einheitlichen Naturtheorie zentrale Ebene der Erkenntnis
kann diesbezüglich in der Quantentheorie, insbesondere in der Quantentheorie der Ur-Alternativen, unter Einbeziehung
der Kantischen Erkenntnistheorie erreicht werden. Aber um diese zu erreichen und wirklich zu verstehen,
müssen zunächst die vorhergehenden Ebenen systematisch durchlaufen werden. Die Frage nach der Natur des
Raumes bewegt sich durch die folgenden Thesen hindurch:

\textbf{These A - Newtonsche klassische Mechanik:} Es gibt einen realen absoluten physikalischen Raum als
fundamentaler Realität und er hat in der Natur diejenigen Eigenschaften, die sich uns auch in unserer
unmittelbaren Erfahrung darstellen. Der Raum ist von der ebenfalls in der Natur existierenden absoluten
Zeit vollkommen unabhängig. Vor allem aber existiert er unabhängig von den in ihm sich befindenden Objekten.
Auch wenn man alle Objekte aus dem Raum entfernen würde, so würde der Raum weiterhin als in sich existierende
unabhängige Entität bestehen. Dieser absolute Raum definiert aber umgekehrt einen absoluten Bewegungsbegriff
für die in ihm sich befindenden Objekte. Diese Position bezüglich der Raum"=Frage entspricht unserem
natürlichen Urteil und wir unterstellen sie eigentlich gewöhnlich solange als wahr, als wir sie
keiner tieferen philosophischen Reflexion unterziehen.

\textbf{These B - Leibnizscher Relationalismus:} Die Leibnizsche Anschauung bezüglich der Raum"=Frage stellt
gewissermaßen die dialektische Gegenposition, die Antithese, zur Newtonschen Auffassung im Rahmen der
klassischen Mechanik dar. Gemäß Leibniz gibt es keinen absoluten Raum. Vielmehr stellt der Raum nur so
etwas wie eine Beziehungsstruktur zwischen den Körpern dar. Wenn man also alle Körper aus dem Raum
entfernen würde, so würde damit zugleich auch der Raum selbst verschwinden. Dem Raum kommt also gemäß
Leibniz gar keine eigenständige für sich bestehende sondern lediglich eine durch die Existenz der
Objekte indirekt sich konstituierende Realität zu.

\textbf{These C - Kantische Transzendentalphilosophie:} In der Kantischen Erkenntnistheorie ist der
Raum eine in unserem Geist liegende Realität, eine Grundform der Anschauung. Diese ist zwar für jegliche
menschliche Erfahrung über die Natur konstitutiv. Insofern gehört der Raum notwendigerweise zur Natur,
insofern sie sich im menschlichen Geist spiegelt. Aber der Realität der Natur selbst, dem Ding an sich,
kommt die Räumlichkeit nicht zu. Diese Anschauung wurde im letzten Abschnitt bereits dargelegt und sie
stellt die im Vergleich zur Leibnizschen noch deutlich grundsätzlichere Antithese zur Newtonschen dar,
indem sie dem Raum nicht nur eine lediglich indirekt über die Objekte verliehene Existenz zuschreibt,
sondern ihm seine Existenz außerhalb des menschlichen Geistes überhaupt abspricht. Dem Raum kommt zwar
eine wirkliche Realität zu, aber diese liegt ausschließlich im menschlichen Geist und seinem Bezug
zur Wirklichkeit.

\textbf{These D - Einsteinsche spezielle Relativitätstheorie:} Der Raum existiert doch als wirkliche Realität in
der Natur, aber er hat andere Eigenschaften als diejenigen, die unser Geist ihm zunächst zuschreibt. Demnach muss
eine Unterscheidung vorgenommen werden zwischen der in unserem Geist a priori gegebenen Anschauungsform des Raumes
im Kantischen Sinne (These C) und dem physikalischen Raum in der Realität an sich. Hierbei ist der Raum
als Anschauungsform in unserem Geist dem realen Raum in der Natur nur in der Näherung der klassischen
Mechanik isomorph. In der Natur selbst ist der Raum gemäß der speziellen Relativitätstheorie mit der
Zeit zur Raum"=Zeit verbunden (zumindest ist eine solche Beschreibung formal möglich) und raum"=zeitliche
Beziehungen sind entgegen unserer natürlichen Anschauung vom Bezugssystem abhängig.

\textbf{These E - Einsteinsche allgemeine Relativitätstheorie:} In der allgemeinen Relativitätstheorie werden
die Erkenntnisse bezüglich des Raumes gemäß der speziellen Relativitätstheorie (These D) beibehalten aber zugleich
um neue erweitert. Der Raum trägt hier eine in sich gekrümmte metrische Struktur. Diese metrische Struktur
ist selbst dynamisch und muss deshalb selbst wie ein Objekt behandelt werden. Die Raumkoordinaten hingegen
beschreiben keine absolute Realität, sondern lediglich eine Beziehung zwischen dynamischen Entitäten, zu denen
auch die metrische Struktur gehört, also das Gravitationsfeld. Real sind nur Koinzidenzen im Raum, was sich
formal in der Diffeomorphismeninvarianz ausdrückt beziehungsweise in der Tatsache, dass es in der allgemeinen
Relativitätstheorie keine absoluten räumlichen Größen gibt. Das Phänomen der Beschleunigung existiert zwar
wirklich, aber in Bezug auf das Gravitationsfeld als einer dynamischen Größe. Damit bestätigt die allgemeine
Relativitätstheorie den Leibnizschen Raum"=Relationslismus (These B). Dies wird in sehr exakter Weise
in \cite{Rovelli:2004} und \cite{Lyre:2004} thematisiert.

\textbf{These F - Heisenbergsche Quantentheorie und Kopenhagener Deutung:} Mit der Quantentheorie
wird entdeckt, dass auf der fundamentalen Ebene räumliche Kausalstrukturen ihre Gültigkeit
vollkommen verlieren. Auf der fundamentalen Ebene gibt es überhaupt keinen Raum mehr und die Kategorie
der Lokalität verliert ihren Sinn. Hier wird Kant (These C) also insofern vollkommen bestätigt, als ein
Quantenobjekt als Ding an sich keinerlei räumliche Eigenschaften hat. Diese erhält es erst dadurch,
dass es in unserer Raumanschauung dargestellt wird. Und die Paradoxien, wie sie sich etwa im
Doppelspalt"=Experiment oder im EPR"=Paradoxon zeigen, entstehen dadurch, dass die Raumanschauung
auf eine Realität angewandt wird, auf die sie schlicht und einfach nicht passt. Die Grenzen,
innerhalb derer klassische räumliche Begriffe näherungsweise angewandt werden können, sind durch
die Heisenbergsche Unbestimmtheitsrelation mathematisch exakt definiert. Die spezielle
Relativitätstheorie (These D) und die allgemeine Relativitätstheorie (These E) korrigieren
die Kantische Erkenntnistheorie (These C) also nur insofern, als der Raum auf der klassischen Oberflächenebene
doch auch in der Natur selbst existiert. Aber auf der tieferen durch die Quantentheorie beschriebenen Ebene
verlieren räumliche Kausalstrukturen in der Natur selbst letztendlich doch ihre Gültigkeit.
Dadurch wird Kant (These C) also nicht nur darin bestätigt, das dem menschlichen Geist a priori
gegebene Denkstrukturen innewohnen, welche menschliche Erfahrung überhaupt erst möglich machen,
sondern auch darin, dass die Räumlichkeit dem Ding an sich, also in diesem Zusammenhang
dem Quantenobjekt, auf der fundamentalen Ebene tatsächlich nicht zukommt. Auf die
entsprechenden Schriften, in denen dies behandelt wird, wurde im letzten Abschnitt bereits verwiesen
\cite{Heisenberg:1958},\cite{Heisenberg:1969},\cite{Heisenberg:1979},\cite{Weizsaecker:1971},\cite{Weizsaecker:1985},\cite{Weizsaecker:1992},\cite{Weizsaecker:1999}.

\textbf{These G - Von Weizsäckersche Quantentheorie der Ur"=Alternativen:} Die Quantentheorie der Ur"=Alternativen
nimmt diese der Quantentheorie letztendlich innewohnende zentrale Erkenntnis bezüglich der Natur des Raumes (These F)
wirklich ernst, indem sie sich von räumlich"=feldtheoretischen und damit klassisch"=mechanistischen
Begriffen endgültig löst. Statt konkreten Objekten im Raum, welche ihren nicht"=lokalen Charakter nur
indirekt über die Unbestimmtheitsrelation erhalten, der zudem in den Quantenfeldtheorien über die Definition
der Wechselwirkung über punktweise Produkte von Feldern doch wieder aufgehoben wird, postuliert sie mit
Alternativen rein abstrakt"=logische Objekte als fundamentalster Darstellung der Realität der Natur in
unserem Geist. Diese Objekte existieren nicht in einem vorgegebenen Raum, sie gestatten aber umgekehrt die
Begründung der Existenz eines solchen Raumes als einer möglichen indirekten Darstellung dieser Alternativen
und ihrer Beziehungen. Damit wird nicht nur der in der allgemeinen Relativitätstheorie (These E) enthaltenen
Leibnizschen Erkenntnis (These B) in wirklich konsequenter Weise Rechnung getragen, dass der Raum eine sich
nur indirekt aus der Existenz von Objekten konstituierende Realität ist, sondern auch der in der
Quantentheorie (These F) enthaltenen Kantischen Erkenntnis (These C), dass der Raum keine fundamentale
Realität der Natur ist, sondern auf der fundamentalen Ebene der Natur überhaupt keine räumlichen Kausalstrukturen
mehr existieren. Die Quantentheorie der Ur"=Alternativen wird in \cite{Weizsaecker:1971},\cite{Weizsaecker:1985}
und \cite{Weizsaecker:1992} in ihren physikalischen und philosophischen Grundlagen ausführlich
dargestellt und entwickelt.

\subsection{Empirische Gründe für die nicht-Lokalität der Quantentheorie}

Es sollen nun die konkreten empirischen Gründe genannt werden, welche die These der nicht"=Lokalität der Natur
eindeutig stützen, also die These, dass der physikalische Ortsraum mit den ihm innewohnenden Kausalstrukturen
keine fundamentale Realität der Natur sein kann, sondern nur eine als Näherung sich ergebende Realität.
Denn dies folgt in Wirklichkeit direkt aus unabweisbaren empirischen Tatsachen. Der Inhalt der folgenden
Erörterungen ist zwar eigentlich allgemein bekannt, aber bisher wurden in Bezug auf die Suche nach einer
einheitlichen Naturtheorie nicht die entsprechenden und unumgänglichen Schlüsse gezogen. Deshalb werden
hier das Doppelspaltexperiment einerseits und das EPR"=Paradoxon andererseits in ihrer Relevanz in Bezug
auf die Raum"=Frage zu Rate gezogen.\\
\\
\noindent
\textbf{Doppelspaltexperiment:} Beim Doppelspaltexperiment wird eine Laser"=Licht"=Quelle vor eine
Blende mit zwei Schlitzen gestellt und dahinter befindet sich eine Photoplatte, die durch die Einwirkung
von Licht geschwärzt wird. Wenn das Experiment zunächst in der Weise durchgeführt wird, dass jeder der
beiden Spalten einzeln für sich geöffnet wird, so ergibt sich jeweils ein spezifisches Interferenzmuster.
Wenn man dann anschließend beide Spalten öffnet, so ergibt sich allerdings ein spezifisches Interferenzmuster
auf der Photoplatte, dass nicht der Überlagerung der Interferenzmuster für die beiden einzeln geöffneten
Spalte entspricht. Man kann dieses Experiment auch mit einer solch schwachen Lichtintensität durchführen,
dass das Eintreffen und Erzeugen von Schwärzungspunkten durch einzelne Photonen beobachtet werden kann,
also einzelne Photonen durch die Blende auf die Photoplatte gelangen. Nun kann man aber unter dieser
Voraussetzung in der folgenden Weise argumentieren: Wenn ein Photon sich nur durch einen der beiden
Spalte bewegt, so ist dieser Vorgang unabhängig davon, ob der andere Spalt geöffnet ist oder nicht.
Demnach dürfte sich das Interferenzmuster für die Durchführung des Experimentes bei Öffnung beider
Spalten aber nicht von demjenigen für die Durchführung des Experimentes unterscheiden, bei der nacheinander
jeweils einer der beiden Spalte geöffnet wird und der jeweils andere geschlossen bleibt. Vielmehr müsste sich im Falle,
dass beide Spalte gleichzeitig geöfffnet sind, als Interferenzmuster eine direkte Überlagerung der beiden Interferenzmuster
für die einzeln geöffneten Spalte ergeben. Tatsächlich aber unterscheidet sich das Interferenzmuster für den Fall,
dass beide Spalte gleichzeitig geöffnet sind, vom Fall, dass sie nacheinander einzeln geöffnet werden, denn es tritt
ein zusätzlicher Interferenzterm auf. Dies aber kann ja nur bedeuten, dass die Annahme, dass sich ein einzelnes Photon
entweder nur durch den einen oder nur durch den anderen Spalt bewegt, an sich grundsätzlich nicht wahr ist. Dem Photon
kann also keine Teilchenbahn zugeordnet werden und es ist delokalisiert. Gleichzeitig aber handelt es sich bei einem
Photon um ein in sich zumindest räumlich unteilbares Quantenobjekt, dass keine innere Kausalstruktur in dem Sinne trägt,
dass einzelne Teile der Welle, durch die das Photon beschrieben wird, aufeinander einwirken könnten. Das Photon besteht also
nicht aus verschiedenen Bereichen oder Teilen, in die es gedanklich zerlegt werden könnte, wie dies bei einer klassischen
Welle der Fall ist, deren einzelne Bereiche in kontinuierlicher Weise kausal aufeinander einwirken. Vielmehr handelt
es sich bei einem Photon um eine in sich im schlechthinnigen Sinne einheitliche Realität, die aber dennoch vollkommen
delokalisiert ist, zumindest im Rahmen dessen, was durch die Unbestimmtheitsrelationen definiert ist. Als ausgedehnte
Welle im Raum müsste sich ein Quantenobjekt eigentlich in unterschiedliche kausal eigenständige Teile zerlegen lassen,
aber dies ist in Wirklichkeit nicht der Fall, da es ein Quantum der Wirkung ist und daher keine kausale Substruktur in
sich trägt. Diese Eigenschaft bezeichnet Niels Bohr mit dem Begriff der Individualität. Und hieraus ergibt sich zwingend,
dass die Eigenschaft der Räumlichkeit nur eine indirekte Art der Darstellung ist, aber dem Quantenobjekt an sich
definitiv nicht zukommt.\\
\\
\noindent
\textbf{EPR-Paradoxon:} Im Jahre 1935 erdachte Albert Einstein gemeinsam mit seinen beiden Kollegen Boris Podolsky und
Nathan Rosen ein Gedankenexperiment, das von zentraler Bedeutung für das Verständnis der eigentlichen Natur der Realität
gemäß der Quantentheorie ist, und dessen Gehalt als EPR"=Paradoxon Berühmtheit erlangte \cite{Einstein:1935}.
Hierbei werden zwei Quantenobjekte betrachtet, die miteinander in Wechselwirkung treten und sich anschließend
weit voneinander wegbewegen, zumindest auf der räumlichen Oberflächenebene unserer Betrachtung. Einstein zeigte nun,
dass wenn man eine Messung des Ortes des ersten Teilchens durchführt, aus der Quantentheorie folgt, dass auch der
Ort des zweiten Teilchens exakt bestimmt sein muss, und dass wenn man eine Messung des Impulses des ersten
Teilchens vornimmt, zugleich auch der Impuls des zweiten Teilchens exakt bestimmt sein muss. Da aber gemäß
dem Postulat der lokalen Kausalität, demgemäß sich Wirkungen mit maximal Lichtgeschwindigkeit durch die
Raum"=Zeit ausbreiten, die Messung am ersten Teilchen das zweite Teilchen nicht instantan beeinflussen kann,
so scheint hieraus zu folgen, dass der Ort und der Impuls des zweiten Teilchens gleichzeitig exakt bestimmt
sein müssen. Dies aber wäre eine direkte Verletzung der Heisenbergschen Unbestimmtheitsrelation zwischen Ort
und Impuls, $\Delta x \Delta p \geq \frac{\hbar}{2}$, und damit wäre gezeigt, dass die Quantentheorie
in sich selbst inkonsistent und damit als fundamentale Beschreibung der Natur ungeeignet ist.
Einsteins Argumentation ist unter der Voraussetzung vollkommen konsistent und unausweichlich,
dass man das Prinzip der lokalen Kausalität als wahr unterstellt, also dass sich Wirkungen in
einem feldtheoretischen Sinne mit maximal Lichtgeschwindigkeit durch den Raum bewegen. Was Einstein
aber in der ihm eigenen argumentativen Härte und Stringenz eigentlich zeigte, das ist nicht
die Inkonsistenz der Quantentheorie, sondern dass die Quantentheorie mit dem Prinzip der lokalen
Kausalität nicht vereinbar ist. Und das bedeutet, dass es genau zwei Möglichkeiten gibt, die sich
allerdings gegenseitig ausschließen, was bedeutet, dass wenn die eine wahr ist, die andere notwendig
unwahr sein muss:\\
\\
\textbf{1) Das Prinzip der lokalen Kausalität ist gültig.}\\
\\
\textbf{2) Die Quantentheorie ist in sich konsistent und der ihr innewohnende nicht"=lokale
Charakter beschreibt die Natur korrekt.}\\
\\
Niels Bohr antwortete bereits im gleichen Jahr mit einer Arbeit, in welcher er die Quantentheorie
verteidigte \cite{Bohr:1935}. Mittlerweile wurde das Einsteinsche Gedankenexperiment aber in anderer
Weise vielfach realisiert, meistens in Bezug auf den Spin der Teilchen als Messgröße. Hierbei spielen
die Betrachtungen Bells eine wichtige Rolle \cite{Bell:1964}. Und diese Experimente entschieden bisher
immer zu Gunsten der Quantentheorie. Es besteht also eine direkte Korrelation zwischen
dem Messergebnis zweier aus einer räumlichen Perspektive betrachtet weit voneinander entfernter
Quantenobjekte, was bedeutet, dass der Zustand des einen Objektes durch den Vorgang der Messung
am anderen Objekt instantan beeinflusst wird. Genaugenommen handelt es sich eigentlich gar nicht
um zwei voneinander getrennte Objekte, sondern um ein einziges Objekt, dass in der Wahrnehmung
nur künstlich als in zwei Objekte aufgespalten erscheint. Dies aber bedeutet nichts anderes,
als dass das Prinzip der lokalen Kausalität, auf dem alle Feldtheorien basieren, zu Gunsten
einer nicht"=lokalen quantentheoretischen Beschreibungsweise der Realität aufgegeben werden
muss und die räumliche Wirklichkeit nur einer indirekten äußeren Darstellung des Geschehens
in der Natur entspricht. Dies gilt zumindest auf der fundamentalen Ebene. Zu einem Kulminationspunkt
gelangte diese Art des Experimentes schließlich bei Anton Zeilinger, dem es gelang, auf diese Weise
quantentheoretische Zustände zu teleportieren \cite{Zeilinger:1999},\cite{Zeilinger:2008}.

\subsection{Mathematische Gründe für die nicht-Lokalität der Quantentheorie}

Die allgemeinste und abstrakteste Formulierung der Quantentheorie im Sinne Paul Adrien Maurice Diracs
und Johann von Neumanns \cite{Dirac:1958},\cite{Neumann:1932} als einer Theorie des Hilbert"=Raumes
setzt keinerlei konkrete physikalische Begriffe voraus, weder einen Massenbegriff, noch einen
gewöhnlichen Wechselwirkungsbegriff, noch räumlich"=feldtheoretisch definierte Objekte oder
überhaupt einen physikalischen Ortsraum. In dieser abstrakten Fassung basiert die Quantentheorie
lediglich auf den hoch abstrakten Begriffen des Zustandes als Vektor und der Observable als
Operator in einem Hilbert"=Raum. Alle anderen Begriffe werden in der gewöhnlichen
Elementarteilchenphysik nur über das der Natur der Quantentheorie eigentlich vollkommen fremde
feldtheoretische Denken in die Beschreibung der Natur gebracht, also über klassische Theorien,
auf die man das abstrakte mathematische Schema der Quantentheorie erst nachträglich durch
den Vorgang der Quantisierung überträgt. Aber die Quantentheorie an sich selbst ist davon
in keiner Weise abhängig. Einzig und alleine die Zeit als fundamentalstem Begriff der
Natur und des menschlichen Denkens muss auch in der Quantentheorie erhalten bleiben.
Denn auch in der Kantischen Philosophie kommt der Zeit fundamentalerer Charakter zu als
dem Raum, da die Zeit als Grundform der inneren Anschauung im Gegensatz zum Raum als Grundform
der äußeren Anschauung sogar zur seelischen Welt der Empfindungen gehört. Die Zeit liegt sowohl
allen seelischen Phänomenen alsauch allen Phänomenen in der Natur zu Grunde und ihr muss daher ein
Sonderstatus beigemessen werden. Daran ändert auch die spezielle Relativitätstheorie nichts,
welche den Unterschied zwischen Zeit und Raum nicht aufhebt, sondern diese beiden
wesensfremden Realitäten nur formal zu einer Raum"=Zeit verbunden zu beschreiben gestattet.
Aber auch hier läuft die Zeit nur in eine Richtung und die Raum- beziehungsweise
Zeitartigkeit des Abstandes zweier Ereignisses ist eine Lorentz"=invariante Größe.
Zudem weist ein physikalisches Objekt bezüglich der Raum"=Zeit immer nur drei
Freiheitsgrade auf, da der vierte über die dynamische Grundgleichung weggenommen wird.
Es gibt also faktisch immer eine zeitliche Bewegung, die etwas anderes ist als eine räumliche Ausdehnung,
nur dass diese zeitliche Bewegung sich von unterschiedlichen Bezugssystemen aus anders darstellt.
Aber alle anderen gewöhnlichen physikalischen Begriffe im klassischen Sinne außer der Zeit spielen
in der abstrakten Quantentheorie überhaupt keine Rolle mehr. Werner Heisenberg formulierte die These,
dass man einen bestimmten Bereich der Natur dann verstanden habe, wenn man die richtigen Begriffe
gefunden habe, mit denen man ihn beschreiben muss. Aber er meinte, dass das Schwierigste bei
diesem Prozess des Übergangs zu neuen Begriffen eigentlich nicht das Auffinden der neuen
Begriffe sei, sondern die gedankliche Loslösung von den alten Begriffen. In diesem Sinne scheint
es mir nahe zu liegen, dass eine der zentralen Herausforderungen bei der Suche nach einer einheitlichen
Beschreibung der Natur, in der die Quantentheorie, die verschiedenen Objekte und deren Wechselwirkungen
in der Elementarteilchenphysik und die allgemeine Relativitätstheorie eine wirkliche Synthese eingehen,
in der radikalen Überwindung klassischer Begriffe besteht und nur noch die Quantentheorie in ihrer
abstrakten Begrifflichkeit nicht aber irgendwelche klassischen Relikte feldtheoretisch"=mechanistischen
Denkens enthalten sind. Aber genau dies leistet die Quantentheorie der Ur"=Alternativen. Denn sie versucht
auch in ihrer physikalischen Begriffsbildung den rein abstrakt"=logischen Charakter als eigentlicher Essenz
der Quantentheorie herauszuarbeiten wie er in der Dirac"=von Neumannschen allgemeinen Formulierung der
Quantentheorie deutlich zum Vorschein kommt. In diesem Zusammenhang sollte auch darauf hingewiesen werden,
dass eine Quantenzahl, durch welche nicht nur der Spin sondern auch die sogenannten internen
Freiheitsgrade wie Isospin und Farbe beschrieben werden, ein sehr viel abstrakterer rein
quantentheoretischer Begriff ist und von jedem feldtheoretisch oder räumlich definierten Begriff in
grundsätzlicher Weise unterschieden ist. In der gewöhnlichen Elementarteilchenphysik herrscht also
bereits eine Dualität aus einer rein quantentheoretischen Begrifflichkeit, wie sie mit den
Quantenzahlen auftritt und einer primär feldtheoretischen Denkweise, indem man diese als
Eigenschaften von räumlich definierten Objekten ansieht. Es ist also schon um der diesbezüglichen
begrifflichen Einheit willen davon auszugehen, dass auch die kontinuierliche räumliche Realität
letztendlich auf eine rein quantentheoretische Realität zurückgeführt werden muss. Umgekehrt kann
es schon deshalb nicht gehen, weil Quantenzahlen abstrakter sind als konkrete räumliche Objekte und
eine begriffliche Rückführung und Einordnung, die das eigentliche Wesen aller strukturellen
Erkenntnis darstellt, immer nur in der Richtung einer Begründung des Konkreten aus dem Abstrakten
geschehen kann, bei dem konkrete Begriffe unter abstrakten Begriffen zusammengefasst werden.
Gemäß Eugene Wigner werden zwar Elementarteilchen bereits rein mathematisch durch Symmetrien
charakterisiert \cite{Wigner:1939},\cite{Weinberg:1995}, als irreduzible Darstellungen der Poincare"=Gruppe,
also als etwas sehr Abstraktes. Aber diese sind wie die Dreiecke in Platons Timaios noch auf die Raum"=Zeit
bezogen und damit zumindest indirekt geometrisch definiert. Die abstrakten Ur"=Alternativen in der
von Weizsäckerschen Theorie hingegen basieren nur auf reiner quantentheoretischer Logik als
fundamentaler Realität, die der Natur letztendlich zu Grunde liegt.

\subsection{Die Kopernikanische Wende bezüglich der Raum-Frage als Konsequenz}

Die in den letzten beiden Unterabschnitten dargelegten Argumente führen also unausweichlich zu der Einsicht, dass die
physikalische Realität auf der basalen Ebene nicht"=räumlich ist. Unsere Vorstellung ist aber gemäß Immanuel Kant an
die a priori in unserem Geist angelegte Anschauungsform des Raumes gebunden. Dies ist der Grund, warum es gemäß der
Kopenhagener Deutung der Quantentheorie notwendig ist, alle Experimente in klassischen Begriffen zu beschreiben,
einfach weil unser Denken und unsere Wahrnehmung an sie gebunden ist. Die Grenze dieser klassischen,
also räumlich"=feldtheoretischen, Beschreibungsweise in Bezug auf die Realität an sich ist aber durch
die Unbestimmtheitsrelation exakt definiert. Wenn wir allerdings basierend auf der Räumlichkeit versuchen,
quantentheoretische Objekte und Vorgänge darzustellen und zu verstehen, so kommt es zu den bekannten Paradoxien.
Die Auflösung dieser Paradoxien besteht also in der Erkenntnis, dass die Quantentheorie von einer Realitätsebene handelt,
auf der räumlich"=feldtheoretische Begriffe noch überhaupt keine Gültigkeit besitzen. Dies aber bedeutet, dass auch die
fundamentalen Objekte noch nicht an diese Begrifflichkeit gebunden sein müssen und können. Da es aber auf der
Oberflächenebene den Raum mit seinen lokalen Kausalbeziehungen wirklich gibt, muss sich diese physikalische
Struktur als Näherung indirekt im Sinne einer Konsequenz ergeben. Und ebendies führt zur zentralen und alles
entscheidenden Kopernikanischen Wende in Bezug auf die Raum"=Frage, die in folgender Weise formuliert werden kann:\\
\\
\fbox{\parbox{163 mm}{\textbf{Nicht geometrisch definierte Objekte befinden sich in einem vorgegebenen Raum mit einer
lokalen Kausalstruktur, sondern abstrakt"=logische Objekte konstituieren umgekehrt die Existenz des Raumes, der sich als
eine bestimmte Art der Darstellung indirekt als Konsequenz ergibt, dem aber keine fundamentale Natur zukommt.}}}\\
\\
Diese Kopernikanische Wende als Konsequenz der nicht"=Räumlichkeit der Natur auf basaler Ebene gemäß der Quantentheorie,
wie sie in diesem Abschnitt eingehend thematisiert wurde, wird durch den Begriff der Ur"=Alternative in konsequenter
Weise ausgedrückt. Denn die Ur"=Alternativen als fundamentalen Informationseinheiten setzen den Raumbegriff noch
nicht voraus, sondern nur die reine Logik und befinden sich daher nicht im Raum. Sie ermöglichen aber umgekehrt,
wie später in dieser Arbeit gezeigt werden wird, die Begründung der Existenz eines dreidimensionalen reellen Raumes,
der über die Dynamik dann mit der Zeit zu einer reellen (3+1)"=dimensionalen Raum"=Zeit als Darstellungsmedium der
Zustände vieler Ur"=Alternativen erweitert werden kann. Damit kann die Existenz von Objekten in einer Raum"=Zeit
also aus den Ur"=Alternativen und damit aus der abstrakten Quantentheorie aufgefasst als einer Theorie der Information
in der Zeit mathematisch begründet werden. Dies ist der entscheidende Grund, warum die Quantentheorie der Ur"=Alternativen
rein begrifflich der überzeugendste bisher existierende Ansatz zu einer einheitlichen Naturtheorie ist. Denn hier ist die
Kopernikanische Wende bezüglich der Raumfrage wirklich bezüglich der grundlegenden Begrifflichkeit enthalten. Deshalb soll
sich im nächsten Abschnitt dem Programm Carl Friedrich von Weizsäckers zugewandt werden, die Quantentheorie als eine
Theorie logischer Alternativen zu verstehen. In diesem Zusammenhang ist es aber wichtig, darauf hinzuweisen, dass das
wirklich klassische und feldtheoretische Element, und damit wohl auch die Wurzel der Entstehung der unendlichen Werte,
im Rahmen relativistischer Quantenfeldtheorien eigentlich erst mit der Definition der Wechselwirkung über punktweise
Produkte von Feldern zurück in die Beschreibung der Natur gelangt. Denn eine quantisierte freie Feldtheorie entspricht
bekanntlich der Quantenmechanik vieler Teilchen. Diese enthält aber die nicht"=Lokalität zumindest implizit über die
Unbestimmtheitsrelation, wenn auch vielleicht in begrifflich nicht ganz konsequenter Weise. Aus diesem Grunde wird
es vor allem entscheidend sein, gerade den Wechselwirkungsbegriff basierend auf den Ur"=Alternativen in einer
abstrakten rein quantentheoretischen Weise zu fassen.

\section{Rekonstruktion der abstrakten Quantentheorie}

\subsection{Das Programm und der Bezug zu anderen Ansätzen}

Das Programm der Rekonstruktion der Quantentheorie als einer einheitlichen Naturtheorie des Carl Friedrich von Weizsäcker
basiert auf dem rein logischen Begriff der Alternative als fundamentalster Darstellung der Realität der Natur in unserem
Geist und besteht grundsätzlich aus zwei Schritten, die man in der folgenden Weise charakterisieren kann:\\
\\
\noindent
\textbf{1) Zunächst muss die allgemeine Quantentheorie in ihrer abstrakten Gestalt als Theorie des Hilbert"=Raumes gemäß
Paul Adrien Maurice Dirac und Johann von Neumann \cite{Dirac:1958},\cite{Neumann:1932} begründet werden. Dies geschieht
über den Begriff der abstrakten logischen Alternative in der Zeit als fundamentalster möglicher Darstellung und
Schematisierung physikalischer Realitäten in unserem Geist. Die in diesem abstrakten Sinne verstandene Quantentheorie
ohne zusätzliche spezielle Annahmen einer weiteren Theorie der Objekte oder des physikalischen Ortsraumes wird dann
als der einheitliche Rahmen zur Beschreibung der Einheit der Natur postuliert.}\\
\\
\noindent
\textbf{2) Aus der abstrakten Quantentheorie als Theorie logischer Einheiten in der Zeit, also letztendlich aus
quantentheoretisch verstandener Information, muss dann die konkrete Physik mit der Existenz ihrer speziellen Objekte
und deren Dynamik und Wechselwirkungen einschließlich des Raumes begründet werden. Dies geschieht mit Hilfe der
logischen Möglichkeit der Aufspaltung jeder Alternative in eine Kombination binärer Alternativen, die in ihrer
grundsätzlichen Rolle als den logisch fundamentalsten Objekten der Naturbeschreibung als Ur"=Alternativen
bezeichnet werden.}\\
\\
\noindent
Carl Friedrich von Weizsäcker erdachte, entwickelte und formulierte dieses grundlegende Programm vor allem in \cite{Weizsaecker:1971},\cite{Weizsaecker:1985},\cite{Weizsaecker:1992},\cite{Weizsaecker:1955},\cite{Weizsaecker:1958},\cite{WeizsaeckerScheibeSüssmann:1958},\cite{Weizsaecker:1974},\cite{Weizsaecker:1978},\cite{Weizsaecker:1984}.
Teilweise Darstellungen beziehungsweise Weiterentwicklungen des von Weizsäckerschen Programmes sind beispielsweise zu finden in \cite{Castell:1974},\cite{Kuenemund:1984},\cite{Goernitz:1986},\cite{Goernitz:1987},\cite{Drieschner:1988},\cite{Goernitz:1992},\cite{Lyre:1994},\cite{Lyre:1995},\cite{Lyre:1996},\cite{Lyre:1998},\cite{Drieschner:2002},\cite{Goernitz:2002},\cite{Castell:2003},\cite{Goernitz:2010},
\cite{Kober:2010},\cite{Kober:2011},\cite{Goernitz:2012},\cite{Goernitz:2014},\cite{Goernitz:2016}.

Die grundlegende Idee des ersten Schrittes des Programms des Carl Friedrich von Weizsäcker basiert auf dem
Kantischen Gedanken, dass die grundlegenden Naturgesetze sich aus den Bedingungen der Möglichkeit von Erfahrung
ergeben. Die hierin enthaltene Argumentation besteht darin, dass man nur dann wirklich sicher sein kann,
dass Naturgesetze grundsätzlich in der Erfahrung gelten, wenn sie eine notwendige Bedingungen der Möglichkeit
von Erfahrung sind, Erfahrung ohne sie also gar nicht möglich wäre. Die universelle Gültigkeit der
allgemeinen Quantentheorie soll sich demnach also aus den Bedingungen der Möglichkeit von Erfahrung ergeben.
Dies geht insofern noch über Kant hinaus als Kant nur den grundlegenden begrifflichen Rahmen nicht
aber die Einzelgesetze aus den Bedingungen der Möglichkeit von Erfahrung begründen wollte. Raum, Zeit
und Kausalität lassen sich also gemäß Kant als Bedingungen der Möglichkeit von Erfahrung deuten, da sie ganz
grundlegend und allgemein sind. Spezielle Gesetze wie das Gravitationsgesetz aber müssen zusätzlich
durch spezielle Erfahrung in unseren Geist gelangen. Bei von Weizsäcker hingegen sollen sich auch alle
Einzelheiten letztendlich aus der abstrakten Quantentheorie ergeben, wenn sie ersteinmal aus den
grundlegenden Bedingungen der Möglichkeit von Erfahrung konstruiert ist. Dies ist natürlich aufgrund
des extrem hohen Anspruches, den dieses Programm in sich enthält, bisher noch nicht annähernd erreicht worden.
Die Rekonstruktion der abstrakten Quantentheorie kann man basierend auf einer Reihe von Postulaten recht
gut vollziehen, wenngleich zumindest nicht von allen diesen Postulaten zwingend argumentativ gezeigt
werden kann, dass es sich um Bedingungen der Möglichkeit von Erfahrung handelt. Aber die Begründung
der konketen Physik aus der abstrakten Quantentheorie über die Ur"=Alternativen ist eben doch ein
ungeheures Programm, das bisher nur ansatzweise vollzogen werden konnte. Was man aber immerhin schon
herleiten kann, das ist die Existenz eines dreidimensionalen reellen Raumes, der über die in der
abstrakten Form der Schrödinger"=Gleichung enthaltene allgemeine Dynamik der Quantentheorie zudem
mit der Zeit zu einer (3+1)"=dimensionalen Raum"=Zeit verbunden werden kann. Dies soll in dieser
Arbeit in vielleicht konsequenterer Weise als bisher geschehen, indem nicht der Zustandsraum einzelner
Ur"=Alternativen als $\mathbb{S}^3$, also als dreidimensionale Sphäre, mit dem Ortsraum identifiziert wird,
sondern die Zustände im Tensorraum vieler Ur"=Alternativen, in dem Erzeugungs- und Vernichtungsoperatoren wirken,
als Funktionen in einem dreidimensionalen reellen Ortsraum dargestellt werden, was über die Definition eines
entsprechenden Hamilton"=Operators die formale Beschreibung in einer (3+1)"=dimensionale Raum"=Zeit gemäß
der speziellen beziehungsweise der allgemeinen Relativitätstheorie zulässt. Damit ist gezeigt,
dass Zustände aus vielen Ur"=Alternativen sich indirekt als Objekte in der Raum"=Zeit darstellen. Dies ermöglicht
grundsätzlich die Darstellung jedes beliebigen dynamischen Vorganges, der sich auf Objekte bezieht,
die aus Ur"=Alternativen aufgebaut sind, in einer (3+1)"=dimensionalen Raum"=Zeit, die sich damit
also argumentativ zwingend als direkte Konsequenz der abstrakten Quantentheorie als einer Theorie
der Information ergibt. Bei der weiteren Begründung der Existenz der speziellen Objekte,
also Elementarteilchen mit zusätzlichen konkreten Attributen wie Spin, Masse und inneren
Symmetrien sowie deren Wechselwirkungen, sollte man zwar als endgültiges Ziel die von Weizsäckersche
Ambition nicht vergessen. Man sollte sich aber wohl zunächst auf den immer noch sehr grundsätzlichen aber nicht
ganz so ambitionierten Versuch beschränken, die real existierenden Objekte und Wechselwirkungen durch
Ur"=Alternativen auszudrücken, also Strukturen aus Ur"=Alternativen zu konstruieren, die näherungsweise
zu den konkreten Objekten und Wechselwirkungen in der Raum"=Zeit führen ohne aber schon begründen zu können,
dass ausschließlich diese Strukturen möglich sind. Und damit kann man dann in der Tat zu einer rein
quantentheoretischen auf Ur"=Alternativen basierenden Beschreibung der Dynamik freier Objekte, der inneren
Symmetrien und auch zumindest eines ersten Modells des Elektromagnetismus und der Gravitation gelangen,
was in den späteren Abschnitten entwickelt wird. Die Abbildung von Zuständen vieler Ur"=Alternativen in die
Raum"=Zeit, der Versuch einer Bildung eines rein auf Ur"=Alternativen sich gründenden abstrakt"=quantentheoretischen
Wechselwirkungsbegriffes, die Einbindung innerer Symmetrien über die Betrachtung von oktonionischen Strukturen
im Rahmen der Ur"=Alternativen und die Konstruktion eines ersten Modells zur Beschreibung der elektromagnetischen
sowie der gravitativen Wechselwirkung im begrifflichen Rahmen der Quantentheorie der Ur"=Alternativen stellen
meinen eigenen spezifischen neuen Beitrag zur von Weizsäckerschen Theorie dar, der im Rahmen dieser
Arbeit behandelt wird.

Die Idee des Raum"=Relationalismus im Sinne der Einsteinschen allgemeinen Relativitätstheorie ist auch in der sehr
vielversprechenden und mathematisch auf sehr hohem Niveau formulierten Schleifenquantengravitation bereits
verwirklicht, die seitens Carlo Rovelli und Lee Smolin entwickelt wurde \cite{Rovelli:2004},\cite{Rovelli:1989},\cite{Rovelli:1994},\cite{Rovelli:1995}. Aber hier geht man trotzdem noch von der allgemeinen Relativitätstheorie als einer Feldtheorie aus, wenn auch in
einer anderen mathematischen Formulierung mit Spin"=Konnektionen und Holonomien, und überträgt dann die Mathematik
der Quantentheorie auf diese feldtheoretische Denkweise. Und auch die Spin"=Netzwerke \cite{Rovelli:2004},\cite{Rovelli:1995}
gehen von einem zwar abstrakten Netz von Punkten aus, das aber an sich selbst nun eine quantentheoretische Beschreibung
der Raum"=Zeit liefern soll, während im Ansatz Carl Friedrich von Weizsäcker auf der fundamentalen Ebene
überhaupt keine Raum"=Zeit existiert, auch nicht in einem diskretisierten Sinne oder im Sinne eines Netzwerkes.
Vielmehr besteht die Natur, welche überhaupt nur in Bezug auf unseren menschlichen Geist beschrieben
werden kann, nur aus abstrakter quantentheoretischer Information. Die Idee, die physikalische
Realität basierend auf dem Begriff der Quanteninformation zu gründen, wird in anderer Weise
auch seitens Fotini Markopoulou und einiger ihrer Kollegen in den folgenden Arbeiten vertreten
\cite{Konopka:2006},\cite{Konopka:2008},\cite{Hamma:2009},\cite{Caravelli:2011},\cite{Caravelli:2012},\cite{CaravelliMarkopoulou:2012},\cite{Markopoulou:2012}. Allerdings werden dort die philosophischen Grundlagen nicht analysiert.
Zudem wird die Idee auch hier nicht in der gleichen begrifflichen Stringenz formuliert. Denn auch
hier geht man zumindest noch von einem Netz abstrakter Punkte aus, die Information enthalten und austauschen,
während sich in der von Weizsäckerschen Theorie der physikalische Informationsbegriff nur noch aus den Ur"=Alternativen
selbst konstituiert. Alle anderen Beziehungsstrukturen müssen sich in der Quantentheorie der Ur"=Alternativen
hieraus ergeben. In \cite{Damour:2007} wird die Idee, dass die Raum"=Zeit sich aus einer fundamentaleren
Realität konstituieren könnte, im Rahmen eines weiteren mathematisch sehr anspruchsvollen Ansatzes diskutiert.

Die Twistor"=Theorie, die von Roger Penrose stammt \cite{Penrose:1960},\cite{Penrose:1977},\cite{Penrose:1985},\cite{Penrose:1986},
macht zwar wie die Quantentheorie der Ur"=Alternativen von der Tatsache Gebrauch, dass eine direkte mathematische
Beziehung, eine Isomorphie zwischen Raum"=Zeit"=Vektoren und Spinoren besteht, sodass sie aufgrund der Mathematik
zunächst sehr verwandt wirkt. In Wirklichkeit aber geht es hier nur um eine mathematische Reformulierung
der allgemeinen Relativitätstheorie als klassischer Feldtheorie, indem man die Raum"=Zeit"=Vektoren in eine
Schreibweise basierend auf Spinoren überführt. In der von Weizsäckerschen Theorie hingegen stellen die Spinoren
eine mathematische Beschreibung quantentheoretischer Informationseinheiten dar, die überhaupt nicht auf
eine Raum"=Zeit bezogen sind und deren Zustände sich dann lediglich in einer Raum"=Zeit darstellen lassen.

Bezüglich der philosophischen Grundintention besteht eine gewisse Verwandschaft der von Weizsäckerschen
Theorie zur Philosophie Ludwig Wittgensteins \cite{Wittgenstein:1921}, der ebenfalls davon ausgeht,
dass so etwas wie elementare Tatsachen fundamental sind.

\subsection{Postulate einer Quantentheorie logischer Alternativen}

Es sollen nun die grundlegenden Postulate als die Bedingungen der Möglichkeit von Erfahrung
vorgestellt und erläutert werden, auf denen das ganze von Weizsäckersche Programm basiert.
Die Logik ist eine absolute Vorbedingung für die Möglichkeit von Erkenntnis. Um welche Art von Logik
es sich handelt, ist dabei noch nicht definiert, aber es muss ein abstraktes strukturiertes
Schema geben, das als eine Art grundlegendes Medium einer formalen strukturierten Argumentation und
Erkenntnis fungiert und erlaubt, Aussagen in eine sinnvolle Beziehung zueinander zu setzen. Zudem ist
die Zeit fundamental, denn Erfahrung machen bedeutet aus der Vergangenheit für die Zukunft lernen.
Überhaupt kann sich nur irgendetwas ereignen, etwas passieren, das man beobachten kann, wenn grundsätzlich
Zeit ist. Die Quantentheorie setzt als fundamentale Entitäten daher letztlich nur die auf einem rein
logischen Grundschema basierende Information sowie die Zeit voraus. Natürlich erhalten damit Logik und
Information zugleich eine über eine rein epistemologische Rolle hinausgehende ontologische Bedeutung.
Die grundlegendste und fundamentalste Art und Weise, über eine beliebige empirisch untersuchbare
oder überhaupt rein logisch erfassbare Realität Wissen besitzen zu können, drückt sich in einer
Alternative aus, welche die Information über diese Realität enthält. Nun kann man aber noch
weiter gehen und die Realität so abstrakt und grundsätzlich auffassen, dass sie überhaupt
nur eine Beziehungsstruktur aus abstrakter Information darstellt. Dies aber ist gleichbedeutend
mit dem Postulat der Identität des Objektbegriffes mit dem Informationsbegriff. Um in einer
solchen Informationsstruktur etwas erkennen zu können, muss man eine einzelne Alternative
zumindest näherungsweise in dem Sinne isolieren können, dass man sie unabhängig von anderen
Alternativen definieren und entscheiden kann. Der Preis für diese Näherung ist dann die Einführung
der Wechselwirkung als Konsequenz der Brechung einer einheitlichen Wirklichkeit durch eine aus
einem Vorgang der Separation hervorgehenden logischen Beschreibungsweise im Sinne der Darstellung
und Aufspaltung in Alternativen. Wenn man weiter davon ausgeht, dass die Welt endlich ist, so ergibt sich,
dass es grundsätzlich nur endliche Alternativen geben kann, also Alternativen mit endlich vielen Elementen,
obwohl es zunächst keine konkret definierbare Obergrenze für die Zahl der Elemente einer Alternative gibt.
Wenn man nun die Struktur der Zeit hineinbringt, so ist das tertium non datur nicht mehr erfüllt,
das besagt, dass eine Aussage entweder wahr oder falsch ist, denn Aussagen, die sich auf die
Zukunft beziehen, müssen im Allgemeinen nicht determiniert sein. Allerdings kann man das tertium
non datur als ein Postulat der Quantenlogik auch unabhängig von der Zeit einführen. Von Weizsäcker
selbst begründete die Verletzung des tertium non datur explizit über die Struktur der Zeit mit
Vergangenheit, Gegenwart und Zukunft. Allerdings erhält man dann eine begriffliche Schwierigkeit.
Denn die Zeit taucht in der Quantentheorie ja zugleich als ein kontinuierlicher Parameter auf,
über den auch die Dynamik gemäß der allgemeinen Schrödinger"=Gleichung definiert ist, welche eine
deterministische Struktur der Zeit impliziert. Obwohl mir diese Begründung der Quantenlogik über
die Zeit insofern als sehr feingeistig und tiefgründig erscheint, als man hier so viel wie möglich
nur aus der Zeit selbst als der fundamentalsten Realität der Natur und des menschlichen Geistes
zu begründen versucht, glaube ich, dass es aufgrund der dann auftretenden Dualität des Zeitbegriffes
zunächst besser ist, die Quantenlogik als von der Zeit unabhängiges zusätzliches Postulat
zu behandeln. Jedenfalls führt die Quantenlogik auf Wahrheitswerte für die Elemente der Alternativen,
die dann faktisch Wahrscheinlichkeiten dafür darstellen, dass die einzelnen Elemente der Alternative,
die sich kontinuierlich mit der Zeit entwickeln, bei einer Messung aufgefunden werden. Basierend auf
diesen Überlegungen kann man die Grundpostulate in der folgenden Weise kategorisieren:\\
\\
\noindent
\textbf{A) Postulat der Alternativen als basaler Realität:} Eine logische Alternative
besteht aus $N$ Möglichkeiten, bei denen alle anderen Möglichkeiten falsch sind,
wenn eine wahr ist. Logische Alternativen sind die fundamentalste Schematisierung
der Realität der Natur in unserem Geist und damit als deren Objekte zugleich die
basale Entität der Natur.\\
\\
\noindent
\textbf{B) Postulat des Finitismus:} Die Alternativen enthalten nur endlich viele Elemente.
Die Zahl der Elemente ist zunächst nicht begrenzt, aber endlich.\\
\\
\noindent
\textbf{C) Postulat der Trennbarkeit:} Die Alternativen können in einer gewissen Näherung
voneinander getrennt entschieden werden, was bedeutet, dass die Entscheidung einer
Alternative unabhängig von der Entscheidung aller anderen Alternativen ist. Die Korrektur
dieser Näherung definiert die Wechselwirkung zwischen den Alternativen. Eigentlich ist die
Realität holistisch aufzufassen, aber wenn man als Bedingung der Möglichkeit von Erfahrung
einen bestimmten Teil der Realität als Objekt isoliert, dann ist dies eine Näherung, deren
Begrenztheit sich indirekt als eine Wechselwirkung mit anderen Objekten darstellt.\\
\\
\noindent
\textbf{D) Postulat der Symmetrie:} Die einzelnen Elemente der Alternative sind gleichberechtigt,
innerhalb der Alternative ununterscheidbar und können daher innerhalb einer Alternative aufeinander
abgebildet werden. Eine Unterscheidung ist nur relativ zu anderen Alternativen möglich. Dies konstituiert
eine Symmetrie der Alternative und des Vektorraumes, der durch die möglichen Wahrscheinlichkeitsverteilungen
definiert ist. Bei einer Alternative mit $N$ Elementen ist dies die $SO(N)$"=Gruppe.\\
\\
\noindent
\textbf{E) Postulat der Quantenlogik:} Die logischen Alternativen in der Natur folgen einer
Quantenlogik, in welcher das tertium non datur verletzt ist, der Satz vom ausgeschlossenen
Dritten neben wahr und unwahr, was bedeutet, dass einzelne Aussagen nicht entweder wahr
oder falsch sind. Deshalb müssen den einzelnen Elementen der Alternativen Wahrscheinlichkeiten
zugeordnet werden, was einen Vektorraum aller Wahrscheinlichkeitsverteilungen als quantentheoretischen
Zuständen über diesen Alternativen definiert. Carl Friedrich von Weizsäcker interpretiert
diese Quantenlogik wie oben bereits thematisiert als eine zeitliche Logik, in welcher sich die
Offenheit der Zukunft manifestiert. Allerdings entsteht dadurch das Problem, dass dann aufgrund
der dynamischen und kontinuierlichen Zeitentwicklung der Zustände in der Quantentheorie die Zeit in
einer dualen Weise auftritt. Dies ist kein spezifisches Problem der von Weizsäckerschen
Begründung der Quantentheorie, sondern ist grundsätzlich in der Quantentheorie in
Gestalt des berühmten Messproblems enthalten. Denn in der Quantentheorie kann sich ein
Zustand auf zwei Weisen ändern: Kontinuierlich gemäß der Schrödinger"=Gleichung,
solange keine Messung durchgeführt wird, und sprunghaft, wenn eine Messung durchgeführt wird.
Aber da Messung ja Wechselwirkung mit einem Messapparat bedeutet, muss sich eigentlich auch
der Messvorgang deterministisch entwickeln, und zwar auf Basis der Schrödinger"=Gleichung des
aus Messobjekt und Messapparat zusammengesetzten Objektes. Diese Inkonsistenz lässt sich
wohl nur durch die Annahme auflösen, dass sich auch der Übergang bei einem Messprozess in Wirklichkeit
deterministisch entwickelt, man den Ausgang dieses Prozesses nur aufgrund der vielen unbekannten
mikroskopischen Freiheitsgrade des Messapparates nicht vorhersagen kann. Dies bedeutet dass
die nicht"=Determiniertheit in der Quantentheorie im Gegensatz zur in der Unbestimmtheitsrelation
sich eigentlich ausdrückenden nicht"=Lokalität keinen ontologischen Charakter hat, da auch der
Messprozess im Prinzip der Dynamik der Quantentheorie gehorchen muss, welche absolut
deterministisch ist \cite{Kober:2009A}. Dies scheint mir in Einklang mit der Behandlung
des Messproblems in \cite{Mittelstaedt:1963} zu stehen. Es wird hier offen gelassen,
ob und in welcher Weise die Quantenlogik als zeitliche Logik zu interpretieren ist.
Daher wird die Zeit hier nur separat in Bezug auf die Zeitentwicklung der
Zustände eingeführt.\\
\\
\noindent
\textbf{F) Postulat der Zeitentwicklung:} Die Alternativen unterliegen der Zeit als neben
der Struktur der Logik zweiter fundamentaler Realität im menschlichen Geist und in der Natur
und und verändern ihren Zustand kontinuierlich mit der Zeit. Die Zeit wird demnach als
kontinuierlicher Parameter in die Quantentheorie der Information eingeführt, der mit einer
einparametrigen Untergruppe der Symmetriegruppe der Alternativen in Zusammenhang steht.
Dies führt als möglicher Darstellung zu einem komplexen Vektorraum mit einer durch eine
Schrödinger"=Gleichung definierten $U(1)$"=Gruppe, welche die Zeitentwicklung determiniert.\\
\\
\noindent
\textbf{G) Postulat der Aufspaltung in Ur"=Alternativen:} Jede Alternative kann als das
Cartesische Produkt von Subalternativen und die entsprechenden Vektorräume der Wahrscheinlichkeiten
können dementsprechend als das Tensorprodukt von Untervektorräumen dargestellt werden. Dies führt
in der Konsequenz zur Möglichkeit der Darstellung aller Alternativen durch eine Kombination
zweidimensionaler Alternativen beziehungsweise aller komplexen Wahrscheinlichkeitsvektoren
durch zweidimensionale Spinoren. Diese prinzipielle Grenze der logischen Teilbarkeit
konstituiert die Ur"=Alternativen als die fundamentalsten Objekte in der Natur im Sinne
schlechthinniger Unteilbarkeit und damit als die eigentlichen Atome. Damit wird auch das
grundsätzliche schon bei Kant analysierte Problem umgangen, dass kleinste räumliche
Objekte in sich nicht begrifflich konsistent formuliert werden können.\\
\\
\noindent
\textbf{Erläuterung zu den Postulaten:} Diese Postulate setzen außer der Zeit und einem
abstrakten Informationsbegriff, der sich aus einer abstrakten logischen Struktur begründet
und in den Alternativen darstellt, keinerlei physikalische Entität oder darauf bezogene
Begriffe voraus, also weder den Raumbegriff, noch einen von den Alternativen getrennten Objektbegriff,
damit also auch keine räumlich definierten Objekte oder überhaupt irgendeinen feldtheoretischen
oder mechanistischen Begriff. Die Realität der Natur ist damit rein abstrakt quantentheoretisch
logisch und demnach geistig definiert. Die Alternativen enthalten nicht Information über eine davon
unabhängig existierende Realität der Natur, sie existieren auch nicht in einem vorgegebenen Raum,
sondern sie sind selbst die basalste Art der Darstellung der Realität der Natur in unserem Geist.
Alle gewöhnlichen physikalischen Begriffe und Realitäten außer der Zeit müssen sich als Konsequenz
der logischen Alternativen und der sich auf sie beziehenden obigen Postulate nachträglich erst ergeben.

\subsection{Alternativen als Darstellung einer abstrakten Quantenlogik}

Ausgehend von dem Begriff einer einfachen und zumindest näherungsweise für sich selbst entscheidbaren
logischen Alternative und basierend auf der Quantenlogik und der Zeitentwicklung als zentralen Postulaten,
die vielleicht miteinander in Zusammenhang stehen, kann nun zunächst die allgemeine Quantentheorie
konstruiert werden. Hierzu soll zunächst der Begriff einer empirisch entscheidbaren Alternative
exakt definiert werden:\\
\\
\fbox{\parbox{163 mm}{\textbf{Definition einer Alternative: Eine $N$"=fache empirisch entscheidbare
Alternative ist ein Tupel aus $N$ Möglichkeiten, bei der dann, wenn eine der Möglichkeiten wahr ist,
alle anderen falsch sind und von denen sich bei einer empirischen Prüfung genau eine als wahr erweist.}}}\\
\\
Es sei also eine Alternative $a$ als Tupel bestehend aus $n$ Möglichkeiten definiert:

\begin{equation}
a=\left(a_1, ... ,a_n\right).
\label{Alternative}
\end{equation}
Wenn man nun postuliert, dass diese Alternative einer Quantenlogik unterliegt, in der das tertium non datur
verletzt ist, so kann sie dann wenn noch keine empirische Entscheidung vorgenommen wurde, in einem Zustand
sein, bei dem gar nicht entschieden ist, welche der $n$ Möglichkeiten tatsächlich wahr ist. Das bedeutet,
dass den verschiedenen Möglichkeiten reelle Wahrheitswerte zugeordnet werden, die einen Zustand in einem
$n$-dimensionalen reellen Vektorraum definieren:

\begin{equation}
\varphi\left(a\right)=\left(\varphi\left(a_1\right), ... ,\varphi\left(a_n\right)\right)
=\left(\varphi_1, ... ,\varphi_n\right).
\label{AlternativeWahrheitswerte}
\end{equation}
Natürlich muss der Vektor auf $1$ normiert werden. Dies bedeutet:

\begin{equation}
\sqrt{\sum_{j=1}^n \varphi_j^2}=1.
\end{equation}
Wenn man nun diesen durch die Wahrheitswerte $\varphi_j=\varphi\left(a_j\right)$ definierten Zustand in
der Weise interpretiert, dass diese Wahrheitswerte die Tendenz angeben, dass die Alternative bei einer
empirischen Überprüfung den entsprechenden Wert annimmt, so kann man die Quadrate der Wahrheitswerte
als Wahrscheinlichkeiten $w_j$ in Bezug auf den Ausgang einer Messung interpretieren, sodass gilt:

\begin{equation}
w_j=\varphi_j^2.
\end{equation}
Der Wahrscheinlichkeitsbegriff setzt eigentlich einen Zeitbegriff voraus wie ihn von Weizsäcker
als Basis einer zeitlichen Logik zu Grunde legt und der eine offene Zukunft impliziert. Dieser steht
aber wie im letzten Unterabschnitt bereits angesprochen wurde in einem Gegensatz zur Dynamik der
Quantentheorie, die eine deterministische Entwicklung der Zustände enthält. Und dies gilt bereits
in der gewöhnlichen Quantentheorie, ist also kein spezifisches Problem der Quantentheorie der
Ur"=Alternativen. Die quantentheoretische Logik mit der Verletzung des tertium non datur
wurde hier deshalb nicht mit der Struktur der Zeit verknüpft, die hier ausschließlich als ein mit
der Dynamik unmittelbar verbundener reeller Parameter eingeführt wird, sondern als ein eigenständiges
Postulat behandelt. Also muss auch der Wahrscheinlichkeitsbegriff in einer solchen Weise
gedeutet werden, dass er keine offene Zukunft voraussetzt. Und dies soll gemäß dem letzten Unterabschnitt
in der Weise geschehen, dass davon ausgegangen wird, dass zwar die Entwicklung aller Zustände und ihrer
Wechselwirkungen durch die Dynamik der Quantentheorie determiniert wird, aber bei einer Messung
der Zustand in einer unvorhersehbaren Weise durch die Wechselwirkung mit dem Messapparat
beeinflusst wird. Dies liegt einfach an der unglaublichen Komplexität der mikroskopischen
Freiheitsgrade eines Messapparates, der natürlich letztendlich auch aus näherungsweise
separierbaren Alternativen besteht. Demnach hat man es ganz gemäß dem zweiten Hauptsatz der
Thermodynamik faktisch mit einem irreversiblen Prozess zu tun, der aber nicht von einem
prinzipiellen Indeterminismus, sondern nur von der Komplexität des Systems herrührt.
Die Tendenz der Alternative, einen bestimmten Wert anzunehmen, wird aber umso größer sein,
je größer die Komponente des Zustandes in Bezug auf diesen Wert ist, weil der Zustand
seine Ausrichtung im Zustandsraum dann immer weniger verändern muss. In diesem Sinne
einer Übergangstendenz, die mit der Beeinflussung eines Messapparates mit unbekanntem
Ausgangszustand in Zusammenhang steht, können die Wahrscheinlichkeiten auch in einem
deterministischen Grundrahmen interpretiert werden. Es sollte eigentlich aus den bisherigen
Ausführungen deutlich geworden sein, dass eine solche deterministische Deutung der Quantentheorie
die nicht-Lokalität der Quantentheorie in keiner Weise antastet. Denn der Sinn der Unbestimmtheitsrelation
besteht eben gerade nicht in einer Aufhebung des Prinzips der Kausalität, sondern in der Aufhebung des
Prinzips der Lokalität und damit feldtheoretischer Prinzipien. Man kann nun noch ein inneres Produkt
im Vektorraum der Wahrheitswerte der Möglichkeiten der Alternativen definieren. Wenn $\varphi^A$ und
$\varphi^B$ zwei Vektoren im Vektorraum der Wahrheitswerte über den Alternativen sind, dann kann ein
inneres Produkt in der folgenden Weise definiert werden:

\begin{equation}
\langle \varphi^A|\varphi^B \rangle=\sum_{j=1}^n \varphi^A_j \varphi^B_j,
\end{equation}
wobei die Wahrscheinlichkeit $w\left(\varphi^A,\varphi^B\right)$, einen Zustand $\varphi^B$ nach einer
Messung vorzufinden, wenn ein Zustand $\varphi^A$ vorliegt, dem Betragsquadrat des inneren
Produktes entspricht:

\begin{equation}
w\left(\varphi^A,\varphi^B\right)=|\langle \varphi^A|\varphi^B \rangle|^2.
\end{equation}
Damit wird der Vektorraum zu einem Hilbert"=Raum.

\subsection{Zeitentwicklung und Dynamik freier Alternativen}

Nun setzt aber nicht nur der Begriff der Wahrscheinlichkeit und die Möglichkeit einer Messung den Begriff
der Zeit voraus, sondern die Zeit und die mit ihr sich vollziehende Veränderung der physikalischen Realität
sind eine unabweisbare Erfahrungstatsache. Zudem ist ein Zeitfluss die Grundvoraussetzung dafür, dass überhaupt
irgendetwas geschehen kann. Ohne Zeit ist die Welt nur ein toter existierender Zustand ohne
Leben und Dynamik. Gemäß der Kantischen Erkenntnistheorie ist die Zeit als a priori gegebene Grundform der
inneren Anschauung nicht nur eine Bedingung der Möglichkeit von Erfahrung, sondern sogar die fundamentalste
Realität in unserem Geist und ist damit auch für die Beschreibung der Natur fundamental, obwohl sie zumindest
gemäß Kant in der Natur selbst nicht vorkommt. Bei Martin Heidegger drückt sich schon im Titel
seines Hauptwerkes "`Sein und Zeit"' dieser für das Sein fundamentale und konstitutive Charakter der
Zeit aus. Von Weizsäcker folgt dieser Anschauung, der Zeit einen solch prinzipiellen Charakter zuzuschreiben,
indem er die Zeit ebenfalls an den Beginn der Begründung der Quantentheorie stellt und aus ihr
die Quantenlogik als einer Logik zeitlicher Aussagen begründet. Aber die Dynamik der Quantentheorie
muss dennoch unabhängig über die Einführung eines Zeitparameters geschehen. Und in dieser Arbeit
wird die Zeit ausschließlich auf diese Weise eingeführt, da ansonsten das in den letzten beiden Unterabschnitten
bereits erwähnte Problem einer dualen Rolle der Zeit entsteht. Die Transformation einer Alternative
gemäß der zeitlichen Entwicklung muss einer Abbildung der Alternative auf sich selbst entsprechen, also einem
Automorphismus des Zustandsraumes, da die Alternative ja ihre Identität nicht verlieren, sondern lediglich
ihren Zustand ändern soll. Dies bedeutet, dass die zum Zeitparameter gehörige Transformationsgruppe
eine Untergruppe der $SO(n)$"=Gruppe sein muss, welche natürlich einparametrig sein muss, da die Zeit
in sich selbst nur einen einzigen Freiheitsgrad darstellt. Damit ergibt sich in natürlicher Weise
die $SO(2)$"=Gruppe mit der Zeit als Drehparameter, welche im Raum der Wahrheitswerte
einer Alternative durch die folgende Gleichung definiert wird:

\begin{equation}
\partial_t \varphi_j\left(t\right)=H_{jk} \varphi_k\left(t\right),
\label{SchroedingerGleichungAlternative}
\end{equation}
wobei der Operator $H$ im Hilbert"=Raum der Wahrheitswerte wie folgt definiert werden muss:

\begin{equation}
H_{kl}=\sum_{j=1}^{n/2}\omega_j \kappa^{j}_{kl},
\end{equation}
und der Tensor $\kappa^{j}_{kl}$ wiederum definiert ist gemäß:

\begin{equation}
\kappa_{kl}^j=\left\{\begin{matrix}&+1&\ \text{für}\ k=2j-1,\ l=2j \\&-1&\ \text{für}\ k=2j,\ l=2j-1\\&0&\ \text{ansonsten}\end{matrix}\right.
\end{equation}
Über doppelt auftretende Indizes wird wie gewöhnlich von $1$ bis $n$ summiert, wenn sie nur auf
einer Seite auftreten, ansonsten nicht. Wenn man den Operator $H$ als Matrix darstellt, so nimmt
er die folgende Gestalt an:

\begin{equation}
H=\left(\begin{matrix}0 & \omega_1 & 0 & 0 &  ... & 0 & 0\\
                      -\omega_1 & 0 & 0 & 0 & ... & 0 & 0\\
                      0 & 0 & 0 & \omega_2 & ... & ... & ...\\
                      0 & 0 & -\omega_2 & 0 & ... & ... & ...\\
                      ... & ... & ... & .... & ... & ... & ...\\
                      0 & 0 & ... & ... & ... & 0 & \omega_j\\
                      0 & 0 & ... & ... & ... & -\omega_j & 0\\ \end{matrix}\right).
\end{equation}
Um dies nun in kompakterer Weise auszudrücken, kann man die folgende komplexe Größe definieren:

\begin{equation}
\tilde{\varphi}_j=\varphi_{2j-1}+i\varphi_{2j},\quad j=1,...,n/2.
\label{Alternativekomplex}
\end{equation}
Mit der Definition ($\ref{Alternativekomplex}$) kann die dynamische Gleichung ($\ref{SchroedingerGleichungAlternative}$)
wie folgt ausgedrückt werden:

\begin{equation}
i\partial_t \tilde{\varphi}_j\left(t\right)=\omega_{j} \tilde{\varphi}_j\left(t\right).
\end{equation}
Damit erscheint nun die $SO(2)$"=Gruppe als $U(1)$"=Gruppe. Mit der Definition:

\begin{equation}
\tilde{H}_{kl}=\omega_k \delta_{kl},
\end{equation}
ergibt sich die Gleichung:

\begin{equation}
i\partial_t \tilde{\varphi}_j\left(t\right)=\tilde{H}_{jk} \tilde{\varphi}_k\left(t\right)\quad\Leftrightarrow\quad
i\partial_t \tilde{\varphi}\left(t\right)=\tilde{H}\tilde{\varphi}\left(t\right).
\label{Schroedinger_Gleichung}
\end{equation}
Diese hat nun die Gestalt der allgemeinen Schrödinger"=Gleichung, wobei $\tilde{H}$ natürlich als Hamilton"=Operator
interpretiert werden muss. Damit ist die allgemeine Struktur der abstrakten Quantentheorie statuiert.
Die zeitliche Entwicklung des auf die abstrakte Alternative bezogenen Zustandes $\tilde{\varphi}\left(t\right)$
ist dann gegeben durch den folgenden Ausdruck:

\begin{equation}
\tilde{\varphi}\left(t\right)=e^{-i\tilde{H}t}\tilde{\varphi}\left(t_0\right).
\label{Zeitentwicklung_Alternative}
\end{equation}

\section{Die Ur-Alternative als Grundbegriff der Quantentheorie}

\subsection{Konstituierung eines Tensorraumes vieler Ur-Alternativen}

Um nun zu der konkreten Physik zu gelangen, welche nicht nur die Existenz eines Ortsraumes als Darstellungsmedium
der abstrakten Informationsbeziehungen und seine Verbindung mit der Zeit zur Raum"=Zeit, sondern auch die
Existenz der speziellen konkreten Objekte einschließlich ihrer verschiedenen Wechselwirkungen enthält,
muss zunächst von der logischen Möglichkeit Gebrauch gemacht werden, eine beliebige logische
Alternative, wie sie in ($\ref{Alternative}$) charakterisiert wurde, in ein Cartesisches Produkt
binärer Alternativen aufzuspalten:

\begin{equation}
a=\bigotimes_n u_n,\quad u_n=\left(u_{n1},u_{n2}\right).
\end{equation}
Die binären Alternativen werden aufgrund des prinzipiellen Charakters, der ihnen zukommt, als Ur"=Alternativen bezeichnet.
Wenn man den Alternativen nun gemäß ($\ref{AlternativeWahrheitswerte}$) Wahrheitswerte zuordnet, die man gemäß der
zu einer zweckmäßigen Darstellung der Dynamik in ($\ref{Alternativekomplex}$) vorgenommenen Definition als komplexe
Wahrheitswerte definieren muss, so erhält man normierte zweidimensionale Spinoren:

\begin{equation}
u=\left(\begin{matrix} u_{1}\\ u_{2}\end{matrix}\right)\quad\longrightarrow\quad \varphi
=\left(\begin{matrix}\varphi\left(u_1\right)\\ \varphi\left(u_2\right)\end{matrix}\right)
=\left(\begin{matrix}\varphi_a+i\varphi_b\\ \varphi_c+i\varphi_d\end{matrix}\right),\quad
\sqrt{\varphi_a^2+\varphi_b^2+\varphi_c^2+\varphi_d^2}=1,
\label{Ur-Alternative}
\end{equation}
wobei $\varphi_a$, $\varphi_b$, $\varphi_c$ und $\varphi_d$ reellwertig sind. Ein solcher Spinor weist gemäß
dem Postulat der Symmetrie eine Symmetrie bezüglich der $SU(2)$"=Gruppe auf, die spezielle unitäre Gruppe
in einem zweidimensionalen komplexen Raum, welche isomorph zur $SO(3)$"=Gruppe ist, der Drehgruppe des
dreidimensionalen reellen Raumes. Zu den Spinoren können nun hermitesch adjungierte Größen definiert werden:

\begin{equation}
\varphi^{\dagger}=\left(\varphi_a-i\varphi_b, \varphi_c-i\varphi_d\right),\quad \sqrt{\varphi_a^2+\varphi_b^2+\varphi_c^2+\varphi_d^2}=1.
\end{equation}
Die Ur"=Alternativen sind keine beliebig ausgewählten Objekte. Sie sind nicht willkürlich postuliert, sondern sie repräsentieren
die einfachsten in einer beliebigen quantentheoretischen Beschreibungsweise überhaupt denkbaren Objekte. Ur"=Alternativen sind
deshalb zudem fundamentale Objekte, Atome im eigentlichen Sinne schlechthinniger Unteilbarkeit, denn sie sind als elementare
Informationseinheiten nicht in einem räumlichen oder nur physikalischen, sondern in einem logischen Sinne unteilbar.
Die Ur"=Alternativen sind hier nämlich gerade nicht als Objekte in einer bereits bestehenden physikalischen Realität
zu verstehen, so wie das Konzept der Quanteninformation in der Regel verwendet wird. Vielmehr konstituieren sie die
physikalische Realität überhaupt erst. Es gibt gemäß der Quantentheorie der Ur"=Alternativen überhaupt keine
physikalische Realität, die ohne diese fundamentale ontologische Basis überhaupt bestehen könnte, welche durch
die Ur"=Alternativen in unserem Geist dann indirekt als Quanteninformation dargestellt wird.

Durch die Definition ($\ref{Ur-Alternative}$) ist zunächst natürlich nur die mathematische Struktur einer einzelnen
Ur"=Alternative als einem zweidimensionalen Spinor definiert. Um nun die in größeren Alternativen enthaltene Information
durch Ur"=Alternativen darstellen zu können, ist es notwendig, einen Tensorraum vieler Ur"=Alternativen zu definieren.
Dessen Basiszustände sind durch die jeweilige Anzahl an Ur"=Alternativen in den vier Basiszuständen einer einzelnen
Alternative definiert. Um in diesem Raum operieren zu können und zwischen den verschiedenen Zuständen zu vermitteln,
ist es sinnvoll, Erzeugungs- und Vernichtungsoperatoren für Ur"=Alternativen in den Basiszuständen einer einzelnen
Ur"=Alternative zu definieren. Dies kann rein formal durch eine zweite Quantisierung gemäß der Bose"=Statistik geschehen,
durch welche die mit komplexen Wahrheitswerten belegte Ur"=Alternative in einen Operator überführt wird. Dieser
Übergang zu einem Operator:

\begin{equation}
\varphi \quad\longrightarrow\quad \hat \varphi,
\end{equation}
geschieht durch Forderung der folgenden Vertauschungsrelationen für die beiden Komponenten der Ur"=Alternative:

\begin{equation}
\left[\hat \varphi_m,\hat \varphi_n^{\dagger}\right]=\delta_{mn},\quad m,n=1,2.
\label{UrAlternativenKommutator}
\end{equation}
Es wird hier mit einer Vertauschungsrelation Bose"=Statistik und nicht mit einer Antivertauschungsrelation
Fermi"=Statistik postuliert, weil es im Falle der Fermi"=Statistik aufgrund der Gültigkeit des Paulischen
Ausschließungsprinzips für bezüglich Vertauschung antisymmetrische Zustände überhaupt nur vier Alternativen
geben könnte, nämlich jeweils eine in den vier Basiszuständen einer Ur"=Alternative. Die Bose"=Statistik
erlaubt beliebig viele Ur"=Alternativen in einem Zustand. Natürlich sind die Ur"=Alternativen
ununterscheidbar voneinander, da man für eine Unterscheidung ja erneut Ur"=Alternativen zu Grunde
legen müsste, denn die Ur"=Alternativen werden ja als die fundamentalste Realität in der Natur angesehen.
Daher muss in jedem Falle irgendeine Art der Symmetrie unter Vertauschung der Ur"=Alternativen in einem
Zustand zu Grunde gelegt werden. Um mehrere Objekte in einer Vielteilchentheorie und einer Theorie der
Wechselwirkung zu beschreiben, muss aber eine allgemeinere Statistik eingeführt werden,
nämlich die Parabose"=Statistik, welche sich nicht auf nur symmetrische oder nur antisymmetrische
Zustände bezüglich der Vertauschung von Ur"=Alternativen beschränkt und in einem späteren Abschnitt
eingeführt wird. Dies liegt natürlich daran, dass ein Zustand mehrerer freier Objekte, wenn diese
jeweils für sich unter Vertauschung der zu ihnen gehörigen Ur"=Alternativen symmetrisch sind,
also ein symmetrisches Produkt symmetrischer Zustände, seinerseits nicht mehr symmetrisch bezüglich aller
Ur"=Alternativen ist. Vielmehr sind dann die Ur"=Alternativen überhaupt nur durch die jeweilige Subsymmetrie
in ihrer Zugehörigkeit zu den einzelnen Objekten bestimmt, wodurch sich die in einem Gesamtzustand im Tensorraum
vieler Ur"=Alternativen enthaltenen Zustände den einzelnen voneinander verschiedenenen Teilobjekten zuordnen lassen.
Aber da in diesem Abschnitt nur einzelne freie Objekte betrachtet werden sollen, genügt es zunächst, sich hier
auf Bose"=Statistik zu beschränken. Die Vertauschungsrelation ($\ref{UrAlternativenKommutator}$) kann
erfüllt werden, indem die einzelnen Komponenten der Ur"=Alternative zu Operatoren werden:

\begin{equation}
\hat \varphi=\left(\begin{matrix}\hat \varphi_a+i\hat \varphi_b\\ \hat \varphi_c+i\hat \varphi_d\end{matrix}\right).
\end{equation}
Dies bedeutet für den Operator der hermitesch adjungierten Ur"=Alternative:

\begin{equation}
\hat \varphi^{\dagger}=\left(\hat \varphi_a^{\dagger}-i\hat \varphi_b^{\dagger}, \hat \varphi_c^{\dagger}-i\hat \varphi_d^{\dagger}\right).
\end{equation}
Die zu den Komponenten der Ur"=Alternative gehörenden Operatoren müssen folgende Vertauschungsrelationen erfüllen:

\begin{equation}
\left[\hat \varphi_a,\hat \varphi_a^{\dagger}\right]=\left[\hat \varphi_b,\hat \varphi_b^{\dagger}\right]=
\left[\hat \varphi_c,\hat \varphi_c^{\dagger}\right]=\left[\hat \varphi_d,\hat \varphi_d^{\dagger}\right]=1,
\end{equation}
wobei die nicht aufgeführten Kommutatoren einfach gleich null sind. Man kann diese Operatoren nun
in der folgenden Weise umbenennen:

\begin{equation}
a \equiv \hat \varphi_a ,\quad b \equiv \hat \varphi_b,\quad c \equiv \hat \varphi_c,\quad d \equiv \hat \varphi_d.
\label{OperatorenUmbenennung}
\end{equation}
Dann lauten die Vertauschungsrelationen wie folgt:

\begin{equation}
\left[a,a^{\dagger}\right]=\left[b,b^{\dagger}\right]=\left[c,c^{\dagger}\right]=\left[d,d^{\dagger}\right]=1.
\end{equation}
Damit stellen die Operatoren $a$, $b$, $c$, $d$ Vernichtungsoperatoren und die Operatoren
$a^{\dagger}$, $b^{\dagger}$, $c^{\dagger}$, $d^{\dagger}$ Erzeugungsoperatoren im Tensorraum
vieler Ur"=Alternativen dar. Ein Basiszustand im Tensorraum der Ur"=Alternativen ist, wie erwähnt,
durch die Anzahl der Ur"=Alternativen in jedem der vier Basiszustände einer einzelnen Ur"=Alternative
gegeben: $|N_a,N_b,N_c,N_d\rangle$. Die Erzeugungs- und Vernichtungsoperatoren wirken auf die
Basiszustände des Tensorraumes der Ur"=Alternativen in der folgenden Art und Weise:

\begin{eqnarray}
a|N_a,N_b,N_c,N_d\rangle&=&\sqrt{N_a}|N_a-1,N_b,N_c,N_d\rangle,\nonumber\\
a^{\dagger}|N_a,N_b,N_c,N_d\rangle&=&\sqrt{N_a+1}|N_a+1,N_b,N_c,N_d\nonumber\rangle,\\
b|N_a,N_b,N_c,N_d\rangle&=&\sqrt{N_b}|N_a,N_b-1,N_c,N_d\nonumber\rangle,\\
b^{\dagger}|N_a,N_b,N_c,N_d\rangle&=&\sqrt{N_b+1}|N_a,N_b+1,N_c,N_d\nonumber\rangle,\\
c|N_a,N_b,N_c,N_d\rangle&=&\sqrt{N_c}|N_a,N_b,N_c-1,N_d\nonumber\rangle,\\
c^{\dagger}|N_a,N_b,N_c,N_d\rangle&=&\sqrt{N_c+1}|N_a,N_b,N_c+1,N_d\nonumber\rangle,\\
d|N_a,N_b,N_c,N_d\rangle&=&\sqrt{N_d}|N_a,N_b,N_c,N_d-1\nonumber\rangle,\\
d^{\dagger}|N_a,N_b,N_c,N_d\rangle&=&\sqrt{N_d+1}|N_a,N_b,N_c,N_d+1\rangle.
\label{WirkungOperatoren}
\end{eqnarray}
Dies bedeutet, dass $a^{\dagger}a$, $b^{\dagger}b$, $c^{\dagger}c$ und $d^{\dagger}d$ als Besetzungszahloperatoren
die folgenden Eigenwertgleichungen erfüllen:

\begin{eqnarray}
&a^{\dagger}a|N_a,N_b,N_c,N_d\rangle=N_a|N_a,N_b,N_c,N_d\rangle,\quad &b^{\dagger}b|N_a,N_b,N_c,N_d\rangle=N_b|N_a,N_b,N_c,N_d\rangle,\nonumber\\ &c^{\dagger}c|N_a,N_b,N_c,N_d\rangle=N_c|N_a,N_b,N_c,N_d\rangle,\quad &d^{\dagger}d|N_a,N_b,N_c,N_d\rangle=N_d|N_a,N_b,N_c,N_d\rangle.\nonumber\\
\label{Eigenwertgleichungen}
\end{eqnarray}
Ein allgemeiner Zustand im Tensorraum der Ur"=Alternativen kann dann als eine Superposition der Basiszustände
definiert werden:

\begin{equation}
|\Psi\rangle=\sum_{N_a}\sum_{N_b}\sum_{N_c}\sum_{N_d}\psi\left(N_a,N_b,N_c,N_d\right)|N_a,N_b,N_c,N_d\rangle.
\label{Zustand_Tensorraum_abcd}
\end{equation}
Es soll nun zur weiteren Konstruktion der auf den Ur"=Alternativen basierenden Strukturen zunächst die
Darstellung der Zustände geändert werden, indem basierend auf den Spinoren aus ($\ref{Ur-Alternative}$),
also $\varphi$, welche die mit komplexen Wahrheitswerten belegten Ur"=Alternativen beschreiben,
ein dazu analoger Majorana"=Spinor definiert wird:

\begin{equation}
\chi=\frac{1}{\sqrt{2}}\left(\begin{matrix}\varphi \\ i \sigma^2 \varphi^{*} \end{matrix}\right)
=\frac{1}{\sqrt{2}}\left(\begin{matrix} \varphi_a+\varphi_b i \\ \varphi_c+\varphi_d i \\
\varphi_c-\varphi_d i \\-\varphi_a+\varphi_b i \end{matrix}\right)\equiv
\left(\begin{matrix} \chi_A \\ \chi_B \\ \chi_C \\ \chi_D \end{matrix}\right),
\end{equation}
wobei $\sigma^2$ die zweite Pauli"=Matrix ist. Durch Ersetzung der Komponenten der Ur"=Alternative $\varphi$
durch die entsprechenden Operatoren, geht auch der Spinor $\chi$ in einen Operator über:

\begin{equation}
\chi \quad\longrightarrow\quad \hat \chi,
\end{equation}
der die folgenden Vertauschungsrelationen erfüllt:

\begin{equation}
\left[\hat \chi_m,\hat \chi_n^{\dagger}\right]=\delta_{mn},\quad m,n=A,B,C,D,
\end{equation}
wobei der hermitesch adjungierte Operator $\chi^{\dagger}$ in der folgenden Weise definiert ist:

\begin{equation}
\chi^{\dagger}=\left(\chi_A^{\dagger}, \chi_B^{\dagger}, \chi_C^{\dagger}, \chi_D^{\dagger}\right).
\end{equation}
Wenn man nun in Analogie zu ($\ref{OperatorenUmbenennung}$) die folgende Umbenennung vornimmt:

\begin{equation}
A\equiv \hat \chi_A,\quad B\equiv \hat \chi_B,\quad C\equiv \hat \chi_C,\quad D\equiv \hat \chi_D,
\end{equation}
so gelten die folgenden Relationen:

\begin{eqnarray}
&&A=\frac{1}{\sqrt{2}}\left(a+ib\right),\quad B=\frac{1}{\sqrt{2}}\left(c+id\right),\quad
C=\frac{1}{\sqrt{2}}\left(c-id\right),\quad D=\frac{1}{\sqrt{2}}\left(-a+ib\right),\label{ErzeugungsVernichtungsOperatoren}\\
&&A^{\dagger}=\frac{1}{\sqrt{2}}\left(a^{\dagger}-ib^{\dagger}\right),\quad B^{\dagger}=\frac{1}{\sqrt{2}}\left(c^{\dagger}-id^{\dagger}\right),\quad
C^{\dagger}=\frac{1}{\sqrt{2}}\left(c^{\dagger}+id^{\dagger}\right),\quad D^{\dagger}=\frac{1}{\sqrt{2}}\left(-a^{\dagger}-ib^{\dagger}\right),\nonumber
\end{eqnarray}
und damit die folgenden Kommutatoren:

\begin{equation}
\left[A,A^{\dagger}\right]=\left[B,B^{\dagger}\right]=\left[C,C^{\dagger}\right]=\left[D,D^{\dagger}\right]=1.
\end{equation}
In dieser neuen Darstellung sind in Analogie zu den Basiszuständen $|N_a,N_b,N_c,N_d\rangle$ die Basiszustände
$|N_A,N_B,N_C,N_D\rangle$ definiert. Und die Operatoren $A$, $B$, $C$, $D$, $A^{\dagger}$, $B^{\dagger}$, $C^{\dagger}$, $D^{\dagger}$
wirken auf diese Basiszustände analog zu ($\ref{WirkungOperatoren}$) in der folgenden Weise:

\begin{eqnarray}
A|N_A,N_B,N_C,N_D\rangle&=&\sqrt{N_A}|N_A-1,N_B,N_C,N_D\rangle,\nonumber\\
A^{\dagger}|N_A,N_B,N_C,N_D\rangle&=&\sqrt{N_A+1}|N_A+1,N_B,N_C,N_D\rangle,\nonumber\\
B|N_A,N_B,N_C,N_D\rangle&=&\sqrt{N_B}|N_A,N_B-1,N_C,N_D\rangle,\nonumber\\
B^{\dagger}|N_A,N_B,N_C,N_D\rangle&=&\sqrt{N_B+1}|N_A,N_B+1,N_C,N_D\rangle,\nonumber\\
C|N_A,N_B,N_C,N_D\rangle&=&\sqrt{N_C}|N_A,N_B,N_C-1,N_D\rangle,\nonumber\\
C^{\dagger}|N_A,N_B,N_C,N_D\rangle&=&\sqrt{N_C+1}|N_A,N_B,N_C+1,N_D\rangle,\nonumber\\
D|N_A,N_B,N_C,N_D\rangle&=&\sqrt{N_D}|N_A,N_B,N_C,N_D-1\rangle,\nonumber\\
D^{\dagger}|N_A,N_B,N_C,N_D\rangle&=&\sqrt{N_D+1}|N_A,N_B,N_C,N_D+1\rangle.
\label{WirkungOperatorenABCD}
\end{eqnarray}
In dieser neuen Darstellung bedeutet dies zudem, dass $A^{\dagger}A$, $B^{\dagger}B$, $C^{\dagger}C$ und $D^{\dagger}D$
als Besetzungszahloperatoren analog zu ($\ref{Eigenwertgleichungen}$) die folgenden Eigenwertgleichungen erfüllen:

\begin{eqnarray}
&A^{\dagger}A|N_A,N_B,N_C,N_D\rangle=N_A|N_A,N_B,N_C,N_D\rangle,\quad &B^{\dagger}B|N_A,N_B,N_C,N_D\rangle=N_B|N_A,N_B,N_C,N_D\rangle,\nonumber\\ &C^{\dagger}C|N_A,N_B,N_C,N_D\rangle=N_C|N_A,N_B,N_C,N_D\rangle,\quad &D^{\dagger}D|N_A,N_B,N_C,N_D\rangle=N_D|N_A,N_B,N_C,N_D\rangle.\nonumber\\
\label{EigenwertgleichungenABCD}
\end{eqnarray}
In der auf diese Basis bezogenen Darstellung lässt sich ein allgemeiner Zustand im Tensorraum dann in Analogie
zu ($\ref{Zustand_Tensorraum_abcd}$) natürlich erneut als Superposition darstellen:

\begin{equation}
|\Psi\rangle=\sum_{N_A}\sum_{N_B}\sum_{N_C}\sum_{N_D}\psi\left(N_A,N_B,N_C,N_D\right)|N_A,N_B,N_C,N_D\rangle.
\end{equation}
Um eine kompaktere Schreibweise zu erhalten, sei die folgende Definition vorgenommen:

\begin{equation}
N_{ABCD} \equiv \left(N_A,N_B,N_C,N_D\right).
\label{kompakteDarstellung}
\end{equation}
Damit kann dann ein Basiszustand wie folgt geschrieben werden:

\begin{equation}
|N_{ABCD}\rangle \equiv |N_A,N_B,N_C,N_D\rangle,
\label{Basiszustand_Tensorraum}
\end{equation}
und ein allgemeiner Zustand im Tensorraum wie folgt geschrieben werden:

\begin{equation}
|\Psi\rangle=\sum_{N_{ABCD}}\psi\left(N_{ABCD}\right)|N_{ABCD}\rangle.
\label{Zustand_Tensorraum}
\end{equation}
Zudem soll $N$ die Gesamtzahl aller Ur"=Alternativen in einem Zustand beschreiben,
also die Gesamtmenge an Information:

\begin{equation}
N=N_A+N_B+N_C+N_D.
\end{equation}

\subsection{Die Gründe für die zentrale Bedeutung des Begriffes der Ur-Alternative}

Es sollen nun die zentralen Gründe kategorisiert werden, welche den Begriff der Ur"=Alternative
als fundamentalen Begriff der Physik besonders plausibel erscheinen lassen.

\noindent
\textbf{A) Ur"=Alternativen basieren nicht auf einer räumlich"=feldtheoretischen Begrifflichkeit:}

Ur"=Alternativen setzen keinen physikalischen Raum und daher auch keinerlei feldtheoretische Begriffe voraus.
Denn Ur"=Alternativen sind keine Objekte, die in einem Raum oder überhaupt irgendeiner bereits bestehenden
physikalischen Realität existieren würden. Umgekehrt konstituiert sich aus ihnen und der durch sie begründeten
abstrakten logischen Beziehungsstruktur überhaupt erst alle physikalische Realität einschließlich des Raumes
und der darin sich befindlichen Objekte. In diesem Sinne unterscheiden sie sich auch grundlegend
von einem Begriff der Quanteninformation, wie er in den unterschiedlichsten Zusammenhängen und anderen
physikalischen Ansätzen immer wieder verwendet wird, nämlich von einem solchen, bei dem Quanteninformation
sich auf andere physikalische Realitäten bezieht, im Rahmen einer bestimmten bereits bestehenden physikalischen
Realität ausgetauscht wird oder in eine solche Beschreibung eingebunden wird. Die Ur"=Alternativen hingegen sind
auf der fundamentalen Ebene die einzige physikalische Realität, die überhaupt existiert. Außer ihnen und der
Zeit darf nichts anderes vorausgesetzt werden.

\noindent
\textbf{B) Ur"=Alternativen sind aus logischen Gründen im schlechthinnigen Sinne unteilbar:}

Gemäß der zweiten Kantischen Antinomie kann es keine kleinsten räumlichen Objekte geben,
da jedes Volumen zumindest im Prinzip weiter in Teilvolumina geteilt werden kann. Eine
Ur"=Alternative ist aber kein räumliches Objekt, sondern ein rein logisches Objekt, und zwar
das fundamentalste logische Objekt, dass es im Rahmen einer Quantenlogik überhaupt geben kann.
Es ist aus logischen Gründen nicht weiter teilbar und in diesem Sinne eine atomare Einheit
im Sinne schlechthinniger Unteilbarkeit. Damit ist mit den Ur"=Alternativen erstmals ein
fundamentales Objekt postuliert, von dem sich logisch begründen lässt, warum es nicht nur
fundamental sein kann, sondern in einer quantentheoretischen Beschreibungsweise auch
fundamental sein muss.

\noindent
\textbf{C) Ur"=Alternativen sind die einfachsten denkbaren quantentheoretischen Objekte:}

Eine Ur"=Alternative ist kein willkürlich gewähltes Objekt, sondern es ist das fundamentalste
und einfachste Objekt, dass im Rahmen einer beliebigen Quantentheorie überhaupt denkbar ist.
Denn es ist mathematisch durch den einfachsten in der Quantentheorie überhaupt denkbaren
Zustandsraum definiert. Aufgrund der ungeheuren Abstraktheit der allgemeinen Quantentheorie
basiert dieses Objekt aber eigentlich sogar auf dem einfachsten überhaupt denkbaren
logischen Begriff, nämlich einer binären Alternative. Eine Ur"=Alternative geht bezüglich
ihres logischen Gehaltes nur insofern über eine gewöhnliche einfache binäre Alternative hinaus,
als sie eine Quantenlogik voraussetzt, in der das tertium non datur verletzt ist, eine Aussage
also nicht einfach nur wahr oder falsch sein kann, sondern Zwischenwerte besitzen kann, die nur
eine bestimmte Tendenz bezüglich wahr und falsch definiert. Zudem legt sie komplexe
Wahrheitswerte zu Grunde, wobei dies letztendlich auch nur eine Frage der Darstellung ist,
die durch die Einbindung einer Zeitentwicklung nahegelegt wird. Jedenfalls verleiht diese
im Rahmen der Quantentheorie größtmögliche Abstraktheit und Einfachheit den Ur"=Alternativen
auch deshalb eine große Überzeugungskraft, weil Begriffe, die sehr viele unterschiedliche
Realitäten und Strukturen in sich vereinheitlichen sollen, also sehr allgemein sein sollen,
naturgemäß sehr abstrakt und einfach sein müssen. Denn das grundlegende Wesen des Verstehens
in der Naturwissenschaft besteht in der Einordnung möglichst vieler Realitäten unter
einheitliche, grundlegende und daher möglichst allgemeingültige Begriffe. Allgemeingültigkeit
und Einheitlichkeit bedeutet Abstraktheit und Einfachheit. Je abstrakter und einfacher und daher
fundamentaler aber die Grundbegriffe werden, desto schwieriger wird natürlich zugleich ihr
Verständnis und desto schwieriger wird die Herleitung der Komplexität, die weiter an der
Oberfläche der Realität existiert.

\noindent
\textbf{D) Ur"=Alternativen enthalten implizit die Symmetriestruktur des realen Raumes:}

Gerade weil eine Ur"=Alternative kein willkürliches, sondern das grundlegendste quantentheoretische
Objekt darstellt, ist es umso erstaunlicher, dass gerade dieses einfachste Objekt aufgrund der Isomorphie
zwischen $SU(2)$"=Gruppe und $SO(3)$"=Gruppe implizit die Symmetriestruktur des empirisch gefundenen physikalischen
Raumes in sich trägt. Die abstrakte Struktur der Quantentheorie und die Struktur des Raumes scheinen a priori
überhaupt nichts miteinander zu tun zu haben. Der Raum ist aber dennoch eine für die gewöhnliche Physik ganz
grundlegende und allgemeine Realität, denn alle gewöhnlichen physikalischen Objekte befinden sich zumindest
in der Oberflächenbetrachtung der klassischen Physik im physikalischen Anschauungsraum. Deshalb erscheint
die Tatsache, dass eine direkte mathematische Beziehung zwischen der Symmetriestruktur des einfachsten
möglichen Objektes einer so grundlegenden und abstrakten Theorie wie der Quantentheorie und derjenigen
des Raumes besteht, mehr als nur ein Zufall zu sein. Sie stellt vielmehr ein großes Indiz dafür dar,
dass mit den Ur"=Alternativen eine zentrale Wahrheit über die Natur berührt ist.

\noindent
\textbf{E) Ur"=Alternativen konstituieren einen diskreten Zustandsraum:}

Dadurch, dass die Ur"=Alternativen in natürlicher Weise diskrete Zustandsräume konstituieren, wird die
Möglichkeit eröffnet, das Auftreten von Unendlichkeiten und Divergenzen von vorneherein zu verhindern.
Die Probleme des Kontinuums waren der Grund, warum die Quantentheorie überhaupt entwickelt wurde.
Eben jenes Kontinuum ist aber im Rahmen relativistischer Quantenfeldtheorien wieder mit in die Beschreibung
der Natur hineingekommen. Die Kontinuumsproblematik tritt im Rahmen relativistischer Quantenfeldtheorien speziell
durch die Wechselwirkungen auf, welche durch kontinuierliche punktweise Produkte von Feldern beschrieben werden,
denn eine freie Quantenfeldtheorie entspricht ja einer einfach nur einer Vielteilchentheorie. Die durch
die Beschreibung der Wechselwirkung basierend auf einem feldtheoretischen Kontinuum auftretenden Unendlichkeiten
werden im Rahmen der gewöhnlichen Quantenfeldtheorien durch das Verfahren der Renormierung beseitigt,
welches im Falle der Gravitation bekanntlich nicht erfolgreich angewandt werden kann. Es ist zu erwarten,
dass sich dieses Problem nur dann an der Wurzel packen lässt, wenn man die Physik von Grund auf diskret aufbaut.
Aber dies geschieht in der Quantentheorie der Ur"=Alternativen gerade nicht durch eine künstliche
Diskretisierung der Raum"=Zeit oder die bloße Einführung einer kleinsten Länge, sondern dadurch, dass man
mit einer Realität beginnt, nämlich diskreten Alternativen, die überhaupt noch nicht in Bezug auf die
Raum"=Zeit definiert sind, sondern die Existenz der Raum"=Zeit erst im Nachhinein konstituieren.
Aber sie erklären dann dennoch, warum durch diese Darstellungsmöglichkeit das Kontinuum überhaupt in die
Beschreibung der Natur hineingelangt. Dadurch kann auch die in der gewöhnlichen Elementarteilchenphysik
bestehende Dualität zwischen abstrakten Quantenzahlen einerseits und räumlichen Objekten andererseits
aufgehoben und überwunden werden.

\section{Die Konstituierung freier Objekte im Raum}

In den bisherigen Abschnitten wurden die Grundlagen der Quantentheorie der Ur"=Alternativen dargestellt und es
wurde begründet, warum sie als einheitliche Naturtheorie so vielversprechend ist. Die nun folgenden Abschnitte
enthalten eigene neue spezielle Konzepte und mathematische Modelle, um in diesem begrifflichen Rahmen die konkrete
Physik zu konstruieren. In diesem Abschnitt werden die symmetrischen Zustände im Tensorraum vieler Ur"=Alternativen
in einen dreidimensionalen reellen Raum abgebildet, der dann mit dem realen physikalischen Ortsraum identifiziert
werden kann. In diesem Sinne können diese Zustände als räumlich darstellbare Quantenobjekte interpretiert werden,
die wir gewöhnlich im Sinne des Welle"=Teilchen"=Dualismus als Teilchen oder Wellen bezeichnen. Über die Definition
eines Hamilton"=Operators wird dann durch die allgemeine Schrödinger"=Gleichung eine Zeitentwicklung induziert.
Man kann die Zeitkoordinate wie in der Relativitätstheorie formal als vierte Dimension in einer Raum"=Zeit auffassen
und den dreidimensionalen Raum als Hyperfläche in einer Raum"=Zeit. Aber wie in der Relativitätstheorie auch enthält
die Zeitdimension keinen unabhängigen dynamischen Freiheitsgrad, einfach aufgrund der Existenz einer dynamischen
Grundgleichung. Die Tatsache, dass die Zeitentwicklung in diesem Rahmen als Automorphismus des Zustandsraumes
aufgefasst wird, dessen Zustände ihrerseits in einem dreidimensionalen Raum dargestellt werden, ist also in
Einklang mit der Relativitätstheorie, solange der Hamilton"=Operator basierend auf der relativistischen
Energie"=Impuls"=Beziehung definiert wird. In \cite{Goernitz:1992} wird bereits eine Beschreibung von
Quantenfeldtheorien im Rahmen der Ur"=Alternativen vorgeschlagen, aber hier wird keine Abbildung des
Tensorraumes der Ur"=Alternativen in den physikalischen Ortsraum vollzogen. Dies wird in \cite{Kober:2011}
versucht, aber hier wird die Zeit in falscher Weise eingeführt, nicht über einen Automorphismus des
Zustandsraumes, sondern als zusätzlicher Freiheitsgrad, was auch zur Folge hat, dass die Unitarität
und damit die Wahrscheinlichkeitserhaltung der Zustände nicht gewährleistet ist. In dem Ansatz dieses
Abschnittes sind diese Konzeptionsschwächen in einem modifizierten Ansatz behoben.

\subsection{Die Konstruktion von Orts- und Impulsoperatoren}

Man kann aus den in ($\ref{ErzeugungsVernichtungsOperatoren}$) definierten Operatoren, welche sich auf
Ur"=Alternativen beziehen, die folgenden neuen Operatoren konstruieren:

\begin{eqnarray}
A_x&=&\frac{1}{2}\left(A+B-C-D\right),\quad A_x^{\dagger}=\frac{1}{2}\left(A^{\dagger}+B^{\dagger}-C^{\dagger}-D^{\dagger}\right),\nonumber\\
A_y&=&\frac{1}{2}\left(A-B+C-D\right),\quad A_y^{\dagger}=\frac{1}{2}\left(A^{\dagger}-B^{\dagger}+C^{\dagger}-D^{\dagger}\right),\nonumber\\
A_z&=&\frac{1}{2}\left(A-B-C+D\right),\quad A_z^{\dagger}=\frac{1}{2}\left(A^{\dagger}-B^{\dagger}-C^{\dagger}+D^{\dagger}\right).
\label{ErzeugungsVernichtungsOperatorenXYZ}
\end{eqnarray}
Diese erfüllen die gleichen Vertauschungsrelationen:

\begin{equation}
\left[A_x,A_x^{\dagger}\right]=\left[A_y,A_y^{\dagger}\right]=\left[A_z,A_z^{\dagger}\right]=1.
\label{VertauschungsrelationenXYZ}
\end{equation}
Über diese neu konstruierten Erzeugungs- und Vernichtungsoperatoren kann man nun weitere Operatoren erzeugen:

\begin{eqnarray}
X&=&\frac{1}{\sqrt{2}}\left(A_x+A_x^{\dagger}\right),\quad P_x=-\frac{i}{\sqrt{2}}\left(A_x-A_x^{\dagger}\right),\nonumber\\
Y&=&\frac{1}{\sqrt{2}}\left(A_y+A_y^{\dagger}\right),\quad P_y=-\frac{i}{\sqrt{2}}\left(A_y-A_y^{\dagger}\right),\nonumber\\
Z&=&\frac{1}{\sqrt{2}}\left(A_z+A_z^{\dagger}\right),\quad P_z=-\frac{i}{\sqrt{2}}\left(A_z-A_z^{\dagger}\right),
\label{Ort_Impuls_Operatoren}
\end{eqnarray}
welche hermitesch sind und zudem die Algebra von Orts- und Impulsoperatoren in drei
unabhängigen Dimensionen erfüllen, mit denen sie daher identifiziert werden sollen:

\begin{equation}
\left[X,P_x\right]=\left[Y,P_y\right]=\left[Z,P_z\right]=i.
\end{equation}
Es sei aber noch einmal in aller Eindringlichkeit darauf hingewiesen, dass sich diese Operatoren mit ihrer Algebra,
da sie aus den Erzeugungs- und Vernichtungsoperatoren im Tensorraum konstruiert werden, in keiner Weise
auf einen bereits bestehenden Ortsraum beziehen. Sie beziehen sich ausschließlich auf den abstrakten
Informationsraum der Ur"=Alternativen und sie gestatten lediglich, wie wir sehen werden, die nachträgliche
Begründung der Existenz des Ortsraumes, der im Einklang mit den Ausführungen in den früheren Abschnitten
in der Quantentheorie der Ur"=Alternativen keine fundamentale Realität darstellt. Diese nachträgliche
Begründung des Ortsraumes als bloße Art der Darstellung abstrakter Zustände ist ja die Intention dieses
Abschnittes. Nun kann man aus den über die Erzeugungs- und Vernichtungsoperatoren definierten
Impulsoperatoren in ($\ref{Ort_Impuls_Operatoren}$) desweiteren einen Energieoperator konstruieren,
der dann zugleich der Hamilton"=Operators ist. Dieser soll hier in einer solchen Weise definiert werden,
dass er mit der speziellen Relativitätstheorie in Einklang steht:

\begin{equation}
E^2=P_x^2+P_y^2+P_z^2\quad \Leftrightarrow\quad E=\pm \sqrt{P_x^2+P_y^2+P_z^2}.
\label{Energie-Impuls-Relation}
\end{equation}
Ausgedrückt durch die ursprünglichen Erzeugungs- und Vernichtungsoperatoren in ($\ref{ErzeugungsVernichtungsOperatoren}$)
haben die Orts- und Impulsoperatoren die folgende Gestalt:

\begin{eqnarray}
X&=&\frac{1}{2\sqrt{2}}\left(A+B-C-D+A^{\dagger}+B^{\dagger}-C^{\dagger}-D^{\dagger}\right),\nonumber\\
Y&=&\frac{1}{2\sqrt{2}}\left(A-B+C-D+A^{\dagger}-B^{\dagger}+C^{\dagger}-D^{\dagger}\right),\nonumber\\
Z&=&\frac{1}{2\sqrt{2}}\left(A-B-C+D+A^{\dagger}-B^{\dagger}-C^{\dagger}+D^{\dagger}\right),\nonumber\\
P_x&=&-\frac{i}{2\sqrt{2}}\left(A+B-C-D-A^{\dagger}-B^{\dagger}+C^{\dagger}+D^{\dagger}\right),\nonumber\\
P_y&=&-\frac{i}{2\sqrt{2}}\left(A-B+C-D-A^{\dagger}+B^{\dagger}-C^{\dagger}+D^{\dagger}\right),\nonumber\\
P_z&=&-\frac{i}{2\sqrt{2}}\left(A-B-C+D-A^{\dagger}+B^{\dagger}+C^{\dagger}-D^{\dagger}\right).
\label{Ort_Impuls_Operatoren_ABCD}
\end{eqnarray}
Die Quadrate der Impulsoperatoren lauten dann wie folgt:

\begin{eqnarray}
P_x^2&=&\frac{1}{2}\left(-A_x A_x+A_x A_x^{\dagger}+A_x^{\dagger}A_x-A_x^{\dagger}A_x^{\dagger}\right)
=\frac{1}{2}\left(-A_x A_x+2 A_x^{\dagger}A_x-A_x^{\dagger}A_x^{\dagger}+1\right)\nonumber\\
&=&\frac{1}{8}\left[-AA-BB-CC-DD-2AB+2AC+2AD+2BC+2BD-2CD
\right.\nonumber\\ &&\left.
-A^{\dagger}A^{\dagger}-B^{\dagger}B^{\dagger}-C^{\dagger}C^{\dagger}-D^{\dagger}D^{\dagger}
-2A^{\dagger}B^{\dagger}+2A^{\dagger}C^{\dagger}+2A^{\dagger}D^{\dagger}
+2B^{\dagger}C^{\dagger}+2B^{\dagger}D^{\dagger}-2C^{\dagger}D^{\dagger}
\right.\nonumber\\ &&\left.
+2\left(A^{\dagger}A+B^{\dagger}B+C^{\dagger}C+D^{\dagger}D
+A^{\dagger}B-A^{\dagger}C-A^{\dagger}D+B^{\dagger}A-B^{\dagger}C-B^{\dagger}D
\right.\right.\nonumber\\ &&\left.\left.
-C^{\dagger}A-C^{\dagger}B+C^{\dagger}D-D^{\dagger}A-D^{\dagger}B+D^{\dagger}C\right)+1\right],
\end{eqnarray}

\begin{eqnarray}
P_y^2&=&\frac{1}{2}\left(-A_y A_y+A_y A_y^{\dagger}+A_y^{\dagger}A_y-A_y^{\dagger}A_y^{\dagger}\right)
=\frac{1}{2}\left(-A_y A_y+2 A_y^{\dagger}A_y-A_y^{\dagger}A_y^{\dagger}+1\right)\nonumber\\
&=&\frac{1}{8}\left[-AA-BB-CC-DD+2AB-2AC+2AD+2BC-2BD+2CD
\right.\nonumber\\ &&\left.
-A^{\dagger}A^{\dagger}-B^{\dagger}B^{\dagger}-C^{\dagger}C^{\dagger}-D^{\dagger}D^{\dagger}
+2A^{\dagger}B^{\dagger}-2A^{\dagger}C^{\dagger}+2A^{\dagger}D^{\dagger}
+2B^{\dagger}C^{\dagger}-2B^{\dagger}D^{\dagger}+2C^{\dagger}D^{\dagger}
\right.\nonumber\\ &&\left.
+2\left(A^{\dagger}A+B^{\dagger}B+C^{\dagger}C+D^{\dagger}D
-A^{\dagger}B+A^{\dagger}C-A^{\dagger}D-B^{\dagger}A-B^{\dagger}C+B^{\dagger}D
\right.\right.\nonumber\\ &&\left.\left.
+C^{\dagger}A-C^{\dagger}B-C^{\dagger}D-D^{\dagger}A+D^{\dagger}B-D^{\dagger}C\right)+1\right],
\end{eqnarray}

\begin{eqnarray}
P_z^2&=&\frac{1}{2}\left(-A_z A_z+A_z A_z^{\dagger}+A_z^{\dagger}A_z-A_z^{\dagger}A_z^{\dagger}\right)
=\frac{1}{2}\left(-A_z A_z+2 A_z^{\dagger}A_z-A_z^{\dagger}A_z^{\dagger}+1\right)\nonumber\\
&=&\frac{1}{8}\left[-AA-BB-CC-DD+2AB+2AC-2AD-2BC+2BD+2CD
\right.\nonumber\\ &&\left.
-A^{\dagger}A^{\dagger}-B^{\dagger}B^{\dagger}-C^{\dagger}C^{\dagger}-D^{\dagger}D^{\dagger}
+2A^{\dagger}B^{\dagger}+2A^{\dagger}C^{\dagger}-2A^{\dagger}D^{\dagger}
-2B^{\dagger}C^{\dagger}+2B^{\dagger}D^{\dagger}+2C^{\dagger}D^{\dagger}
\right.\nonumber\\ &&\left.
+2\left(A^{\dagger}A+B^{\dagger}B+C^{\dagger}C+D^{\dagger}D
-A^{\dagger}B-A^{\dagger}C+A^{\dagger}D-B^{\dagger}A+B^{\dagger}C-B^{\dagger}D
\right.\right.\nonumber\\ &&\left.\left.
-C^{\dagger}A+C^{\dagger}B-C^{\dagger}D+D^{\dagger}A-D^{\dagger}B-D^{\dagger}C\right)+1\right].
\end{eqnarray}
Das über ($\ref{Energie-Impuls-Relation}$) definierte Quadrat des Energieoperators erhält
damit die folgende Gestalt:

\begin{eqnarray}
E^2&=&\frac{1}{2}\left(-A_x A_x+A_x A_x^{\dagger}+A_x^{\dagger}A_x-A_x^{\dagger}A_x^{\dagger}
-A_y A_y+A_y A_y^{\dagger}+A_y^{\dagger}A_y-A_y^{\dagger}A_y^{\dagger}\right.\nonumber\\
&&\left.-A_z A_z+A_z A_z^{\dagger}+A_z^{\dagger}A_z-A_z^{\dagger}A_z^{\dagger}\right)\nonumber\\
&=&\frac{1}{2}\left(-A_x A_x+2 A_x^{\dagger}A_x-A_x^{\dagger}A_x^{\dagger}-A_y A_y+2 A_y^{\dagger}A_y-A_y^{\dagger}A_y^{\dagger}
-A_z A_z+2 A_z^{\dagger}A_z-A_z^{\dagger}A_z^{\dagger}+3\right)\nonumber\\
&=&\frac{1}{8}\left[-3AA-3BB-3CC-3DD-3A^{\dagger}A^{\dagger}-3B^{\dagger}B^{\dagger}-3C^{\dagger}C^{\dagger}-3D^{\dagger}D^{\dagger}
\right.\nonumber\\ &&\left.
+2AB+2AC+2AD+2BC+2BD+2CD
\right.\nonumber\\ &&\left.
+2A^{\dagger}B^{\dagger}+2A^{\dagger}C^{\dagger}+2A^{\dagger}D^{\dagger}
+2B^{\dagger}C^{\dagger}+2B^{\dagger}D^{\dagger}+2C^{\dagger}D^{\dagger}
\right.\nonumber\\ &&\left.
+2\left(3A^{\dagger}A+3B^{\dagger}B+3C^{\dagger}C+3D^{\dagger}D
\right.\right.\nonumber\\ &&\left.\left.
-A^{\dagger}B-A^{\dagger}C-A^{\dagger}D-B^{\dagger}A-B^{\dagger}C-B^{\dagger}D
\right.\right.\nonumber\\ &&\left.\left.
-C^{\dagger}A-C^{\dagger}B-C^{\dagger}D-D^{\dagger}A-D^{\dagger}B-D^{\dagger}C\right)+3\right],
\end{eqnarray}
und die positive Komponente des Energieoperators lautet damit:

\begin{eqnarray}
E&=&\frac{1}{\sqrt{2}}\left(-A_x A_x+A_x A_x^{\dagger}+A_x^{\dagger}A_x-A_x^{\dagger}A_x^{\dagger}
-A_y A_y+A_y A_y^{\dagger}+A_y^{\dagger}A_y-A_y^{\dagger}A_y^{\dagger}\right.\nonumber\\
&&\left.-A_z A_z+A_z A_z^{\dagger}+A_z^{\dagger}A_z-A_z^{\dagger}A_z^{\dagger}+3\right)^{\frac{1}{2}}\nonumber\\
&=&\frac{1}{\sqrt{2}}\left(-A_x A_x+2 A_x^{\dagger}A_x-A_x^{\dagger}A_x^{\dagger}-A_y A_y+2 A_y^{\dagger}A_y-A_y^{\dagger}A_y^{\dagger}
-A_z A_z+2 A_z^{\dagger}A_z-A_z^{\dagger}A_z^{\dagger}+3\right)^{\frac{1}{2}}\nonumber\\
&=&\frac{1}{2\sqrt{2}}\left[-3AA-3BB-3CC-3DD-3A^{\dagger}A^{\dagger}-3B^{\dagger}B^{\dagger}-3C^{\dagger}C^{\dagger}-3D^{\dagger}D^{\dagger}
\right.\nonumber\\ &&\left.
+2AB+2AC+2AD+2BC+2BD+2CD
\right.\nonumber\\ &&\left.
+2A^{\dagger}B^{\dagger}+2A^{\dagger}C^{\dagger}+2A^{\dagger}D^{\dagger}
+2B^{\dagger}C^{\dagger}+2B^{\dagger}D^{\dagger}+2C^{\dagger}D^{\dagger}
\right.\nonumber\\ &&\left.
+2\left(3A^{\dagger}A+3B^{\dagger}B+3C^{\dagger}C+3D^{\dagger}D
\right.\right.\nonumber\\ &&\left.\left.
-A^{\dagger}B-A^{\dagger}C-A^{\dagger}D-B^{\dagger}A-B^{\dagger}C-B^{\dagger}D
\right.\right.\nonumber\\ &&\left.\left.
-C^{\dagger}A-C^{\dagger}B-C^{\dagger}D-D^{\dagger}A-D^{\dagger}B-D^{\dagger}C\right)+3\right]^{\frac{1}{2}}.
\label{Energie_Operator}
\end{eqnarray}
Mit Hilfe des Energieoperators ($\ref{Energie_Operator}$) und der Impulsoperatoren aus ($\ref{Ort_Impuls_Operatoren}$)
beziehungsweise ($\ref{Ort_Impuls_Operatoren_ABCD}$) kann man nun einen Vierer"=Impuls im Tensorraum der
Ur"=Alternativen definieren:

\begin{equation}
P_{ABCD}=\left(E,P_x,P_y,P_z\right).
\label{Viererimpuls_Tensorraum}
\end{equation}
Um basierend auf den in ($\ref{ErzeugungsVernichtungsOperatorenXYZ}$) definierten Operatoren eine neue
Darstellung der Zustände im Tensorraum zu erhalten, müssen die zu den Operatoren in
($\ref{ErzeugungsVernichtungsOperatorenXYZ}$) gehörigen Besetzungszahlzustände betrachtet werden.
Allerdings ist in diesen Operatoren ($\ref{ErzeugungsVernichtungsOperatorenXYZ}$) ein Freiheitsgrad
noch nicht enthalten, da der Tensorraum ja insgesamt vier Besetzungszahl"=Unterraüme enthält.
Man kann nämlich einen weiteren Operator $A_n$ und dessen hermitesch adjungierten Operatoren
$A_n^{\dagger}$ wie folgt definieren:

\begin{equation}
A_n=\frac{1}{2}\left(A+B+C+D\right),\quad A_n^{\dagger}=\frac{1}{2}\left(A^{\dagger}+B^{\dagger}+C^{\dagger}+D^{\dagger}\right),
\label{ErzeugungsVernichtungsOperatorn}
\end{equation}
wobei $A_n$ mit $A_x$, $A_y$ und $A_z$ kommutiert und analog zu ($\ref{VertauschungsrelationenXYZ}$) gilt:

\begin{equation}
\left[A_n, A^{\dagger}_n\right]=1.
\end{equation}
Die entsprechenden Besetzungszahloperatoren haben ausgedrückt durch die Operatoren
in ($\ref{ErzeugungsVernichtungsOperatoren}$) die folgende Gestalt:

\begin{eqnarray}
A_{x}^{\dagger}A_x&=&\frac{1}{4}\left(A^{\dagger}A+A^{\dagger}B-A^{\dagger}C-A^{\dagger}D
+B^{\dagger}A+B^{\dagger}B-B^{\dagger}C-B^{\dagger}D\right.\nonumber\\
&&\left.-C^{\dagger}A-C^{\dagger}B+C^{\dagger}C+C^{\dagger}D
-D^{\dagger}A-D^{\dagger}B+D^{\dagger}C+D^{\dagger}D\right),\nonumber\\
A_{y}^{\dagger}A_y&=&\frac{1}{4}\left(A^{\dagger}A-A^{\dagger}B+A^{\dagger}C-A^{\dagger}D
-B^{\dagger}A+B^{\dagger}B-B^{\dagger}C+B^{\dagger}D\right.\nonumber\\
&&\left.+C^{\dagger}A-C^{\dagger}B+C^{\dagger}C-C^{\dagger}D
-D^{\dagger}A+D^{\dagger}B-D^{\dagger}C+D^{\dagger}D\right),\nonumber\\
A_{z}^{\dagger}A_z&=&\frac{1}{4}\left(A^{\dagger}A-A^{\dagger}B-A^{\dagger}C+A^{\dagger}D
-B^{\dagger}A+B^{\dagger}B+B^{\dagger}C-B^{\dagger}D\right.\nonumber\\
&&\left.-C^{\dagger}A+C^{\dagger}B+C^{\dagger}C-C^{\dagger}D
+D^{\dagger}A-D^{\dagger}B-D^{\dagger}C+D^{\dagger}D\right),\nonumber\\
A_{n}^{\dagger}A_n&=&\frac{1}{4}\left(A^{\dagger}A+A^{\dagger}B+A^{\dagger}C+A^{\dagger}D
+B^{\dagger}A+B^{\dagger}B+B^{\dagger}C+B^{\dagger}D\right.\nonumber\\
&&\left.+C^{\dagger}A+C^{\dagger}B+C^{\dagger}C+C^{\dagger}D
+D^{\dagger}A+D^{\dagger}B+D^{\dagger}C+D^{\dagger}D\right).
\label{Besetzungszahloperatorenxyzn}
\end{eqnarray}
Im Unterschied zu den anderen Besetzungszahloperatoren enthält $A_n^{\dagger}A_n$ nur Terme mit positivem Vorzeichen.
Basierend auf den entsprechenden Besetzungszahlen können nun Besetzungszahlzustände $|N_x,N_y,N_z,N_n\rangle$
definiert werden, auf welche die Operatoren ($\ref{ErzeugungsVernichtungsOperatorenXYZ}$) und
($\ref{ErzeugungsVernichtungsOperatorn}$) analog zu ($\ref{WirkungOperatoren}$) und ($\ref{WirkungOperatorenABCD}$)
in der folgenden Weise wirken:

\begin{eqnarray}
A_x|N_x,N_y,N_z,N_n\rangle&=&\sqrt{N_x}|N_x-1,N_y,N_z,N_n\rangle,\nonumber\\
A_x^{\dagger}|N_x,N_y,N_z,N_n\rangle&=&\sqrt{N_x+1}|N_x+1,N_y,N_z,N_n\rangle,\nonumber\\
A_y|N_x,N_y,N_z,N_n\rangle&=&\sqrt{N_y}|N_x,N_y-1,N_z,N_n\rangle,\nonumber\\
A_y^{\dagger}|N_x,N_y,N_z,N_n\rangle&=&\sqrt{N_y+1}|N_x,N_y+1,N_z,N_n\rangle,\nonumber\\
A_z|N_x,N_y,N_z,N_n\rangle&=&\sqrt{N_z}|N_x,N_y,N_z-1,N_n\rangle,\nonumber\\
A_z^{\dagger}|N_x,N_y,N_z,N_n\rangle&=&\sqrt{N_z+1}|N_x,N_y,N_z+1,N_n\rangle,\nonumber\\
A_n|N_x,N_y,N_u,N_n\rangle&=&\sqrt{N_n}|N_x,N_y,N_z,N_n-1\rangle,\nonumber\\
A_n^{\dagger}|N_x,N_y,N_z,N_n\rangle&=&\sqrt{N_n+1}|N_x,N_y,N_z,N_n+1\rangle.
\label{WirkungOperatorenxyzn}
\end{eqnarray}
Die Eigenwertgleichungen der Besetzungszahloperatoren ($\ref{Besetzungszahloperatorenxyzn}$) lauten
analog zu ($\ref{Eigenwertgleichungen}$) und ($\ref{EigenwertgleichungenABCD}$) wie folgt:

\begin{eqnarray}
&A_x^{\dagger}A_x|N_x,N_y,N_z,N_n\rangle=N_x|N_x,N_y,N_z,N_n\rangle,\quad
&A_y^{\dagger}A_y|N_x,N_y,N_z,N_n\rangle=N_y|N_x,N_y,N_z,N_n\rangle,\nonumber\\
&A_z^{\dagger}A_z|N_x,N_y,N_z,N_n\rangle=N_z|N_x,N_y,N_z,N_n\rangle,\quad
&A_n^{\dagger}A_n|N_x,N_y,N_z,N_n\rangle=N_n|N_x,N_y,N_z,N_n\rangle.\nonumber\\
\label{Eigenwertgleichungenxyzn}
\end{eqnarray}
Es gilt die folgende Relation:

\begin{equation}
N=A_{x}^{\dagger}A_x+A_{y}^{\dagger}A_y+A_{z}^{\dagger}A_z+A_{n}^{\dagger}A_n=A^{\dagger}A+B^{\dagger}B+C^{\dagger}C+D^{\dagger}D.
\end{equation}
Das bedeutet, dass die Gesamtbesetzungszahl in Bezug auf diese beiden Darstellungen eines bestimmten Zustandes
des Tensorraumes gleich ist. Man kann also einen Zustand im Tensorraum auch durch die Besetzungszahlen
$N_x$, $N_y$, $N_z$ und $N_n$ charakterisieren, womit die Basiszustände aus ($\ref{Basiszustand_Tensorraum}$)
überführt werden:

\begin{equation}
|N_A,N_B,N_C,N_D\rangle \quad\longleftrightarrow\quad |N_x,N_y,N_z,N_n\rangle.
\end{equation}
Wenn man in Analogie zu ($\ref{kompakteDarstellung}$) zu der kompakten Schreibweise übergeht:

\begin{equation}
|N_{xyzn}\rangle=|N_x,N_y,N_z,N_n\rangle,
\label{Basiszustand_Tensorraum_xyzn}
\end{equation}
dann kann man damit einen allgemeinen Zustand im Tensorraum ($\ref{Zustand_Tensorraum}$) basierend
auf diesen neuen Basiszuständen in der folgenden Weise ausdrücken:

\begin{eqnarray}
|\Psi\rangle=\sum_{N_{ABCD}}\psi\left(N_{ABCD}\right)|N_{ABCD}\rangle
=\sum_{N_{xyzn}}\psi\left(N_{xyzn}\right)|N_{xyzn}\rangle.
\label{Zustand_Tensorraum_xyzn}
\end{eqnarray}
Der vierte Freiheitsgrad, der durch den Besetzungszahloperator $A_n^{\dagger}A_n$ repräsentiert wird, ist aber,
wenn die anderen drei Besetzungszahlen fest definiert sind, durch die Gesamtbesetzungszahl mitbestimmt:

\begin{equation}
N_n=N-N_x-N_y-N_z.
\end{equation}
Im vierten Freiheitsgrad ist also implizit der Freiheitsgrad der Informationsmenge $N$ enthalten und das bedeutet,
dass bei konstanter Informationsmenge aufgrund der Gleichung:

\begin{equation}
A_n^{\dagger}A_n=N-A_x^{\dagger}A_x-A_y^{\dagger}A_y-A_z^{\dagger}A_z,
\end{equation}
die Besetzungzahl $A_n^{\dagger}A_n$ keinen unabhängigen Freiheitsgrad mehr darstellt.

Die Informationsmenge $N$ wird hier als eine eigenständige Größe behandelt, die im Mindesten bei einem
freien Objekt nicht in die Dynamik miteinbezogen ist und deshalb konstant bleibt. Überhaupt erscheint
es mir zumindest zweifelhaft, dass die Information mit der Zeit systematisch wächst, wie es von Weizsäcker
postulierte, denn von Weizsäcker verknüpfte diesen Gedanken mit der Entstehung von Fakten bei Messprozessen,
obwohl bei Messprozessen doch eigentlich nur Zustände in Eigenzustände übergehen und bei Ur"=Alternativen
würde das einen Übergang in Basiszustände aber nicht eine Vergrößerung der Zahl der Ur"=Alternativen bedeuten.
In diesem Zusammenhang erscheint es sinnvoll, zwischen zwei Arten der Information zu unterscheiden, und zwar
zwischen der Elementarinformation und der semantischen Information. Die Elementarinformation entspricht den
Informationseinheiten, die in etwas enthalten sind, und die semantische Information dem Bedeutungsgehalt.
Bei einer Textdatei etwa entspricht die in Bits gemessene Größe der Menge an Informationseinheiten,
die sie enthält, und der Inhalt des Textes dem Bedeutungsgehalt. Es können zwei Textdateien gleich lang
sein und die gleiche Menge an Bits enthalten, aber die eine Datei kann trotzdem nur eine sinnlose
Aneinanderreihung von Buchstaben enthalten und die andere eine sehr wichtige Botschaft, sodass sie
sich bezüglich der Menge an semantischem Informationsgehalt erheblich voneinander unterscheiden,
also auch bezüglich der Strukturen, die aus den elementaren Informationseinheiten gebildet werden.
Die Ur"=Alternativen stellen aber keine Information im Sinne eines semantischen Gehaltes dar,
sondern sind elementare Informationseinheiten analog zu den Bits im Computer. Bei einer Messung
oder bei bestimmten anderen physikalischen Prozessen wird allenfalls die semantische Information erhöht,
was bedeutet, dass die Ur"=Alternativen sich in einer solchen Weise reorganisieren, dass sich dabei
Strukturen bilden, die vom Menschen als Dokument für einen bestimmten Vorgang interpretiert werden können.
Aber selbst wenn die Zahl der Ur"=Alternativen mit der Zeit systematisch zunähme, was natürlich
auch aus anderen Gründen im Prinzip denkbar sein könnte, so wäre wahrscheinlich zumindest die
Dynamik freier Objekte von dem Freiheitsgrad der Informationsmenge unabhängig. Daher erscheint
es als plausibel, den vierten Teilraum im Tensorraum der Ur"=Alternativen, welcher also die
Menge an Information in einem Zustand enthält, nicht mit der indirekten Begründung einer
weiteren Dimension im physikalischen Ortsraum in Verbindung zu bringen, wie dies in Bezug
auf die anderen Teilräume durch die im nächsten Unterabschnitt vollzogene Abbildung der
Besetungszahlzustände in den Ortsraum als Darstellungsmedium geschehen soll.
Stattdessen ist davon auszugehen, dass der vierte Teilraum in diesem Modell eine davon
unabhängige Größe beschreibt, die bei der gewöhnlichen Dynamik eines freien Objektes als absolute
Informationsmenge, die in einem Tensorraumzustand vieler Ur"=Alternativen enthalten ist, konstant bleibt.
Die Zeit wird im übernächsten Unterabschnitt in die Beschreibung eingeführt. Dort wird dann auch die
spezifische Weise, in welcher sie im Rahmen dieser Beschreibung auftritt, und die in völligem Einklang
mit der speziellen und der allgemeinen Relativitätstheorie steht, in aller Gründlichkeit diskutiert.

\subsection{Abbildung der Zustände im Tensorraum in den Ortsraum}

Da die Besetzungszahl $N_n$ bei gegebenen Besetzungszahlen $N_x$, $N_y$, $N_z$ direkt über die Gesamtinformationsmenge $N$
definiert ist, stellt sie, wie erwähnt, keinen eigenständigen Freiheitsgrad eines freien Objektes dar, sondern ergibt
sich aus den anderen Freiheitsgraden. Daher werden in einem Basiszustand $|N_x,N_y,N_z,N_n\rangle$ nur die Teilräume,
welche den Basiszuständen $|N_x\rangle$, $|N_y\rangle$ und $|N_z\rangle$ entsprechen, in den Ortsraum abgebildet und
entsprechen dann den drei Raumdimensionen. Dies verhält sich analog zu einer Normierung eines Vektors, bei der die
Gesamtlänge keine unabhängige Dimension des Vektorraumes darstellt. Deshalb ergibt sich auch zunächst ein
dreidimensionaler und kein vierdimensionaler reeller Ortsraum als Darstellungsraum der Zustände im Tensorraum
vieler Ur"=Alternativen. Über die Sonderrolle der Zeitdimension, die in anderer Weise eingeführt werden muss,
wird im nächsten Unterabschnitt noch zu sprechen sein. Man kann nun die Orts- und Impulsoperatoren in
($\ref{Ort_Impuls_Operatoren}$) in der folgenden Weise in Bezug auf einen dreidimensionalen reellen
Ortsraum darstellen:

\begin{eqnarray}
X&=&\frac{1}{\sqrt{2}}\left(A_x+A_x^{\dagger}\right)=x,\quad P_x=-\frac{i}{\sqrt{2}}\left(A_x-A_x^{\dagger}\right)=-i\partial_x,\nonumber\\
Y&=&\frac{1}{\sqrt{2}}\left(A_x+A_x^{\dagger}\right)=y,\quad P_y=-\frac{i}{\sqrt{2}}\left(A_x-A_x^{\dagger}\right)=-i\partial_y,\nonumber\\
Z&=&\frac{1}{\sqrt{2}}\left(A_x+A_x^{\dagger}\right)=z,\quad P_z=-\frac{i}{\sqrt{2}}\left(A_x-A_x^{\dagger}\right)=-i\partial_z,
\label{Ort_Impuls_Darstellung}
\end{eqnarray}
wobei $x$, $y$ und $z$ einfach gewöhnliche reelle Koordinaten sind. Diese Darstellung der abstrakten Orts- und
Impulsoperatoren als Operatoren in einem dreidimensionalen reellen Raum ermöglicht in vollkommener mathematischer
Analogie zur Abbildung der Besetzungszahlzustände des mehrdimensionalen quantentheoretischen harmonischen
Oszillators die folgende räumliche Darstellung der Besetzungszahlzustände:

\begin{eqnarray}
|N_x \rangle \quad\longleftrightarrow\quad w_{N_x}\left(x\right)&=&\frac{1}{N_x!2^{N_x}\pi^2}\left(x-\partial_x\right)^{N_x}\exp\left(-\frac{x^2}{2}\right),\nonumber\\
|N_y \rangle \quad\longleftrightarrow\quad w_{N_y}\left(y\right)&=&\frac{1}{N_y!2^{N_y}\pi^2}\left(y-\partial_y\right)^{N_y}\exp\left(-\frac{y^2}{2}\right),\nonumber\\
|N_z \rangle \quad\longleftrightarrow\quad
w_{N_z}(z)&=&\frac{1}{N_z!2^{N_z}\pi^2}\left(z-\partial_z\right)^{N_z}\exp\left(-\frac{z^2}{2}\right).
\end{eqnarray}
Wenn aber der vierte Freiheitsgrad, wie oben ausgeführt, nur die Informationsmenge enthält und daher keine im
eigentlichen Sinne als eigenständiger Freiheitsgrad in die Dynamik einbezogene Größe darstellt, so bedeutet dies,
dass man einen Basiszustand des Tensorraumes in Bezug auf die bezüglich der Operatoren $A_x$, $A_y$, $A_z$ und $A_n$
bezogene Basis bei gegebener Informationsmenge in der folgenden Weise in eine Darstellung als eine Wellenfunktion
in einem dreidimensionalen Ortsraum überführen kann:

\begin{equation}
|N_{xyzn}\rangle \quad\longleftrightarrow\quad \left[w_{N_x}(x)w_{N_y}(y)w_{N_z}(z)\right]_N
\equiv f_N(N_x,N_y,N_z,x,y,z)\equiv f_{N_{xyzn}}\left(\textbf{x}\right),
\label{Basiszustand_Tensorraum_Ortsdarstellung}
\end{equation}
wobei der Index $N$ bezüglich $f_N(N_x,N_y,N_z,x,y,z)$ natürlich die Gesamtinformationsmenge in diesem Zustand
als viertem Freiheitsgrad andeutet. Es sei darauf hingewiesen, dass diese Wellenfunktionen normiert sind. Dies
ist also vollkommen anders als in der gewöhnlichen Quantenmechanik, in welcher die ebenen Wellen, die sich in
einem Kontinuum möglicher Zustände bewegen und mit Hilfe derer durch Überlagerung normierte Zustände von
Teilchen gebildet werden, an sich selbst gar nicht normiert sind. Im Rahmen der Quantentheorie der Ur"=Alternativen
sind die Basiszustände freier Objekte nicht nur diskret, sondern auch normiert, was wie im Falle des Wasserstoffatoms
miteinander in Zusammenhang zu stehen scheint. Dies ist ein weiteres überzeugendes Indiz für die Quantentheorie der
Ur"=Alternativen. Der Übergang ($\ref{Basiszustand_Tensorraum_Ortsdarstellung}$) bedeutet für einen allgemeinen
Zustand im Tensorraum, dass er in der folgenden Weise in eine Welle im Ortsraum überführt werden kann:

\begin{eqnarray}
&&|\Psi \rangle=\sum_{N_{ABCD}}\psi\left(N_{ABCD}\right)|N_{ABCD}\rangle=\sum_{N_{xyzn}}\psi\left(N_{xyzn}\right)|N_{xyzn}\rangle
\label{Zustand_Darstellung_Ortsraum}\\
&&\longleftrightarrow\quad \sum_{N_x,N_y,N_z,N_n}\psi\left(N_x,N_y,N_z,N_n\right)f_N\left(N_x,N_y,N_z,x,y,z\right)
=\sum_{N_{xyzn}}\psi\left(N_{xyzn}\right)f_{N_{xyzn}}(\mathbf{x})=\Psi\left(\mathbf{x}\right).\nonumber
\end{eqnarray}
Damit entspricht einem allgemeinen freien Zustand im Tensorraum vieler Ur"=Alternativen mit einer konstanten Informationsmenge $N$
eine Wellenfunktion in einem reellen dreidimensionalen Raum. Ein Zustand im Tensorraum ist also durch zwei Dinge charakterisiert:
Durch die Menge der in ihm enthaltenen Information $N$ und durch die Darstellung als Wellenfunktion in einem reellen
dreidimensionalen Raum. Wenn man Zustände dargestellt durch Wellenfunktionen mit unterschiedlicher Informationsmenge überlagert,
was in ($\ref{Zustand_Darstellung_Ortsraum}$) durch die Summierung über $N_n$ natürlich implizit geschieht, so ist das
Ergebnis wieder eine Wellenfunktion im Raum. Aufgrund dieser Isomorphie eines Zustandes im reinen abstrakten
quantentheoretischen Informationsraum der Ur"=Alternativen mit einer Wellenfunktion in einem reellen
dreidimensionalen Raum ist damit die Existenz des physikalischen Raumes aus einer reinen abstrakten
Quantentheorie der Information begründet. Dieser spielt die Rolle eines Mediums, in dem sich die Zustände
der Quantenobjekte indirekt darstellen, obwohl sie eigentlich einer abstrakteren Realität reiner quantentheoretischer
Informationseinheiten angehören. Dies erklärt exakt die Situation, die wir in den Phänomenen wahrnehmen, die sich uns
empirisch darstellen. Alle Objekte scheinen sich in einem dreidimensionalen reellen Raum zu befinden. Wenn wir aber bis
auf die Ebene der Quantenobjekte vordringen, so stellen sich diese Objekte zwar noch als Wellen in unserem Anschauungsraum dar,
aber verhalten sich in einer Weise als seien sie an die räumlichen Kausalstrukturen gar nicht mehr gebunden, was sich in
Phänomenen wie dem Doppelspaltexperiment, dem EPR"=Paradoxon und der Reduktion der Wellenfunktionen zeigt. In der Auffassung
der gewöhnlichen Quantenmechanik oder Quantenfeldtheorie tun wir so, als hätten wir es noch mit räumlichen Objekten wie
Teilchen und Feldern zu tun, deren Gesetze wir dann durch den Vorgang der Quantisierung abändern und die dann die gewöhnlichen
Paradoxien enthält. Die wirkliche Erklärung für diese paradoxe Situation liefert erst die Quantentheorie der Ur"=Alternativen,
indem sie erkennt, dass die Objekte gar keine räumlichen Objekte sind, sondern Zustände abstrakter Informationseinheiten,
die wie abstrakte Quantenzahlen behandelt werden müssen, die sich nur in dieser räumlichen Weise darstellen. Und deshalb
sind sie natürlich auch nicht an die Kausalstrukturen dieses Raumes gebunden. Um dies deutlicher zu machen, könnte man
einen Computer als Gleichnis heranziehen. Die Ur"=Alternativen als hinter der eigentlichen Erscheinung liegende Realität
verhalten sich zur räumlichen Wirklichkeit, die auf der Oberflächenebene angetroffen wird, in der gleichen Weise wie die
abstrakte Struktur elektrischer Signale auf dem Bildschirm. Auf dem Bildschirm befinden sich konkrete anschauliche
lokalisierte Objekte. Aber das ist nicht die eigentliche Realität im Computer, die einem abstrakten Muster elektrischer
Signale in einem Mikrochip entspricht. Wenn sich der Mauszeiger von der einen Seite des Bildschirms auf die andere bewegt,
so bedeutet dies nicht, dass sich der Mikrochip oder die darin enthaltenen Signale räumlich bewegen, sondern sie ändern
nur die Struktur. Dieser Vergleich ist natürlich begrenzt, da ja auch die abstrakten elektrischen Signale irgendwie
noch räumlich lokalisiert sind, während die Ur"=Alterativen der ganzen physikalischen Realität zu Grunde liegen und
sozusagen als reine Logik die physikalische Realität aus dem Nichts heraus erst erzeugen. Aber immerhin kann dieses
Gleichnis metaphorisch sehr deutlich und eindringlich klar machen, dass die konkrete räumliche Wirklichkeit nur eine
Darstellung einer fundamentaleren abstrakten Wirklichkeit ist, welche an die Gesetzmäßigkeiten dieser räumlichen
Oberflächenebene ebenso wenig gebunden ist wie das Verhalten der elektrischen Signale an die Beziehungen
auf dem Bildschirm. Der Mikrochip bestimmt die Gesetze und der Bildschirm stellt durch eine Überführung nur dar.
Das Geschehen auf dem Bildschirm ist abhängig vom der Signalstruktur im Mikrochip aber nicht umgekehrt. Und daher
verwundert es, wenn man diesen Gedanken zurück auf die Quantentheorie der Ur"=Alterativen überträgt, überhaupt
nicht mehr, dass in der Quantentheorie Geschehnisse und Effekte eintreten, die in keiner Weise mehr in
Übereinstimmung mit den lokalen Kausalstrukturen des physikalischen Raumes stehen.

\subsection{Die Dynamik in der Quantentheorie der Ur-Alternativen}

Die Zeit kommt erst über die Dynamik mit in die Beschreibung hinein. Sie kann bei der Darstellung eines Zustandes
in einem reellen Raum zunächst noch gar nicht enthalten sein, denn alleine aufgrund der Existenz einer Dynamik,
die in der Quantentheorie mit der allgemeinen Schrödinger"=Gleichung immer vorausgesetzt werden muss, enthält das
Wissen um die Zeitentwicklung keinerlei zusätzliche Information. Wenn der Zustand zu einem bestimmten Zeitpunkt
bekannt ist, dann ist er aufgrund der Dynamik der Schrödinger"=Gleichung im Prinzip bis in alle Ewigkeit bekannt,
solange keine unbekannte Wechselwirkung einwirkt und keine Messung durchgeführt wird. Daher kann die Zeitdimension
nicht einem unabhängigen Freiheitsgrad im Informationsraum entsprechen, sondern entspricht einem Automorphismus desselben
auf sich selbst. Dies steht in keinerlei Gegensatz zur speziellen Relativitätstheorie, denn deren veränderte Kinematik und
Dynamik mit ihrer Abhängigkeitsbeziehung zwischen Raum und Zeit kann man zwar formal in einer (3+1)"=dimensionalen Raum-Zeit
darstellen, womit die Zeit formal wie eine weitere Raumdimension eingeführt wird. Aber das bedeutet in keiner Weise,
dass damit der fundamentale Wesensunterschied zwischen Raum und Zeit aufgehoben wäre. Dies zeigt sich auch
formal indirekt darin, dass die Zeit in der Minkowski"=Metrik mit einem negativen Vorzeichen versehen und die Raum-
beziehungsweise Zeitartigkeit eines Zustandes Lorentz"=invariant ist. Viel wichtiger aber ist, dass die Zeit nur in
eine Richtung läuft und einer lebendigen Bewegung entspricht. Dies steht auch mit der wichtigen Erkenntnis Einsteins
in Zusammenhang, dass das "`Jetzt"' rein formal in der Physik nicht vorkommt, obwohl es natürlich zur tatsächlichen
Wirklichkeit der Zeit gehört. Die Zeit ist zudem, und das ist das Entscheidende in Bezug auf die Argumentation hier,
generell nicht mit einem unabhängigen Freiheitsgrad in der Physik verbunden wie das beim Raum der Fall ist, auch nicht
in der Relativitätstheorie, denn durch die Dynamik ist die Zeitentwicklung determiniert, was in der Relativitätstheorie der
Tatsache entspricht, dass in der (3+1)"=dimensionalen Raum"=Zeit ein Freiheitsgrad weggenommen wird. Dem möglichen und
zunächst vielleicht naheliegenden Einwand, dass durch eine unabhängige Einführung der Zeitkoordinate die Einheit der
Raum"=Zeit im Sinne der Transformation zwischen Raum- und Zeitkoordinaten nicht gewährleistet sei, kann man natürlich
damit begegnen, dass sobald mit der Schödinger"=Gleichung eine Zeitentwicklung in die Beschreibung gebracht wird,
über eine relativistische Definition des Hamilton"=Operators implizit natürlich die relativistischen Relationen zwischen
Energie und Impuls und damit auch zwischen Zeitkoordinate und Ortskoordinate mit in die Beschreibung hineinkommen.
Genaugenommen entspricht dem dreidimensionalen reellen Raum, welcher zur Darstellung der Zustände im Tensorraum dient,
eine raumartige Hyperfläche im Raum"=Zeit"=Kontinuum, während der über die Schrödinger"=Gleichung eingeführten Zeitentwicklung
dann die vierte Dimension der Gesamtmannigfaltigkeit entspricht, in welche diese Hyperfläche eingebettet ist.
Man ist in der Definition dieser Hyperfläche frei, welche durch die entsprechende Identifikation der Koordinaten der
Raumdarstellung mit den Koordinaten dieser Hyperfläche entspricht. Raum und Zeit sind also faktisch aufgrund der
Interpretation der drei Raumkoordinaten als Hyperfläche in einer formalen Raum"=Zeit, der Entstehung einer vollen
Raum"=Zeit durch die Dynamik gemäß der Schrödinger"=Gleichung und der relativistischen Definition des Hamilton"=Operators,
der die Zeitentwicklung im relativistischen Sinne zu den Impulsoperatoren in Beziehung setzt und damit auch zu den
Ortsoperatoren, doch miteinander verbunden. Und es kommt wie gesagt nicht auf die formale Beschreibung an,
sondern auf die faktische Gegenwart der kinematischen und dynamischen Beziehungen der speziellen Relativitätstheorie,
welche durch eine relativistische Definition des Hamilton"=Operators absolut gewährleistet ist.

Um also die Zeit und die Dynamik in die Beschreibung der Quantentheorie der Ur"=Alternativen zu integrieren,
muss zunächst die Schrödinger"=Gleichung in ihrer allgemeinen abstrakten Form zu Grunde gelegt werden,
wie sie in ($\ref{Schroedinger_Gleichung}$) mit Bezug auf einen allgemeinen auf den Begriff einer
logischen Alternative bezogenen Zustand formuliert wurde. Diese lautet in Bezug auf die Darstellung
einer quantentheoretischen Alternative durch binäre Alternativen, also als Zustand im Tensorraum
der Ur"=Alternativen ($\ref{Zustand_Tensorraum}$), wie folgt:

\begin{equation}
i\partial_t|\Psi(t)\rangle=H|\Psi(t)\rangle,
\label{Schroedinger-Gleichung_TensorraumA}
\end{equation}
wobei die Zustände $|\Psi(t)\rangle$ gemäß ($\ref{Zustand_Tensorraum}$) beziehungsweise ($\ref{Zustand_Tensorraum_xyzn}$)
definiert sind, nur dass sie jetzt zeitabhängig werden. Selbstredend legt dies zunächst nur die allgemeine dynamische
Grundstruktur gemäß der abstrakten Quantentheorie fest. Die Zeitentwicklung ist damit in der folgenden Weise determiniert:

\begin{equation}
|\Psi(t)\rangle=e^{-iHt}|\Psi(t_0)\rangle.
\label{ZeitentwicklungA}
\end{equation}
Über die konkrete Gestalt des Hamilton"=Operators $H$ ist damit noch nichts ausgesagt. Wenn man nun den konkreten
durch Ur"=Alternativen dargestellten freien Hamilton"=Operator mit in die Betrachtung hineinnimmt, welcher auf
der Beziehung des Energieoperators zu den Impulsoperatoren ($\ref{Energie-Impuls-Relation}$) basiert, so kann
man zu einer konkreten Dynamik im Tensorraum der Ur"=Alternativen gelangen. Wenn also $|\Psi(t)\rangle$ als
ein allgemeiner zeitabhängiger Zustand im Tensorraum angesehen wird, und der Hamilton"=Operator über die
Energie"=Impuls"=Relation ($\ref{Energie-Impuls-Relation}$) definiert wird, so bedeutet dies,
dass die Schrödinger"=Gleichung folgende konkrete Gestalt annimmt:

\begin{equation}
i\partial_t\sum_{N_{ABCD}}\psi\left(N_{ABCD},t\right)|N_{ABCD}\rangle
=\sqrt{P_x^2+P_y^2+P_z^2}\sum_{N_{ABCD}}\psi\left(N_{ABCD},t\right)|N_{ABCD}\rangle.
\label{Schroedinger-Gleichung_TensorraumB}
\end{equation}
Als Lösung der Schrödinger"=Gleichung ergibt sich damit die folgende Gestalt
eines allgemeinen zeitabhängigen Zustandes im Tensorraum:

\begin{eqnarray}
|\Psi(t)\rangle&=&\sum_{N_{ABCD}}\psi\left(N_{ABCD},t\right)|N_{ABCD}\rangle\nonumber\\
&=&e^{-i\left(\sqrt{P_x^2+P_y^2+P_z^2}\right)t}\sum_{N_{ABCD}}\psi\left(N_{ABCD},t_0\right)|N_{ABCD}\rangle\nonumber\\
&=&\sum_{N_{ABCD}}\psi\left(N_{ABCD},t_0\right)e^{-i\left(\sqrt{P_x^2+P_y^2+P_z^2}\right)t}|N_{ABCD}\rangle.
\label{ZeitentwicklungB}
\end{eqnarray}
Um nun die Komponenten zur Zeit $t$, $\psi\left(N_{ABCD},t\right)$, in Abhängigkeit der Komponenten
zur Zeit $t_0$, $\psi\left(N_{ABCD},t_0\right)$, zu bestimmen, muss zunächst einmal der in
($\ref{Energie-Impuls-Relation}$) definierte Energieoperator auf die Basiszustände des Tensorraumes
angewandt werden: $E|N_{ABCD}\rangle$. Um den hieraus sich ergebenden Zustand zu bestimmen,
muss zunächst die Anwendung des Quadrates des Energieoperators betrachtet werden:

\begin{eqnarray}
&&E^2|\Psi\rangle=\sum_{N_{ABCD}}\psi(N_{ABCD})E^2|N_{ABCD}\rangle
=\sum_{N_{ABCD}}\psi(N_{ABCD})\left(P_x^2+P_y^2+P_z^2\right)|N_{ABCD}\rangle\nonumber\\
&&=\frac{1}{8}\sum_{N_{ABCD}}\psi(N_{ABCD})\left[
-3AA-3BB-3CC-3DD-3A^{\dagger}A^{\dagger}-3B^{\dagger}B^{\dagger}
-3C^{\dagger}C^{\dagger}-3D^{\dagger}D^{\dagger}
\right.\nonumber\\ &&\left.
+2AB+2AC+2AD+2BC+2BD+2CD
\right.\nonumber\\ &&\left.
+2A^{\dagger}B^{\dagger}+2A^{\dagger}C^{\dagger}+2A^{\dagger}D^{\dagger}
+2B^{\dagger}C^{\dagger}+2B^{\dagger}D^{\dagger}+2C^{\dagger}D^{\dagger}
\right.\nonumber\\ &&\left.
+2\left(3A^{\dagger}A+3B^{\dagger}B+3C^{\dagger}C+3D^{\dagger}D
\right.\right.\nonumber\\ &&\left.\left.
-A^{\dagger}B-A^{\dagger}C-A^{\dagger}D-B^{\dagger}A-B^{\dagger}C-B^{\dagger}D
\right.\right.\nonumber\\ &&\left.\left.
-C^{\dagger}A-C^{\dagger}B-C^{\dagger}D-D^{\dagger}A-D^{\dagger}B-D^{\dagger}C\right)+3\right]|N_{ABCD}\rangle.
\end{eqnarray}
Dies kann weiter umgeformt werden zu:

\begin{align}
&E^2|\Psi\rangle=\frac{1}{8}\sum_{N_{ABCD}}\psi(N_{ABCD})\left[-3\sqrt{N_A}\sqrt{N_A-1}|N_A-2,N_B,N_C,N_D\rangle
\right.\\&\left.-3\sqrt{N_B}\sqrt{N_B-1}|N_A,N_B-2,N_C,N_D\rangle\right.\nonumber\\
&\left.-3\sqrt{N_C}\sqrt{N_C-1}|N_A,N_B,N_C-2,N_D\rangle-3\sqrt{N_D}\sqrt{N_D-1}|N_A,N_B,N_C,N_D-2\rangle\right.\nonumber\\
&\left.-3\sqrt{N_A+2}\sqrt{N_A+1}|N_A+2,N_B,N_C,N_D\rangle-3\sqrt{N_B+2}\sqrt{N_B+1}|N_A,N_B+2,N_C,N_D\rangle\right.\nonumber\\
&\left.-3\sqrt{N_C+2}\sqrt{N_C+1}|N_A,N_B,N_C+2,N_D\rangle-3\sqrt{N_D+2}\sqrt{N_D+1}|N_A,N_B,N_C,N_D+2\rangle\right.\nonumber\\
&\left.+2\sqrt{N_A}\sqrt{N_B}|N_A-1,N_B-1,N_C,N_D\rangle+2\sqrt{N_A}\sqrt{N_C}|N_A-1,N_B,N_C-1,N_D\rangle\right.\nonumber\\
&\left.+2\sqrt{N_A}\sqrt{N_D}|N_A-1,N_B,N_C,N_D-1\rangle+2\sqrt{N_B}\sqrt{N_C}|N_A,N_B-1,N_C-1,N_D\rangle\right.\nonumber\\
&\left.+2\sqrt{N_B}\sqrt{N_D}|N_A,N_B-1,N_C,N_D-1\rangle+2\sqrt{N_C}\sqrt{N_D}|N_A,N_B,N_C-1,N_D-1\rangle\right.\nonumber\\
&\left.+2\sqrt{N_A+1}\sqrt{N_B+1}|N_A+1,N_B+1,N_C,N_D\rangle+2\sqrt{N_A+1}\sqrt{N_C+1}|N_A+1,N_B,N_C+1,N_D\rangle\right.\nonumber\\
&\left.+2\sqrt{N_A+1}\sqrt{N_D+1}|N_A+1,N_B,N_C,N_D+1\rangle+2\sqrt{N_B+1}\sqrt{N_C+1}|N_A,N_B+1,N_C+1,N_D\rangle\right.\nonumber\\
&\left.+2\sqrt{N_B+1}\sqrt{N_D+1}|N_A,N_B+1,N_C,N_D+1\rangle+2\sqrt{N_C+1}\sqrt{N_D+1}|N_A,N_B,N_C+1,N_D+1\rangle\right.\displaybreak\nonumber\\
&\left.+2\left(3N_A|N_A,N_B,N_C,N_D\rangle+3N_B|N_A,N_B,N_C,N_D\rangle+3N_C|N_A,N_B,N_C,N_D\rangle+3N_D|N_A,N_B,N_C,N_D\rangle\right.\right.\nonumber\\
&\left.\left.-\sqrt{N_A+1}\sqrt{N_B}|N_A+1,N_B-1,N_C,N_D\rangle-\sqrt{N_A+1}\sqrt{N_C}|N_A+1,N_B,N_C-1,N_D\rangle\right.\right.\nonumber\\
&\left.\left.-\sqrt{N_A+1}\sqrt{N_D}|N_A+1,N_B,N_C,N_D-1\rangle-\sqrt{N_A}\sqrt{N_B+1}|N_A-1,N_B+1,N_C,N_D\rangle\right.\right.\nonumber\\
&\left.\left.-\sqrt{N_B+1}\sqrt{N_C}|N_A,N_B+1,N_C-1,N_D\rangle-\sqrt{N_B+1}\sqrt{N_D}|N_A,N_B+1,N_C,N_D-1\rangle\right.\right.\nonumber\\
&\left.\left.-\sqrt{N_A}\sqrt{N_C+1}|N_A-1,N_B,N_C+1,N_D\rangle-\sqrt{N_B}\sqrt{N_C+1}|N_A,N_B-1,N_C+1,N_D\rangle\right.\right.\nonumber\\
&\left.\left.-\sqrt{N_C+1}\sqrt{N_D}|N_A,N_B,N_C+1,N_D-1\rangle-\sqrt{N_A}\sqrt{N_D+1}|N_A-1,N_B,N_C,N_D+1\rangle\right.\right.\nonumber\\
&\left.\left.-\sqrt{N_B}\sqrt{N_D+1}|N_A,N_B-1,N_C,N_D+1\rangle-\sqrt{N_C}\sqrt{N_D+1}|N_A,N_B,N_C-1,N_D+1\rangle\right)\right.\nonumber\\
&\left.+3|N_A,N_B,N_C,N_D\rangle\right],\nonumber
\end{align}
und durch Verschiebung der Indizes weiter zu:

\begin{align}
&E^2|\Psi\rangle=\frac{1}{8}\sum_{N_{ABCD}}\left[-3\psi(N_A+2,N_B,N_C,N_D)\sqrt{N_A+2}\sqrt{N_A+1}\right.\\
&\left.-3\psi(N_A,N_B+2,N_C,N_D)\sqrt{N_B+2}\sqrt{N_B+1}\right.\nonumber\\
&\left.-3\psi(N_A,N_B,N_C+2,N_D)\sqrt{N_C+2}\sqrt{N_C+1}-3\psi(N_A,N_B,N_C,N_D+2)\sqrt{N_D+2}\sqrt{N_D+1}\right.\nonumber\\
&\left.-3\psi(N_A-2,N_B,N_C,N_D)\sqrt{N_A}\sqrt{N_A-1}-3\psi(N_A,N_B-2,N_C,N_D)\sqrt{N_B}\sqrt{N_B-1}\right.\nonumber\\
&\left.-3\psi(N_A,N_B,N_C-1,N_D)\sqrt{N_C}\sqrt{N_C-1}-3\psi(N_A+2,N_B,N_C,N_D-2)\sqrt{N_D}\sqrt{N_D-1}\right.\nonumber\\
&\left.+2\psi(N_A+1,N_B+1,N_C,N_D)\sqrt{N_A+1}\sqrt{N_B+1}+2\psi(N_A+1,N_B,N_C+1,N_D)\sqrt{N_A+1}\sqrt{N_C+1}\right.\nonumber\\
&\left.+2\psi(N_A+1,N_B,N_C,N_D+1)\sqrt{N_A+1}\sqrt{N_D+1}+2\psi(N_A,N_B+1,N_C+1,N_D)\sqrt{N_B+1}\sqrt{N_C+1}\right.\nonumber\\
&\left.+2\psi(N_A,N_B+1,N_C,N_D+1)\sqrt{N_B+1}\sqrt{N_D+1}+2\psi(N_A,N_B,N_C+1,N_D+1)\sqrt{N_C+1}\sqrt{N_D+1}\right.\nonumber\\
&\left.+2\psi(N_A-1,N_B-1,N_C,N_D)\sqrt{N_A-1}\sqrt{N_B-1}+2\psi(N_A-1,N_B,N_C-1,N_D)\sqrt{N_A-1}\sqrt{N_C-1}\right.\nonumber\\
&\left.+2\psi(N_A-1,N_B,N_C,N_D-1)\sqrt{N_A-1}\sqrt{N_D-1}+2\psi(N_A,N_B-1,N_C-1,N_D)\sqrt{N_B-1}\sqrt{N_C-1}\right.\nonumber\\
&\left.+2\psi(N_A,N_B-1,N_C,N_D-1)\sqrt{N_B-1}\sqrt{N_D-1}+2\psi(N_A,N_B,N_C-1,N_D-1)\sqrt{N_C-1}\sqrt{N_D-1}\right.\nonumber\\
&\left.+2\left(3\psi(N_A,N_B,N_C,N_D)N_A+3\psi(N_A,N_B,N_C,N_D)N_B+3\psi(N_A,N_B,N_C,N_D)N_C+3\psi(N_A,N_B,N_C,N_D)N_D\right.\right.\nonumber\\
&\left.\left.-\psi(N_A+1,N_B-1,N_C,N_D)\sqrt{N_A+1}\sqrt{N_B}-\psi(N_A+1,N_B,N_C-1,N_D)\sqrt{N_A+1}\sqrt{N_C}\right.\right.\nonumber\\
&\left.\left.-\psi(N_A+1,N_B,N_C,N_D-1)\sqrt{N_A+1}\sqrt{N_D}-\psi(N_A-1,N_B+1,N_C,N_D)\sqrt{N_A}\sqrt{N_B+1}\right.\right.\nonumber\\
&\left.\left.-\psi(N_A,N_B+1,N_C-1,N_D)\sqrt{N_B+1}\sqrt{N_C}-\psi(N_A,N_B+1,N_C,N_D-1)\sqrt{N_B+1}\sqrt{N_D}\right.\right.\nonumber\\
&\left.\left.-\psi(N_A-1,N_B,N_C+1,N_D)\sqrt{N_A}\sqrt{N_C+1}-\psi(N_A,N_B-1,N_C+1,N_D)\sqrt{N_B}\sqrt{N_C+1}\right.\right.\nonumber\\
&\left.\left.-\psi(N_A,N_B,N_C+1,N_D-1)\sqrt{N_C+1}\sqrt{N_D}-\psi(N_A-1,N_B,N_C,N_D+1)\sqrt{N_A}\sqrt{N_D+1}\right.\right.\nonumber\\
&\left.\left.-\psi(N_A,N_B-1,N_C,N_D+1)\sqrt{N_B}\sqrt{N_D+1}-\psi(N_A,N_B,N_C-1,N_D+1)\sqrt{N_C}\sqrt{N_D+1}\right)\right.\nonumber\\
&\left.+3\psi(N_A,N_B,N_C,N_D)\right]|N_{ABCD}\rangle.\nonumber
\end{align}
Diese Umformung ist nur unter der Voraussetzung möglich, dass:

\begin{equation}
\psi(N_A,N_B,N_C,N_D)=0,\quad \textrm{wenn}\quad \left(N_A < 2\right)\ \lor\ \left(N_B < 2\right)
\ \lor\ \left(N_C < 2\right)\ \lor\ \left(N_D < 2\right).
\end{equation}
Dadurch kann man dann die Anwendung des Energieoperators auf einen beliebigen Basiszustand
des Tensorraumes bestimmen zu:

\begin{align}
&E|\Psi\rangle=\sum_{N_{ABCD}}\psi(N_{ABCD})E|N_{ABCD}\rangle=\sum_{N_{ABCD}}\psi(N_{ABCD})\sqrt{P_x^2+P_y^2+P_z^2}|N_{ABCD}\rangle\nonumber\\
&=\frac{1}{2\sqrt{2}}\left\{\left[
-3\psi(N_A+2,N_B,N_C,N_D)\sqrt{N_A+2}\sqrt{N_A+1}-3\psi(N_A,N_B+2,N_C,N_D)\sqrt{N_B+2}\sqrt{N_B+1}\right.\right.\nonumber\\
&\left.\left.-3\psi(N_A,N_B,N_C+2,N_D)\sqrt{N_C+2}\sqrt{N_C+1}-3\psi(N_A,N_B,N_C,N_D+2)\sqrt{N_D+2}\sqrt{N_D+1}\right.\right.\nonumber\\
&\left.\left.-3\psi(N_A-2,N_B,N_C,N_D)\sqrt{N_A}\sqrt{N_A-1}-3\psi(N_A,N_B-2,N_C,N_D)\sqrt{N_B}\sqrt{N_B-1}\right.\right.\nonumber\\
&\left.\left.-3\psi(N_A,N_B,N_C-1,N_D)\sqrt{N_C}\sqrt{N_C-1}-3\psi(N_A+2,N_B,N_C,N_D-2)\sqrt{N_D}\sqrt{N_D-1}\right.\right.\nonumber\\
&\left.\left.+2\psi(N_A+1,N_B+1,N_C,N_D)\sqrt{N_A+1}\sqrt{N_B+1}+2\psi(N_A+1,N_B,N_C+1,N_D)\sqrt{N_A+1}\sqrt{N_C+1}\right.\right.\nonumber\\
&\left.\left.+2\psi(N_A+1,N_B,N_C,N_D+1)\sqrt{N_A+1}\sqrt{N_D+1}+2\psi(N_A,N_B+1,N_C+1,N_D)\sqrt{N_B+1}\sqrt{N_C+1}\right.\right.\nonumber\\
&\left.\left.+2\psi(N_A,N_B+1,N_C,N_D+1)\sqrt{N_B+1}\sqrt{N_D+1}+2\psi(N_A,N_B,N_C+1,N_D+1)\sqrt{N_C+1}\sqrt{N_D+1}\right.\right.\nonumber\\
&\left.\left.+2\psi(N_A-1,N_B-1,N_C,N_D)\sqrt{N_A-1}\sqrt{N_B-1}+2\psi(N_A-1,N_B,N_C-1,N_D)\sqrt{N_A-1}\sqrt{N_C-1}\right.\right.\nonumber\\
&\left.\left.+2\psi(N_A-1,N_B,N_C,N_D-1)\sqrt{N_A-1}\sqrt{N_D-1}+2\psi(N_A,N_B-1,N_C-1,N_D)\sqrt{N_B-1}\sqrt{N_C-1}\right.\right.\nonumber\\
&\left.\left.+2\psi(N_A,N_B-1,N_C,N_D-1)\sqrt{N_B-1}\sqrt{N_D-1}+2\psi(N_A,N_B,N_C-1,N_D-1)\sqrt{N_C-1}\sqrt{N_D-1}\right.\right.\nonumber\\
&\left.\left.+2\left(3\psi(N_A,N_B,N_C,N_D)N_A+3\psi(N_A,N_B,N_C,N_D)N_B+3\psi(N_A,N_B,N_C,N_D)N_C+3\psi(N_A,N_B,N_C,N_D)N_D
\right.\right.\right.\nonumber\\
&\left.\left.\left.-\psi(N_A+1,N_B-1,N_C,N_D)\sqrt{N_A+1}\sqrt{N_B}-\psi(N_A+1,N_B,N_C-1,N_D)\sqrt{N_A+1}\sqrt{N_C}
\right.\right.\right.\nonumber\\
&\left.\left.\left.-\psi(N_A+1,N_B,N_C,N_D-1)\sqrt{N_A+1}\sqrt{N_D}-\psi(N_A-1,N_B+1,N_C,N_D)\sqrt{N_A}\sqrt{N_B+1}
\right.\right.\right.\nonumber\\
&\left.\left.\left.-\psi(N_A,N_B+1,N_C-1,N_D)\sqrt{N_B+1}\sqrt{N_C}-\psi(N_A,N_B+1,N_C,N_D-1)\sqrt{N_B+1}\sqrt{N_D}
\right.\right.\right.\nonumber\\
&\left.\left.\left.-\psi(N_A-1,N_B,N_C+1,N_D)\sqrt{N_A}\sqrt{N_C+1}-\psi(N_A,N_B-1,N_C+1,N_D)\sqrt{N_B}\sqrt{N_C+1}
\right.\right.\right.\nonumber\\
&\left.\left.\left.-\psi(N_A,N_B,N_C+1,N_D-1)\sqrt{N_C+1}\sqrt{N_D}-\psi(N_A-1,N_B,N_C,N_D+1)\sqrt{N_A}\sqrt{N_D+1}
\right.\right.\right.\nonumber\\
&\left.\left.\left.-\psi(N_A,N_B-1,N_C,N_D+1)\sqrt{N_B}\sqrt{N_D+1}-\psi(N_A,N_B,N_C-1,N_D+1)\sqrt{N_C}\sqrt{N_D+1}\right)
\right.\right.\nonumber\\&\left.\left.
+3\psi(N_A,N_B,N_C,N_D)\right]\times \left[\psi(N_{ABCD})\right]^{\frac{1}{2}}\right\}|N_{ABCD}\rangle.
\end{align}
Wenn man definiert:

\begin{align}
&\mathcal{E}_\psi\left(N_{ABCD}\right)
=\frac{1}{2\sqrt{2}}\left[-3\psi(N_A+2,N_B,N_C,N_D)\sqrt{N_A+2}\sqrt{N_A+1}
\right.\nonumber\\&\left.
-3\psi(N_A,N_B+2,N_C,N_D)\sqrt{N_B+2}\sqrt{N_B+1}\right.\nonumber\\
&\left.-3\psi(N_A,N_B,N_C+2,N_D)\sqrt{N_C+2}\sqrt{N_C+1}-3\psi(N_A,N_B,N_C,N_D+2)\sqrt{N_D+2}\sqrt{N_D+1}\right.\nonumber\\
&\left.-3\psi(N_A-2,N_B,N_C,N_D)\sqrt{N_A}\sqrt{N_A-1}-3\psi(N_A,N_B-2,N_C,N_D)\sqrt{N_B}\sqrt{N_B-1}\right.\nonumber\\
&\left.-3\psi(N_A,N_B,N_C-1,N_D)\sqrt{N_C}\sqrt{N_C-1}-3\psi(N_A+2,N_B,N_C,N_D-2)\sqrt{N_D}\sqrt{N_D-1}\right.\nonumber\\
&\left.+2\psi(N_A+1,N_B+1,N_C,N_D)\sqrt{N_A+1}\sqrt{N_B+1}+2\psi(N_A+1,N_B,N_C+1,N_D)\sqrt{N_A+1}\sqrt{N_C+1}\right.\nonumber\\
&\left.+2\psi(N_A+1,N_B,N_C,N_D+1)\sqrt{N_A+1}\sqrt{N_D+1}+2\psi(N_A,N_B+1,N_C+1,N_D)\sqrt{N_B+1}\sqrt{N_C+1}\right.\nonumber\\
&\left.+2\psi(N_A,N_B+1,N_C,N_D+1)\sqrt{N_B+1}\sqrt{N_D+1}+2\psi(N_A,N_B,N_C+1,N_D+1)\sqrt{N_C+1}\sqrt{N_D+1}\right.\displaybreak\nonumber\\
&\left.+2\psi(N_A-1,N_B-1,N_C,N_D)\sqrt{N_A-1}\sqrt{N_B-1}+2\psi(N_A-1,N_B,N_C-1,N_D)\sqrt{N_A-1}\sqrt{N_C-1}\right.\nonumber\\
&\left.+2\psi(N_A-1,N_B,N_C,N_D-1)\sqrt{N_A-1}\sqrt{N_D-1}+2\psi(N_A,N_B-1,N_C-1,N_D)\sqrt{N_B-1}\sqrt{N_C-1}\right.\nonumber\\
&\left.+2\psi(N_A,N_B-1,N_C,N_D-1)\sqrt{N_B-1}\sqrt{N_D-1}+2\psi(N_A,N_B,N_C-1,N_D-1)\sqrt{N_C-1}\sqrt{N_D-1}\right.\nonumber\\
&\left.+2\left(3\psi(N_A,N_B,N_C,N_D)N_A+3\psi(N_A,N_B,N_C,N_D)N_B+3\psi(N_A,N_B,N_C,N_D)N_C+3\psi(N_A,N_B,N_C,N_D)N_D
\right.\right.\nonumber\\
&\left.\left.-\psi(N_A+1,N_B-1,N_C,N_D)\sqrt{N_A+1}\sqrt{N_B}-\psi(N_A+1,N_B,N_C-1,N_D)\sqrt{N_A+1}\sqrt{N_C}
\right.\right.\nonumber\\
&\left.\left.-\psi(N_A+1,N_B,N_C,N_D-1)\sqrt{N_A+1}\sqrt{N_D}-\psi(N_A-1,N_B+1,N_C,N_D)\sqrt{N_A}\sqrt{N_B+1}
\right.\right.\nonumber\\
&\left.\left.-\psi(N_A,N_B+1,N_C-1,N_D)\sqrt{N_B+1}\sqrt{N_C}-\psi(N_A,N_B+1,N_C,N_D-1)\sqrt{N_B+1}\sqrt{N_D}
\right.\right.\nonumber\\
&\left.\left.-\psi(N_A-1,N_B,N_C+1,N_D)\sqrt{N_A}\sqrt{N_C+1}-\psi(N_A,N_B-1,N_C+1,N_D)\sqrt{N_B}\sqrt{N_C+1}
\right.\right.\nonumber\\
&\left.\left.-\psi(N_A,N_B,N_C+1,N_D-1)\sqrt{N_C+1}\sqrt{N_D}-\psi(N_A-1,N_B,N_C,N_D+1)\sqrt{N_A}\sqrt{N_D+1}
\right.\right.\nonumber\\
&\left.\left.-\psi(N_A,N_B-1,N_C,N_D+1)\sqrt{N_B}\sqrt{N_D+1}-\psi(N_A,N_B,N_C-1,N_D+1)\sqrt{N_C}\sqrt{N_D+1}\right)\right.
\nonumber\\&\left.
+3\psi(N_A,N_B,N_C,N_D)\right]\times\left[\psi(N_{ABCD})\right]^{\frac{1}{2}},
\end{align}
so bedeutet dies:

\begin{eqnarray}
E|\Psi\rangle&=&\sum_{N_{ABCD}}\psi(N_{ABCD})E|N_{ABCD}\rangle=\sum_{N_{ABCD}}\psi(N_{ABCD})\sqrt{P_x^2+P_y^2+P_z^2}|N_{ABCD}\rangle\nonumber\\
&=&\sum_{N_{ABCD}}\mathcal{E}_\psi\left(N_{ABCD}\right)|N_{ABCD}\rangle=\sum_{N_{ABCD}}\mathcal{P}^0_\psi\left(N_{ABCD}\right)|N_{ABCD}\rangle
=|\mathcal{P}^0\Psi\rangle.
\label{Anwendung_Energie_Zustand}
\end{eqnarray}
Und mit ($\ref{ZeitentwicklungA}$) und ($\ref{ZeitentwicklungB}$) erhält man über ($\ref{Anwendung_Energie_Zustand}$)
die Zeitentwicklung eines Zustandes im Tensorraum der Ur"=Alternativen. Denn durch eine Iteration
der Anwendung des Energieoperators, also des Hamilton"=Operators, auf den Zustand im Sinne von
($\ref{Anwendung_Energie_Zustand}$), kann man dann im Prinzip alle Terme der Anwendung der
Reihenentwicklung der Exponentialfunktion mit dem Energieoperator berechnen und in der entsprechenden
Näherung die zeitabhängigen Koeffizienten $\psi\left(N_{ABCD},t\right)$ erhalten. Um den Übergang in
den Ortsraum zu vollziehen, muss die Anwendung des Energieoperators in Bezug auf die alternative
Darstellung der Basiszustände gemäß ($\ref{Basiszustand_Tensorraum_xyzn}$) durchgeführt werden.
Die Anwendung des Quadrates des Energieoperators ($\ref{Energie_Operator}$) liefert den
folgenden Ausdruck:

\begin{eqnarray}
&&E^2|\Psi\rangle=\sum_{N_{xyzn}}\psi(N_{xyzn})E^2|N_{xyzn}\rangle=\sum_{N_{xyzn}}\psi(N_{xyzn})\left(P_x^2+P_y^2+P_z^2\right)|N_{xyzn}\rangle\nonumber\\
&&=\frac{1}{2}\left(-A_x A_x+2A_x^{\dagger}A_x-A_x^{\dagger}A_x^{\dagger}
-A_y A_y+2A_y^{\dagger}A_y-A_y^{\dagger}A_y^{\dagger}\right.\nonumber\\&&\left.
-A_z A_z+2A_z^{\dagger}A_z-A_z^{\dagger}A_z^{\dagger}+3\right)|N_{xyzn}\rangle\nonumber\\
&&=\frac{1}{2}\left\{-\psi(N_x+2,N_y,N_z,N_n)\sqrt{N_x+2}\sqrt{N_x+1}-\psi(N_x,N_y+2,N_z,N_n)\sqrt{N_y+2}\sqrt{N_y+1}
\right.\nonumber\\&&\left.
-\psi(N_x,N_y,N_z+2,N_n)\sqrt{N_z+2}\sqrt{N_z+1}-\psi(N_x-2,N_y,N_z,N_n)\sqrt{N_x}\sqrt{N_x-1}
\right.\nonumber\\&&\left.
-\psi(N_x,N_y-2,N_z,N_n)\sqrt{N_y}\sqrt{N_y-1}-\psi(N_x,N_y,N_z-2,N_n)\sqrt{N_z}\sqrt{N_z-1}
\right.\nonumber\\&&\left.
+2\left[\psi(N_x,N_y,N_z,N_n)N_x+\psi(N_x,N_y,N_z,N_n)N_y
\right.\right.\nonumber\\&&\left.\left.
+\psi(N_x,N_y,N_z,N_n)N_z\right]+3\psi(N_x,N_y,N_z,N_n)\right\}|N_{xyzn}\rangle.
\end{eqnarray}
Dadurch kann man dann die Anwendung des Energieoperators auf einen beliebigen Basiszustand
des Tensorraumes bestimmen zu:

\begin{eqnarray}
&&E|\Psi\rangle=\sum_{N_{xyzn}}\psi(N_{xyzn})E|N_{xyzn}\rangle=\sum_{N_{xyzn}}\psi(N_{xyzn})\sqrt{P_x^2+P_y^2+P_z^2}|N_{xyzn}\rangle\nonumber\\
&&=\frac{1}{2\sqrt{2}}\left\{-\psi(N_x+2,N_y,N_z,N_n)\sqrt{N_x+2}\sqrt{N_x+1}-\psi(N_x,N_y+2,N_z,N_n)\sqrt{N_y+2}\sqrt{N_y+1}
\right.\nonumber\\&&\left.
-\psi(N_x,N_y,N_z+2,N_n)\sqrt{N_z+2}\sqrt{N_z+1}-\psi(N_x-2,N_y,N_z,N_n)\sqrt{N_x}\sqrt{N_x-1}
\right.\nonumber\\&&\left.
-\psi(N_x,N_y-2,N_z,N_n)\sqrt{N_y}\sqrt{N_y-1}-\psi(N_x,N_y,N_z-2,N_n)\sqrt{N_z}\sqrt{N_z-1}
\right.\nonumber\\&&\left.
+2\left[\psi(N_x,N_y,N_z,N_n)N_x+\psi(N_x,N_y,N_z,N_n)N_y
\right.\right.\nonumber\\&&\left.\left.
+\psi(N_x,N_y,N_z,N_n)N_z\right]+3\psi(N_x,N_y,N_z,N_n)\right\}^{\frac{1}{2}}|N_{xyzn}\rangle.
\end{eqnarray}
Wenn man definiert:

\begin{eqnarray}
&&\mathcal{E}_{\psi}\left(N_{xyzn}\right)=\frac{1}{2\sqrt{2}}\left\{
-\psi(N_x+2,N_y,N_z,N_n)\sqrt{N_x+2}\sqrt{N_x+1}
\right.\nonumber\\&&\left.
-\psi(N_x,N_y+2,N_z,N_n)\sqrt{N_y+2}\sqrt{N_y+1}
-\psi(N_x,N_y,N_z+2,N_n)\sqrt{N_z+2}\sqrt{N_z+1}
\right.\nonumber\\&&\left.
-\psi(N_x-2,N_y,N_z,N_n)\sqrt{N_x}\sqrt{N_x-1}
-\psi(N_x,N_y-2,N_z,N_n)\sqrt{N_y}\sqrt{N_y-1}
\right.\nonumber\\&&\left.
-\psi(N_x,N_y,N_z-2,N_n)\sqrt{N_z}\sqrt{N_z-1}
+2\left[\psi(N_x,N_y,N_z,N_n)N_x+\psi(N_x,N_y,N_z,N_n)N_y
\right.\right.\nonumber\\&&\left.\left.
+\psi(N_x,N_y,N_z,N_n)N_z\right]+3\psi(N_x,N_y,N_z,N_n)
\right\}^{\frac{1}{2}},
\end{eqnarray}
so bedeutet dies:

\begin{eqnarray}
E|\Psi\rangle&=&\sum_{N_{xyzn}}\psi(N_{xyzn})E|N_{xyzn}\rangle=\sum_{N_{xyzn}}\psi(N_{xyzn})\sqrt{P_x^2+P_y^2+P_z^2}|N_{xyzn}\rangle\nonumber\\
&=&\sum_{N_{xyzn}}\mathcal{E}_{\psi}\left(N_{xyzn}\right)|N_{xyzn}\rangle.
\label{Energieeigenwertxyz}
\end{eqnarray}
Bei einem Übergang zur Darstellung im Ortsraum ergibt sich dann für die Zeitentwicklung eines allgemeinen
symmetrischen Zustandes im Tensorraum der Ur"=Alternativen mit ($\ref{Zustand_Darstellung_Ortsraum}$),
($\ref{ZeitentwicklungA}$), ($\ref{ZeitentwicklungB}$) und ($\ref{Energieeigenwertxyz}$) der folgende Ausdruck,
der aufgrund der relativistischen Definition des Energieoperators ($\ref{Energie-Impuls-Relation}$) formal
auch als ein Zustand in der Raum-Zeit im Sinne der speziellen Relativitätstheorie angesehen werden kann:

\begin{eqnarray}
&&|\Psi(t)\rangle=\sum_{N_{xyzn}}\psi\left(N_{xyzn}\right)e^{-iEt}|N_{xyzn}\rangle=\sum_{N_{xyzn}}\psi\left(N_{xyzn},t\right)|N_{xyzn}\rangle\nonumber\\
&&\quad\longleftrightarrow\quad \sum_{N_{xyzn}}\psi\left(N_{xyzn},t\right)f_{N_{xyzn}}\left(\mathbf{x}\right)
=\Psi\left(\mathbf{x},t\right).
\label{ZeitentwicklungOrtsdarstellung}
\end{eqnarray}
Denn natürlich enthält diese auf der Schrödinger"=Gleichung im Tensorraum ($\ref{Schroedinger-Gleichung_TensorraumA}$)
unter Voraussetzung der Definition eines über ($\ref{Energie-Impuls-Relation}$) definierten Energieoperators
basierende Zeitentwicklung ($\ref{ZeitentwicklungOrtsdarstellung}$) implizit die Dynamik einer Klein"=Gordon"=Gleichung.
Dies ist aufgrund der Relation der Operatoren über die relativistischen Energie"=Impuls"=Relation ($\ref{Energie-Impuls-Relation}$)
unmittelbar ersichtlich. Da der Energieoperator und die Zeitentwicklung sich wechselseitig definieren, ist die
Darstellung des Energieoperators immer gegeben durch:

\begin{equation}
E=-i\partial_t.
\end{equation}
Wenn man nun desweiteren die Darstellung der Impulsoperatoren gemäß ($\ref{Ort_Impuls_Darstellung}$) und zudem die
Definition der Energie über die Impulsoperatoren ($\ref{Energie-Impuls-Relation}$), die in Analogie zur gewöhnlichen
relativistischen Energie"=Impuls"=Relation gewählt wurde, zu Grunde legt, und auf einen Zustand im Tensorraum anwendet,
so bedeutet dies:

\begin{equation}
\left(E^2-P_x^2-P_y^2-P_z^2\right)|\Psi(t)\rangle=0 \quad\longleftrightarrow\quad
\left(\partial_t^2-\partial_x^2-\partial_y^2-\partial_z^2\right)\Psi\left(\mathbf{x},t\right)=0.
\label{Klein-Gordon-Gleichung}
\end{equation}
Und damit ist dann faktisch gezeigt, dass es durch Definition des Energieoperators gemäß ($\ref{Energie-Impuls-Relation}$)
und damit über die Erzeugungs- und Vernichtungsoperatoren im Tensorraum und durch die Voraussetzung der allgemeinen
Schrödinger"=Gleichung der abstrakten Quantentheorie in Bezug auf den Tensorraum ($\ref{Schroedinger-Gleichung_TensorraumA}$)
möglich wird, die allgemeinen freien Zustände im Tensorraum der Ur"=Alternativen mit bosonischer Permutationssymmetrie
darzustellen als Wellenfunktionen in einer formalen Raum"=Zeit, die der Klein"=Gordon"=Gleichung genügen. Die bisherigen
Objekte enthalten noch keine Masse. Wenn man aber wie das Standardmodell der Elementarteilchenphysik mit dem Higgs"=Teilchen
und wie die Heisenbergsche nichtlineare Spinorfeldtheorie davon ausgeht, dass Masse durch Wechselwirkung entsteht, so ist
es nicht verwunderlich, dass freie Objekte masselos sind. Die Masse muss also durch eine Beschreibung der Wechselwirkung
induziert werden. Die Behandlung des Phänomens der Wechselwirkung in der Quantentheorie der Ur"=Alternativen wird im
übernächsten Abschnitt behandelt. Es wird für die späteren Abschnitte noch bedeutsam werden, die Wirkung der
Impulsoperatoren auf einen allgemeinen Zustand zu bestimmen:

\begin{align}
&P^x|\Psi\rangle=-\frac{i}{2\sqrt{2}}\sum_{N_{ABCD}}\psi\left(N_{ABCD}\right)
\left(A+B-C-D-A^{\dagger}-B^{\dagger}+C^{\dagger}+D^{\dagger}\right)|N_{ABCD}\rangle\nonumber\\
&=-\frac{i}{2\sqrt{2}}\sum_{N_{ABCD}}\psi\left(N_{ABCD}\right)
\left[\sqrt{N_A}|N_A-1,N_B,N_C,N_D\rangle+\sqrt{N_B}|N_A,N_B-1,N_C,N_D\rangle\right.\nonumber\\
&\left.-\sqrt{N_C}|N_A,N_B,N_C-1,N_D\rangle-\sqrt{N_D}|N_A,N_B,N_C,N_D-1\rangle\right.\nonumber\\
&\left.-\sqrt{N_A+1}|N_A+1,N_B,N_C,N_D-1\rangle-\sqrt{N_B+1}|N_A,N_B+1,N_C,N_D\rangle\right.\nonumber\\
&\left.+\sqrt{N_C+1}|N_A,N_B,N_C+1,N_D\rangle+\sqrt{N_D+1}|N_A,N_B,N_C,N_D+1\rangle\right]\nonumber\\
&=-\frac{i}{2\sqrt{2}}\sum_{N_{ABCD}}
\left[\psi\left(N_A+1,N_B,N_C,N_D\right)\sqrt{N_A+1}+\psi\left(N_A,N_B+1,N_C,N_D\right)\sqrt{N_B+1}\right.\nonumber\\
&\left.-\psi\left(N_A,N_B,N_C+1,N_D\right)\sqrt{N_C+1}-\psi\left(N_A,N_B,N_C,N_D+1\right)\sqrt{N_D+1}\right.\nonumber\\
&\left.-\psi\left(N_A-1,N_B,N_C,N_D\right)\sqrt{N_A}-\psi\left(N_A,N_B-1,N_C,N_D\right)\sqrt{N_B}\right.\nonumber\\
&\left.+\psi\left(N_A,N_B,N_C-1,N_D\right)\sqrt{N_C}+\psi\left(N_A,N_B,N_C,N_D-1\right)\sqrt{N_D}\right]|N_{ABCD}\rangle\nonumber\\
&\equiv\sum_{N_{ABCD}}\mathcal{P}^x_\psi\left(N_{ABCD}\right)|N_{ABCD}\rangle\nonumber\\
&\equiv|\mathcal{P}^x\Psi\rangle,
\label{Anwendung_Impuls_Zustand_X}
\end{align}

\begin{align}
&P^y|\Psi\rangle=-\frac{i}{2\sqrt{2}}\sum_{N_{ABCD}}\psi\left(N_{ABCD}\right)
\left(A-B+C-D-A^{\dagger}+B^{\dagger}-C^{\dagger}+D^{\dagger}\right)|N_{ABCD}\rangle\nonumber\\
&=-\frac{i}{2\sqrt{2}}\sum_{N_{ABCD}}\psi\left(N_{ABCD}\right)
\left[\sqrt{N_A}|N_A-1,N_B,N_C,N_D\rangle-\sqrt{N_B}|N_A,N_B-1,N_C,N_D\rangle\right.\nonumber\\
&\left.+\sqrt{N_C}|N_A,N_B,N_C-1,N_D\rangle-\sqrt{N_D}|N_A,N_B,N_C,N_D-1\rangle\right.\nonumber\\
&\left.-\sqrt{N_A+1}|N_A+1,N_B,N_C,N_D-1\rangle+\sqrt{N_B+1}|N_A,N_B+1,N_C,N_D\rangle\right.\nonumber\\
&\left.-\sqrt{N_C+1}|N_A,N_B,N_C+1,N_D\rangle+\sqrt{N_D+1}|N_A,N_B,N_C,N_D+1\rangle\right]\nonumber\\
&=-\frac{i}{2\sqrt{2}}\sum_{N_{ABCD}}
\left[\psi\left(N_A+1,N_B,N_C,N_D\right)\sqrt{N_A+1}-\psi\left(N_A,N_B+1,N_C,N_D\right)\sqrt{N_B+1}\right.\nonumber\\
&\left.+\psi\left(N_A,N_B,N_C+1,N_D\right)\sqrt{N_C+1}-\psi\left(N_A,N_B,N_C,N_D+1\right)\sqrt{N_D+1}\right.\nonumber\\
&\left.-\psi\left(N_A-1,N_B,N_C,N_D\right)\sqrt{N_A}+\psi\left(N_A,N_B-1,N_C,N_D\right)\sqrt{N_B}\right.\nonumber\\
&\left.-\psi\left(N_A,N_B,N_C-1,N_D\right)\sqrt{N_C}+\psi\left(N_A,N_B,N_C,N_D-1\right)\sqrt{N_D}\right]|N_{ABCD}\rangle\nonumber\\
&\equiv\sum_{N_{ABCD}}\mathcal{P}^y_\psi\left(N_{ABCD}\right)|N_{ABCD}\rangle\nonumber\\
&\equiv|\mathcal{P}^y\Psi\rangle,
\label{Anwendung_Impuls_Zustand_Y}
\end{align}

\begin{align}
&P^z|\Psi\rangle=-\frac{i}{2\sqrt{2}}\sum_{N_{ABCD}}\psi\left(N_{ABCD}\right)
\left(A-B-C+D-A^{\dagger}+B^{\dagger}+C^{\dagger}-D^{\dagger}\right)|N_{ABCD}\rangle\nonumber\\
&=-\frac{i}{2\sqrt{2}}\sum_{N_{ABCD}}\psi\left(N_{ABCD}\right)
\left[\sqrt{N_A}|N_A-1,N_B,N_C,N_D\rangle-\sqrt{N_B}|N_A,N_B-1,N_C,N_D\rangle\right.\nonumber\\
&\left.-\sqrt{N_C}|N_A,N_B,N_C-1,N_D\rangle+\sqrt{N_D}|N_A,N_B,N_C,N_D-1\rangle\right.\nonumber\\
&\left.-\sqrt{N_A+1}|N_A+1,N_B,N_C,N_D-1\rangle+\sqrt{N_B+1}|N_A,N_B+1,N_C,N_D\rangle\right.\nonumber\\
&\left.+\sqrt{N_C+1}|N_A,N_B,N_C+1,N_D\rangle-\sqrt{N_D+1}|N_A,N_B,N_C,N_D+1\rangle\right]\nonumber\\
&=-\frac{i}{2\sqrt{2}}\sum_{N_{ABCD}}
\left[\psi\left(N_A+1,N_B,N_C,N_D\right)\sqrt{N_A+1}-\psi\left(N_A,N_B+1,N_C,N_D\right)\sqrt{N_B+1}\right.\nonumber\\
&\left.-\psi\left(N_A,N_B,N_C+1,N_D\right)\sqrt{N_C+1}+\psi\left(N_A,N_B,N_C,N_D+1\right)\sqrt{N_D+1}\right.\nonumber\\
&\left.-\psi\left(N_A-1,N_B,N_C,N_D\right)\sqrt{N_A}+\psi\left(N_A,N_B-1,N_C,N_D\right)\sqrt{N_B}\right.\nonumber\\
&\left.+\psi\left(N_A,N_B,N_C-1,N_D\right)\sqrt{N_C}-\psi\left(N_A,N_B,N_C,N_D-1\right)\sqrt{N_D}\right]|N_{ABCD}\rangle\nonumber\\
&\equiv\sum_{N_{ABCD}}\mathcal{P}^z_\psi\left(N_{ABCD}\right)|N_{ABCD}\rangle\nonumber\\
&\equiv|\mathcal{P}^z\Psi\rangle.
\label{Anwendung_Impuls_Zustand_Z}
\end{align}
Man kann ($\ref{Anwendung_Energie_Zustand}$), ($\ref{Anwendung_Impuls_Zustand_X}$),
($\ref{Anwendung_Impuls_Zustand_Y}$) und ($\ref{Anwendung_Impuls_Zustand_Z}$) mit
($\ref{Viererimpuls_Tensorraum}$) in der folgenden Gleichung zusammenfassen:

\begin{equation}
P_{ABCD}^\mu|\Psi\rangle=|\mathcal{P}^\mu \Psi \rangle.
\label{Anwendung_Viererimpuls_Zustand}
\end{equation}

\section{Innere Symmetrien im begrifflichen Rahmen der Ur-Alternativen}

\subsection{Die inneren Symmetrien}

Die in der Natur existierenden Objekte sind nicht nur Objekte, die sich in der Raum-Zeit darstellen,
sondern sie sind zugleich durch zusätzliche diskrete Quantenzahlen gekennzeichnet, welche mit den sogenannten
inneren Symmetrien verbunden sind. Zudem gibt es den Spin, welcher obwohl er eine Quantenzahl darstellt,
auch mit den Raum-Zeit-Symmetrien in Zusammenhang steht, und damit gewissermaßen zwischen den räumlichen
Freiheitsgraden und den Freiheitsgraden der inneren Symmetrien steht. Im Rahmen des Standardmodells der
Elementarteilchenphysik handelt es sich bei den Quantenzahlen außer dem Spin bekanntlich vor allem um den
schwachen Isospin, der mit einer $SU(2)$"=Symmetrie verbunden ist, und den sogenannten Farb-Freiheitsgrad
der Quarks, der mit einer $SU(3)$"=Symmetrie verbunden ist. Es soll hier zunächst gezeigt werden, wie man
diese zusätzlichen Symmetrien in eine Beschreibungsweise im Rahmen der Ur"=Alternativen einbinden kann,
ehe im nächsten Unterabschnitt gezeigt wird, wie der Zusammenhang zu den Tensorraum-Zuständen hergestellt
werden kann, die man ja gemäß dem letzten Abschnitt in der Raum-Zeit darstellen kann. Das interessante
ist diesbezüglich natürlich, dass sowohl die räumlichen Freiheitsgrade alsauch die diskreten Freiheitsgrade
der Quantenzahlen letztendlich beide auf die diskreten Ur"=Alternativen zurückgeführt werden.

Um die inneren Symmetrien der Elementarteilchen in die Quantentheorie der Ur"=Alternativen einzubinden,
soll hier anders als in \cite{Goernitz:2016}, wo diese Frage auch behandelt wird, von der mathematischen
Struktur der Oktonionen Gebrauch gemacht werden, die mit der $G_2$"=Gruppe in der Weise verbunden ist,
dass die $G_2$"=Gruppe die Automorphismengruppe der Oktonionen ist, also die Menge aller Abbildungen
des Raumes der Oktonionen auf sich selbst beschreibt. Im Hinblick auf eine algebraische Erweiterung
der Quantentheorie wurden die Oktonionen seitens Pascual Jordan untersucht, beispielsweise in
\cite{Jordan:1932},\cite{Jordan:1933},\cite{Jordan:1934A},\cite{Jordan:1934B}. Die Ergebnisse dessen
werden in \cite{Liebmann:2017} zusammenfassend dargestellt. Beziehungen zur $SU(3)$"=Symmetriestruktur
der Quarks werden in \cite{Guenaydin:1973},\cite{Guenaydin:1978},\cite{LeBlanc:1988} thematisiert.
Ein allgemeines Element der Oktonionen hat die folgende Gestalt:

\begin{equation}
O=r_0 e_0+\sum_{i=1}^{7} r_i e_i,
\end{equation}
wobei $r_0$ und die $r_i$ reelle Parameter darstellen, $e_0$ einfach den Einheitsvektor der gewöhnlichen reellen
Dimension in Abgrenzung zu den imaginären Dimensionen andeutet und die imaginären Größen $e_i$ die folgende
algebraische Relation erfüllen:

\begin{equation}
e_i e_j=-\delta_{ij}+\tilde{\epsilon}_{ijk} e_k,
\end{equation}
wobei der Tensor $\tilde{\epsilon}_{ijk}$ total antisymmetrisch ist und für ihn gilt:

\begin{equation}
\tilde{\epsilon}_{123}=\tilde{\epsilon}_{246}=\tilde{\epsilon}_{435}=\tilde{\epsilon}_{367}
=\tilde{\epsilon}_{651}=\tilde{\epsilon}_{572}=\tilde{\epsilon}_{471}=1.
\end{equation}
Die verschiedenen imaginären Größen $e_i$ sind nicht"=assoziativ und erfüllen den folgenden Assoziator:

\begin{equation}
\left\{e_i,e_j,e_k\right\}=\left(e_i e_j\right)e_k - e_i\left(e_j e_k\right)=-2\bar \epsilon_{ijkl}e_l,
\end{equation}
wobei der Tensor $\bar \epsilon_{ijkl}$ ebenfalls total antisymmetrisch ist und für ihn gilt:

\begin{equation}
\bar \epsilon_{1247}=\bar \epsilon_{1265}=\bar \epsilon_{2345}=\bar \epsilon_{2376}=\bar \epsilon_{3146}=\bar \epsilon_{3157}=\bar \epsilon_{4576}=1.
\end{equation}
Die konjugierte Größe zu einem Element der Oktonionen ist gegeben durch:

\begin{equation}
O^{*}=r_0 e_0-\sum_{i=1}^{7} r_i e_i.
\end{equation}
Die Automorphismen"=Gruppe der Oktonionen ist wie erwähnt die $G_2$"=Gruppe, eine der exzeptionellen
Lie"=Gruppen, welche 14 Generatoren aufweist. Diese kann in einem vierdimensionalen komplexen
Vektorraum dargestellt werden. Um einen komplexen vierdimensionalen Vektor zu konstruieren,
ist in der Quantentheorie der Ur"=Alternativen die Kombination zweier Ur"=Alternativen notwendig,
die hier als $u$ und $v$ bezeichnet seien:

\begin{equation}
u=\left(\begin{matrix}a+bi\\c+di\end{matrix}\right),\quad v=\left(\begin{matrix}e+fi\\g+hi\end{matrix}\right),\quad \sqrt{a^2+b^2+c^2+d^2}=\sqrt{e^2+f^2+g^2+h^2}=1.
\end{equation}
Mit Hilfe dieser kann man einen vierdimensionalen Spinor konstruieren, der aber hier natürlich keinerlei
Beziehung zur Raum"=Zeit aufweist:

\begin{equation}
\Phi=\left(\begin{matrix} \bar u\\ \bar v \end{matrix}\right)=\left(\begin{matrix}a+bi\\c+di\\e+fi\\g+hi\end{matrix}\right),\quad
\sqrt{a^2+b^2+c^2+d^2+e^2+f^2+g^2+h^2}=1,
\label{doppelte_UrAlternative_Phi}
\end{equation}
wobei hier jetzt natürlich eine neue Normierungsbedingung definiert werden muss, da die beiden Ur"=Alternativen ja in
eine Gesamtalternative eingebettet sind, wodurch faktisch ein weiterer Freiheitsgrad entsteht und weshalb die beiden
in die Gesamtalternative integrierten Ur"=Alternativen mit $\bar u$ und $\bar v$ bezeichnet werden. In Wirklichkeit
enthalten die Ur"=Alternativen ja bereits drei Alternativen oder logische Freiheitsgrade, nämlich denjenigen zwischen
den beiden komplexen Dimensionen und die beiden in jeder der beiden komplexen Dimensionen jeweils für sich schon
enthaltenen. Jede der beiden Ur"=Alternativen enthält also eigentlich drei unabhängige Freiheitsgrade. Und mit der
Alternative der beiden Ur"=Alternativen in Bezug aufeinander ergibt sich dann der siebte Freiheitsgrad. Einen solchen
aus zwei Ur"=Alternativen bestehenden vierdimensionalen komplexen Spinor kann man natürlich auch als einen
achtdimensionalen reellen Vektor darstellen:

\begin{equation}
\Phi_R=\left(\begin{matrix}a\\b\\c\\d\\e\\f\\g\\h\end{matrix}\right).
\end{equation}
Bezogen auf diesen kann man die 14 Generatoren der $G_2$-Gruppe in der folgenden Weise als reelle 8 x 8"=Matrizen
darstellen, die als 4 x 4-Matrizen mit durch Pauli-Matrizen dargestellten reellen 2 x 2"=Matrizen als Einträgen
geschrieben werden können, wobei aufgrund der nicht-Assoziativität der Oktonionen, welche gleichbedeutend mit
der Tatsache ist, dass Rechtsmultiplikation etwas anderes ist als Linksmultiplikation, zu jeder der sieben
imaginären Dimensionen der Oktonionen zwei Generatoren gehören. Die nicht"=Assoziativität stellt sich
also in einer Verdopplung der Zahl der Generatoren von 7 auf 14 dar \cite{Waldron:1992}, wobei in der
folgenden Bezeichnung der Generatoren die mit $R$ und $L$ bezeichneten Generatoren jeweils ein Paar in Bezug
auf Rechts- und Linksmultiplikation darstellen:

\begin{align}
&L_1=\left(\begin{matrix}-i\sigma^2 & 0 & 0 & 0\\ 0 & -i\sigma^2 & 0 & 0\\0 & 0 & -i\sigma^2 & 0\\0 & 0 & 0 & i\sigma^2\end{matrix}\right),\quad
R_1=\left(\begin{matrix}-i\sigma^2 & 0 & 0 & 0\\ 0 & i\sigma^2 & 0 & 0\\0 & 0 & i\sigma^2 & 0\\0 & 0 & 0 & -i\sigma^2\end{matrix}\right),
\nonumber\\
&L_2=\left(\begin{matrix}0 & -\sigma^3 & 0 & 0\\ \sigma^3 & 0 & 0 & 0\\0 & 0 & 0 & -\mathbf{1}\\0 & 0 & \mathbf{1} & 0\end{matrix}\right),\quad
R_2=\left(\begin{matrix}0 & -\mathbf{1} & 0 & 0\\ \mathbf{1} & 0 & 0 & 0\\0 & 0 & 0 & \mathbf{1}\\0 & 0 & -\mathbf{1} & 0\end{matrix}\right),
\nonumber\\
&L_3=\left(\begin{matrix}0 & -\sigma^1 & 0 & 0\\ \sigma^1 & 0 & 0 & 0\\0 & 0 & 0 & -i\sigma^2\\0 & 0 & -i\sigma^2 & 0\end{matrix}\right),\quad
R_3=\left(\begin{matrix}0 & -i\sigma^2 & 0 & 0\\ -i\sigma^2 & 0 & 0 & 0\\0 & 0 & 0 & i\sigma^2\\0 & 0 & i\sigma^2 & 0\end{matrix}\right),
\nonumber\\
&L_4=\left(\begin{matrix}0 & 0 & -\sigma^3 & 0\\ 0 & 0 & 0 & \mathbf{1}\\\sigma^3 & 0 & 0 & 0\\0 & -\mathbf{1} & 0 & 0\end{matrix}\right),\quad
R_4=\left(\begin{matrix}0 & 0 & -\mathbf{1} & 0\\ 0 & 0 & 0 & -\mathbf{1}\\\mathbf{1} & 0 & 0 & 0\\0 & \mathbf{1} & 0 & 0\end{matrix}\right),
\nonumber\\
&L_5=\left(\begin{matrix}0 & 0 & -\sigma^1 & 0\\ 0 & 0 & 0 & i\sigma^2\\\sigma^1 & 0 & 0 & 0\\0 & i\sigma^2 & 0 & 0\end{matrix}\right),\quad
R_5=\left(\begin{matrix}0 & 0 & -i\sigma^2 & 0\\ 0 & 0 & 0 & -i\sigma^2\\\ -i\sigma^2 & 0 & 0 & 0\\0 & -i\sigma^2 & 0 & 0\end{matrix}\right),
\nonumber\\
&L_6=\left(\begin{matrix}0 & 0 & 0 & -\mathbf{1}\\ 0 & 0 & -\sigma^3 & 0\\0 & \sigma^3 & 0 & 0\\\mathbf{1} & 0 & 0 & 0\end{matrix}\right),\quad
R_6=\left(\begin{matrix}0 & 0 & 0 & -\sigma^3\\ 0 & 0 & \sigma^3 & 0\\0 & -\sigma^3 & 0 & 0\\\sigma^3 & 0 & 0 & 0
\end{matrix}\right),
\nonumber\\
&L_7=\left(\begin{matrix}0 & 0 & 0 & -i\sigma^2\\ 0 & 0 & -\sigma^1 & 0\\0 & \sigma^1 & 0 & 0\\-i\sigma^2 & 0 & 0 & 0\end{matrix}\right),\quad
R_7=\left(\begin{matrix}0 & 0 & 0 & -\sigma^1\\ 0 & 0 & \sigma^1 & 0\\0 & -\sigma^1 & 0 & 0\\\sigma^1 & 0 & 0 & 0\end{matrix}\right).
\end{align}
Hierbei beschreiben die $\sigma^\mu$ die Pauli"=Matrizen einschließlich der Einheitsmatrix in zwei Dimensionen:

\begin{equation}
\sigma^0=\mathbf{1}=\left(\begin{matrix}1 & 0\\0 & 1\end{matrix}\right),\quad
\sigma^1=\left(\begin{matrix}0 & 1\\1 & 0\end{matrix}\right),\quad
\sigma^2=\left(\begin{matrix}0 & -i\\i & 0\end{matrix}\right),\quad
\sigma^3=\left(\begin{matrix}1 & 0\\0 & -1\end{matrix}\right).
\label{Pauli-Matrizen}
\end{equation}
Auf diese Weise ist also die $G_2$"=Gruppe auf dem Raum einer doppelten Ur"=Alternative dargestellt. Nun kann man
aber die $G_2$"=Symmetrie zerlegen in eine $SU(3)$"=Symmetrie und zwei $SU(2)$"=Symmetrien \cite{Donaldson:2007},
was also bedeutet:

\begin{equation}
G_2 \quad\Leftrightarrow\quad SU(3) \otimes SU(2) \otimes SU(2).
\end{equation}
Entsprechend den acht Generatoren der $SU(3)$"=Gruppe und den jeweils drei Generatoren der beiden
$SU(2)$"=Gruppen, ist dies gleichbedeutend mit der Relation:

\begin{equation}
\textbf{14}=\textbf{8}+\textbf{3}+\textbf{3}.
\end{equation}
Diese algebraische Substruktur der $G_2$"=Gruppe kann dargestellt werden, indem man die Größen $a_k$, $b_k$ und $g_{kl}$ einführt,
wobei die Indizes jeweils von $1$ bis $3$ laufen. Diese algebraischen Größen sind zugleich die Generatoren der $G_2$"=Gruppe
alsauch diejenigen der Untergruppen, wobei die $a_k$ und die $b_k$ jeweils eine $SU(2)$"=Gruppe konstituieren und die $g_{kl}$
eine $SU(3)$"=Gruppe \cite{Bincer:1993}. Diese Generatoren erfüllen dementsprechend die folgende Lie-Algebra:

\begin{eqnarray}
&&\left[g_k^{\ l}, g_m^{\ n}\right]=\delta_m^{\ l} g_k^{\ n}-\delta_k^{\ n} g_m^{\ l}, \quad
\left[g_k^{\ l}, a_m\right]=\delta_m^{\ l} a_k-\frac{1}{3}\delta_k^{\ l} a_m,\quad
\left[g_k^{\ l}, b^n\right]=-\delta_k^{\ n} b^l+\frac{1}{3}\delta_k^{\ l} b^n,\nonumber\\
&&\left[a_m, b^n\right]=g_m^{\ n},\quad
\left[a_m, a_n\right]=-\frac{2}{\sqrt{3}}\epsilon_{mnl}b^l,\quad
\left[b^m, b^n\right]=\frac{2}{\sqrt{3}}\epsilon^{mnl} a_l,
\end{eqnarray}
wobei $\epsilon_{ijk}$ hier nun den gewöhnlichen total antisymmetrischen Tensor
dritter Stufe bezeichnet. Die Forderung der Relationen:

\begin{equation}
\sum_k g_k^k=0,\quad g_k^{l\ \dagger}=g_k^l,\quad a_m^{\dagger}=b^m,
\end{equation}
schränkt die Zahl der Freiheitsgrade der $g_{kl}$ von neun auf acht ein,
wie es den acht Generatoren der $SU(3)$"=Gruppe entspricht.

Da sich also die $G_2$"=Gruppe im Raum einer doppelten binären quantentheoretischen Alternative darstellen
lässt und diese sich in eine $SU(3)$"=Gruppe und zwei $SU(2)$"=Gruppen aufspalten lässt, besteht die Möglichkeit,
auf diese Weise die inneren Symmetrien des Standardmodells in die Quantentheorie der Ur"=Alternativen zu
integrieren. Man könnte die $SU(3)$"=Gruppe mit dem Farbfreiheitsgrad der Quarks identifizieren, eine der
beiden $SU(2)$"=Gruppen mit dem Isospin und die andere der $SU(2)$"=Gruppen mit dem räumlichen Spin. Nun entsteht
natürlich die Schwierigkeit, den Bezug zum Tensorraum und seiner Darstellung in Bezug auf den physikalischen
Ortsraum herzustellen. Hierbei ist entscheidend, dass der Spin sich auch auf den physikalischen Raum bezieht,
während der Isospin und der Farbfreiheitsgrad rein innere Symmetrien ohne Bezug zum physikalischen Raum darstellen.
Dies legt die Vermutung nahe, dass der mit der einen $SU(2)$"=Symmetrie verknüpfte Spin"=Freiheitsgrad eine relative
Ausrichtung der Ur"=Alternativen des Raumes des zusätzlichen doppelten binären Freiheitsgrades in Bezug auf den
Tensorraum beschreibt, während die anderen Freiheitsgrade dann relativ zu diesem zusätzlichen Freiheitsgrad
als Basis definiert und damit vollkommen unabhängig vom Tensorraum einschließlich dessen Darstellung im
physikalischen Ortsraum sind. Dies würde dann eine Erklärung dafür liefern, warum der Freiheitsgrad des
Spin durch Raum-Zeit-Transformationen beeinflusst wird, aber der Isospin und der Farbfreiheitsgrad nicht.
Das bedeutet, dass eine Transformation aller Ur"=Alternativen des internen Raumes einer Drehung
des Spin entspricht, während eine Transformation der Ur"=Alternativen des internen Raumes relativ
zueinander einer Transformation in Bezug auf die inneren Symmetrien entspricht.
Dies liefert in der Tat eine sehr plausible Rechtfertigung dafür, dass es neben den
Raum"=Zeit"=Symmetrien und den internen Symmetrien den Spin mit seiner beides verbindenden
Natur überhaupt gibt. Diese Grundidee wurde in etwas anderer Weise bereits in
\cite{Kober:2009B} thematisiert.

\subsection{Beziehung der inneren Symmetrien zum Raum}

Natürlich muss der Spin-Freiheitsgrad unter dieser Voraussetzung dynamisch in eine direkte Beziehung
zum Tensorraum gesetzt werden und dies kann wie in der gewöhnlichen Beschreibung geschehen, indem man
die Klein"=Gordon"=Gleichung ($\ref{Klein-Gordon-Gleichung}$) linearisiert und in die entsprechende
Dirac"=Gleichung überführt, wie Dirac dies in \cite{Dirac:1928A},\cite{Dirac:1928B} vollzog. Wenn man
die in der Klein"=Gordon"=Gleichung ($\ref{Klein-Gordon-Gleichung}$) enthaltene Energie"=Impuls"=Relation
($\ref{Energie-Impuls-Relation}$) und als Viererimpulsoperator die Definition in ($\ref{Viererimpuls_Tensorraum}$)
zu Grunde legt, so lautet ($\ref{Energie-Impuls-Relation}$):

\begin{equation}
\left(P_{ABCD}\right)^\mu \left(P_{ABCD}\right)_\mu=0.
\label{Viererimpuls_Relation}
\end{equation}
Um diese Relation umzuformulieren muss man die Dirac"=Matrizen verwenden:

\begin{equation}
\gamma^0=\left(\begin{matrix}0 & \mathbf{1}\\ \mathbf{1} & 0 \end{matrix}\right),\quad
\gamma^i=\left(\begin{matrix}0 & -\sigma^i\\ \sigma^i & 0 \end{matrix}\right),\quad
i=1...3,
\end{equation}
wobei die $\sigma^i$ die in ($\ref{Pauli-Matrizen}$) definierten Pauli"=Matrizen sind und $\mathbf{1}$
die ebenfalls in ($\ref{Pauli-Matrizen}$) definierte Einheitsmatrix in zwei Dimensionen ist.
Mit der Relation:

\begin{equation}
\gamma^\mu \gamma^\nu+\gamma^\nu \gamma^\mu=2\eta^{\mu\nu}\textbf{1},\quad \mu=0...3,
\label{Gamma-Matrizen-Relation}
\end{equation}
wobei $\eta^{\mu\nu}$ die Minkowski"=Metrik und $\textbf{1}$ hier nun die Einheitsmatrix in vier
Dimensionen sein soll, kann man ($\ref{Viererimpuls_Relation}$) ausdrücken als:

\begin{equation}
\gamma^\mu \gamma^\nu \left(P_{ABCD}\right)_\mu \left(P_{ABCD}\right)_\nu=0,
\end{equation}
und dies bedeutet:

\begin{equation}
\gamma^\mu \left(P_{ABCD}\right)_\mu=0.
\label{Linearisierung}
\end{equation}
Um nun von dieser Relation zur masselosen Dirac"=Gleichung als einer bestimmten Manifestation der Schrödinger"=Gleichung
als dynamischer Grundgleichung der Quantentheorie überzugehen, muss der darin enthaltene Operator auf einen Zustand
angewandt werden. Und dieser muss zusätzlich zu einem Vektor im Tensorraum der Ur"=Alternativen auch mindestens noch
einen zusätzlichen Dirac"=Spinor enthalten. Es sollen aber zudem mit Hilfe der mathematischen Betrachtungen des
letzten Unterabschnittes die inneren Symmetrien eingebunden werden, also der schwache Isospin sowie die
Quark"=Freiheitsgrade. Zunächst sei deshalb der folgende zweidimensionale Spinor $\Omega$ als einzelne
Ur"=Alternative definiert, welche die Komponente zu positiver und zu negativer Energie beschreibt:

\begin{equation}
\Omega=\left(\begin{matrix} \Omega_1\\ \Omega_2 \end{matrix}\right)=\left(\begin{matrix} u_1 \\ u_2 \end{matrix}\right).
\end{equation}
Desweiteren sei der Vektor $\Gamma$ als Tensorprodukt der Ur"=Alternative $\Omega$ mit der doppelten
Ur"=Alternative $\Phi$ aus dem letzten Unterabschnitt ($\ref{doppelte_UrAlternative_Phi}$) definiert,
in Bezug auf welche die $G_2$"=Gruppe mit der in ihr enthaltenen Substruktur der beiden
$SU(2)$-Gruppen und der $SU(3)$"=Gruppe dargestellt werden kann:

\begin{equation}
\Gamma=\Omega \otimes \Phi=\left(\begin{matrix} \Omega_1 \Phi \\ \Omega_2 \Phi \end{matrix}\right).
\label{Interne_Freiheitsgrade}
\end{equation}
Der Spinor $\Gamma$ enthält also die Freiheitsgrade eines Dirac"=Spinors, der die Komponenten zu positiver
und negativer Energie sowie den Spin enthält, der mit einer der beiden $SU(2)$"=Gruppen identifiziert werden kann,
die in der $G_2$"=Gruppe enthalten sind, und darüber hinaus einen schwachen Isospin, der mit der anderen der
beiden $SU(2)$"=Gruppen identifiziert werden kann, sowie einen Farbfreiheitsgrad, der mit der $SU(3)$"=Gruppe
identifiziert werden kann. Wenn man nun ein weiteres Tensorprodukt bildet, nämlich dasjenige zwischen dem
Vektor $\Gamma$ und einem allgemeinen Zustand im Tensorraum der Ur"=Alternativen,
formuliert in ($\ref{Zustand_Tensorraum}$) beziehungsweise ($\ref{Zustand_Tensorraum_xyzn}$),
so erhält man einen erweiterten Zustand $|\Psi_\Gamma \rangle$:

\begin{equation}
|\Psi_\Gamma \rangle=\sum_{N_{ABCD}}\psi\left(N_{ABCD}\right)|N_{ABCD}\rangle \otimes \Gamma.
\label{Zustand_Tensorraum_Quantenzahlen}
\end{equation}
Dieser Zustand stellt ein Element des Gesamt"=Hilbert"=Raumes $\mathcal{H}_G$ dar, der sich aus
dem Tensorprodukt des Tensorraumes der Ur"=Alternativen $\mathcal{H}_T$ und dem Hilbert"=Raum
der Menge der $\Gamma$-Zustände $\mathcal{H}_\Gamma$ ergibt:

\begin{equation}
\mathcal{H}_G \quad\Leftrightarrow\quad \mathcal{H}_T \otimes \mathcal{H}_\Gamma.
\end{equation}
Wenn man den Operator aus ($\ref{Linearisierung}$) auf einen allgemeinen Zustand
$|\Psi_\Gamma \rangle$ anwendet, definiert in ($\ref{Zustand_Tensorraum_Quantenzahlen}$),
den man zudem als zeitabhängig ansieht, so ergibt sich die folgende Gleichung:

\begin{equation}
\gamma^\mu (P_{ABCD})_\mu|\Psi_\Gamma (t)\rangle=\left[\gamma^0 E-\gamma^1 P_x-\gamma^2 P_y-\gamma^3 P_z\right]|\Psi_\Gamma (t)\rangle=0,
\label{Dirac-Gleichung}
\end{equation}
wobei $E$ gemäß ($\ref{Energie_Operator}$) und $P_x$, $P_y$ und $P_z$ gemäß ($\ref{Ort_Impuls_Operatoren}$)
beziehungsweise ($\ref{Ort_Impuls_Operatoren_ABCD}$) definiert sind. Diese Gleichung ($\ref{Dirac-Gleichung}$)
stellt ihrer Form nach eine Dirac"=Gleichung mit inneren Symmetrien dar. Die Dirac"=Matrizen wirken im
$\Omega$"=Raum beziehungsweise die in den Dirac"=Matrizen enthaltenen Pauli"=Matrizen auf den $SU(2)$-Freiheitsgrad
innerhalb des Zustandes $\Phi$, der mit dem Spin identifiziert werden soll. Der andere $SU(2)$"=Freiheitsgrad und
der $SU(3)$"=Freiheitsgrad bleiben durch die Dirac"=Matrizen unbeeinflusst, weshalb sie auch von den Operatoren
im Tensorraum vollkommen entkoppelt sind, womit sie auch von der Raum"=Zeit absolut unabhängig sind und daher
als wirkliche innere Freiheitsgrade im Sinne des schwachen Isospin und der Farbe interpretiert werden können.
Allenfalls über eine Wechselwirkung könnten Beziehungen der inneren Symmetrien zum Tensorraum
hergestellt werden. Wenn man die folgende Definition vornimmt:

\begin{equation}
\Lambda=\gamma^\mu (P_{ABCD})_\mu=\left[\gamma^0 E-\gamma^1 P_x-\gamma^2 P_y-\gamma^3 P_z\right],
\end{equation}
so kann man die Gleichung umschreiben zu:

\begin{equation}
\Lambda |\Psi_\Gamma (t)\rangle=0.
\label{Gleichung_Lambda}
\end{equation}
Wenn man nun den Übergang in eine Raum"=Zeit"=Darstellung vornimmt, indem man die räumliche
Darstellung ($\ref{Zustand_Darstellung_Ortsraum}$) von Zuständen im Tensorraum verwendet:

\begin{equation}
|\Psi_\Gamma (t)\rangle \quad\longleftrightarrow\quad \Psi\left(\mathbf{x},t\right) \otimes \Gamma=\Psi_\Gamma\left(\mathbf{x},t\right),
\end{equation}
und zudem die Darstellung der Orts- und Impulsoperatoren im reellen dreidimensionalen Raum ($\ref{Ort_Impuls_Darstellung}$)
in ($\ref{Dirac-Gleichung}$) verwendet, dann ergibt sich die folgende Gestalt der Dirac"=Gleichung in Bezug auf eine
Darstellung in der Raum"=Zeit:

\begin{equation}
i\gamma^\mu \partial_\mu \Psi_\Gamma\left(\mathbf{x},t\right)=0.
\end{equation}
In Gestalt der Schrödinger"=Gleichung ($\ref{Schroedinger-Gleichung_TensorraumA}$) geschrieben, bedeutet dies:

\begin{equation}
i\partial_t \Psi_\Gamma(\textbf{x},t)=i\gamma^0\left(\gamma^1\partial_x+\gamma^2\partial_y+\gamma^3\partial_z\right)\Psi_\Gamma\left(\textbf{x},t\right)
=H_D\Psi_\Gamma\left(\textbf{x},t\right)=0,
\label{Dirac-Schroedinger-Gleichung}
\end{equation}
wenn man $H_D$ definiert als:

\begin{equation}
H_D=-\gamma^0\left(\gamma^1 P_x+\gamma^2 P_y+\gamma^3 P_z\right)
\quad\longleftrightarrow\quad i\gamma^0\left(\gamma^1 \partial_x+\gamma^2 \partial_y+\gamma^3 \partial_z\right).
\label{Dirac-Hamilton-Operator}
\end{equation}
Dass eine Masse erst über eine Theorie der Wechselwirkung in die Quantentheorie der Ur"=Alternativen
Einzug erhalten kann, wurde im letzten Abschnitt bereits erläutert.

\section{Zustände vieler Objekte und Wechselwirkung}

\subsection{Vielteilchentheorie}

Man kann nun von einer Quantentheorie vieler Ur"=Alternativen, deren symmetrischen Zuständen freie Quantenobjekte
entsprechen, die in einer (3+1)"=dimensionalen Raum"=Zeit dargestellt werden können, zu einer Theorie vieler
solcher Quantenobjekte übergehen, die einer Vielteilchentheorie oder freien Quantenfeldtheorie analog ist,
aber natürlich in Wirklichkeit nicht einer Quantenfeldtheorie entspricht, da sie basierend auf Ur"=Alternativen
unabhängig von einem physikalischen Raum und damit von feldtheoretischen Prinzipien ist. In einer solchen
hat man es nicht mit einem Hilbert"=Raum zu tun, dessen Basiszuständen Besetzungszahlen für die verschiedenen
Basiszustände einer einzelnen Ur"=Alternative entsprechen, sondern mit einem Hilbert"=Raum, dessen Basiszuständen
Besetzungszahlen für die Basiszustände des symmetrischen Tensorraumes vieler Ur-Alternativen entsprechen,
also Besetzungszahlen in Bezug auf Besetzungszahlen. Man hat es also mit drei Stufen der Quantisierung
zu tun. Auf der ersten Stufe wird eine binäre Alternative durch Zuordnung von komplexen Wahrheitswerten
in eine quantentheoretische binäre Alternative überführt, die damit zu einer Ur"=Alternative wird.
Auf der zweiten Stufe werden die Komponenten der Ur"=Alternative zu Operatoren, die in einem Tensorraum
vieler Ur"=Alternativen wirken und Ur"=Alternativen in den verschiedenen Basiszuständen erzeugen
beziehungsweise vernichten. Den Zuständen in diesem Tensorraum entsprechen einzelne freie Quantenobjekte,
die man gewöhnlich als Teilchen bezeichnet und als Wellen im Raum darstellt. Auf der dritten Stufe erhält
man Operatoren, die Quantenobjekte in den Basiszuständen des Tensorraumes der Ur"=Alternativen erzeugen
und vernichten. Die Zustände beschreiben ein Ensemble vieler Quantenobjekte. Den drei Stufen der
Quantisierung entsprechen also die folgenden Übergänge:

\begin{eqnarray}
&\textbf{binäre Alternative}\quad &\xrightarrow{Quantisierung}\quad \textbf{Zustand Ur"=Alternative},\\
&\textbf{Zustand Ur"=Alternative}\quad &\xrightarrow{Quantisierung}\quad \textbf{Zustand vieler Ur"=Alternativen},\nonumber\\
&\textbf{Zustand vieler Ur"=Alternativen}\quad &\xrightarrow{Quantisierung}\quad \textbf{Zustand vieler Quantenobjekte}\nonumber.
\end{eqnarray}
Eine Konstruktion des Hilbert"=Raumes der Besetzungszahlen in Bezug auf die Basiszustände im Tensorraum vieler Ur"=Alternativen
entspricht einer Zuordnung von Erzeugungs- und Vernichtungsoperatoren zu den Basiszuständen des Tensorraumes vieler Ur"=Alternativen,
denen ja selbst Besetzungszahlen in den Basiszuständen einzelner Ur"=Alternativen entsprechen ($\ref{Basiszustand_Tensorraum}$).
Diese Erzeugungs- und Vernichtungsoperatoren sind, wenn man Bose"=Statistik zu Grunde legt, gemäß der folgenden
Vertauschungsrelation definiert:

\begin{equation}
\left[\hat \psi\left(N_{ABCD}\right),\hat \psi^{\dagger}\left(N^{'}_{ABCD}\right)\right]
=\delta_{N_A N^{'}_A}\delta_{N_B N^{'}_B}\delta_{N_ C N^{'}_C}\delta_{N_D N^{'}_D}.
\label{Operatoren_Quantenobjekte_Vertauschungsrelationen}
\end{equation}
Man kann nun zu den Besetzungszahlen der Quantenobjekte in den Basiszuständen des Tensorraumes vieler
Ur-Alternativen gehörige Zustände $|\mathcal{N}\left(N_{ABCD}\right)\rangle$ definieren, auf welche
die in ($\ref{Operatoren_Quantenobjekte_Vertauschungsrelationen}$) definierten Operatoren wirken und
über diese kann man einen allgemeinen Zustand vieler Quantenobjekte in folgender Weise ausdrücken:

\begin{eqnarray}
|\Psi\rangle_{\mathcal{N}}&=&\sum_{N_{ABCD}}\sum_{\mathcal{N}\left(N_{ABCD}\right)}
\psi_{\mathcal{N}}\left[\mathcal{N}\left(N_{ABCD}\right)\right]|\mathcal{N}\left(N_{ABCD}\right)\rangle\nonumber\\
&=&\sum_{N_{xyzn}}\sum_{\mathcal{N}\left(N_{xyzn}\right)}\psi_{\mathcal{N}}
\left[\mathcal{N}\left(N_{xyzn}\right)\right]|\mathcal{N}\left(N_{xyzn}\right)\rangle.
\end{eqnarray}
Die zu den Feldoperatoren in gewöhnlichen Quantenfeldtheorien analogen Operatoren im Rahmen der
Quantentheorie der Ur-Alternativen, welche sich auf Zustände vieler Quantenobjekte beziehen,
lauten dann in der Ortsdarstellung wie folgt:

\begin{eqnarray}
\hat \Psi(\textbf{x})&=&\sum_{N_{yxzn}}\hat \psi\left(N_{xyzn}\right)f_{N_{xyzn}}\left(\textbf{x}\right),\nonumber\\
\hat \Psi^{\dagger}(\textbf{x})&=&\sum_{N_{xyzn}}\hat \psi^{\dagger}\left(N_{xyzn}\right)f_{N_{xyzn}}\left(\textbf{x}\right),
\label{Operatoren_Quantenobjekte}
\end{eqnarray}
wobei $\psi\left(N_{xyzn}\right)$ und $\psi^{\dagger}\left(N_{xyzn}\right)$ die zu
($\ref{Operatoren_Quantenobjekte_Vertauschungsrelationen}$) analogen Vertauschungsrelationen
in Bezug auf $N_{xyzn}$ erfüllen und in dieser Ortsdarstellung die Funktionen $f_{N_{xyzn}}\left(\textbf{x}\right)$
gemäß ($\ref{Basiszustand_Tensorraum_Ortsdarstellung}$) definiert sind. Wenn man die Operatoren
($\ref{Operatoren_Quantenobjekte}$) in Abhängigkeit von der Zeit formulieren möchte, so muss man
gemäß der Dynamik im Tensorraum der Ur"=Alternativen ($\ref{ZeitentwicklungOrtsdarstellung}$)
den Zeitentwicklungsoperator $e^{-iEt}$ integrieren:

\begin{eqnarray}
\hat \Psi_N(\textbf{x},t)&=&\sum_{N_{yxzn}}\hat \psi\left(N_{xyzn}\right)e^{-iEt}
f_{N_{xyzn}}\left(\textbf{x}\right),\nonumber\\
\hat \Psi^{\dagger}_N(\textbf{x},t)&=&\sum_{N_{xyzn}}\hat \psi^{\dagger}\left(N_{xyzn}\right)e^{iEt}
f_{N_{xyzn}}\left(\textbf{x}\right),
\label{Feldoperatoren_zeitabhaengig}
\end{eqnarray}
wobei $E$ natürlich gemäß ($\ref{Energie_Operator}$) definiert ist. Ein aus vielen bezüglich der
Vertauschung der Ur"=Alternativen symmetrischen Zuständen, welche einzelne Quantenobjekte repräsentieren,
zusammengesetzter Zustand, also ein symmetrisches Produkt symmetrischer Zustände, welches einen
Zustand vieler Quantenobjekte repräsentiert, stellt an sich selbst nicht wieder einen symmetrischen
Zustand bezüglich der einzelnen Ur"=Alternativen dar. Deshalb ist es notwendig, eine verallgemeinerte
Algebra einzuführen, welche Zustände mit allgemeinerer Symmetrie zu beschreiben gestattet.
Diese gegenüber den Vertauschungsrelationen von Operatoren, die zu symmetrischen Zuständen gehören,
und Anti"=Vertauschungsrelationen von Operatoren, die zu anti"=symmetrischen Zuständen gehören,
verallgemeinerte Algebra konstituiert neben der Bose"=Statistik mit symmetrischen Zuständen und
der Fermi"=Statistik mit antisymmetrischen Zuständen, die Parabose"=Statistik mit Zuständen
beliebiger Symmetrie. Wenn für die zu den Erzeugungs- und Vernichtungsoperatoren
($\ref{ErzeugungsVernichtungsOperatoren}$) im Falle der Bose"=Statistik analogen
Erzeugungs- und Vernichtungsoperatoren der Parabose"=Statistik folgende Definition
vorgenommen wird:

\begin{eqnarray}
&&a_1=A,\quad a_2=B,\quad a_3=C,\quad a_4=D,\nonumber\\
&&a_1^{\dagger}=A^{\dagger},\quad a_2^{\dagger}=B^{\dagger},\quad a_3^{\dagger}=C^{\dagger},\quad a_4^{\dagger}=D^{\dagger},
\label{a1234ABCD}
\end{eqnarray}
so erfüllen die Ur"=Alternativen Parabose"=Statistik, wenn für die Erzeugungs- und Vernichtungsoperatoren
die folgenden algebraischen Relationen gelten:

\begin{equation}
\left[\frac{1}{2}\left\{a_r, a_s^{\dagger}\right\}, a_t\right]=-\delta_{st} a_r,\quad
\left[\left\{a_r, a_s\right\},a_t\right]=\left[\left\{a_r^{\dagger}, a_s^{\dagger}\right\}, a_t^{\dagger}\right]=0,\quad r,s,t=1...4,
\end{equation}
wobei wie gewöhnlich eckige Klammern einen Kommutator beschreiben und geschweifte Klammern einen Antikommutator.
Diese algebraischen Relationen werden dann erfüllt, wenn man die Erzeugungs- und Vernichtungsoperatoren in der
folgenden Weise definiert:

\begin{equation}
a_r=\sum_{\alpha=1}^{p} b_r^{\alpha},\quad a_r^{\dagger}=\sum_{\alpha=1}^p b_r^{\alpha\dagger},\quad r=1...4,
\end{equation}
wobei die folgende Algebra für die neu eingeführten Operatoren $b_r^{\alpha}$ und $b_r^{\alpha\dagger}$ gilt:

\begin{eqnarray}
&&\left[b_r^{\alpha}, b_s^{\alpha \dagger}\right]=\delta_{rs},\quad
\left[b_r^{\alpha},b_s^{\alpha}\right]=\left[b_r^{\alpha\dagger},b_s^{\alpha\dagger}\right]=0,\quad r,s=1...4\nonumber\\
&&\left\{b_r^{\alpha}, b_s^{\beta\dagger}\right\}=\left\{b_r^{\alpha}, b_s^{\beta}\right\}
=\left\{b_r^{\alpha\dagger}, b_s^{\beta\dagger}\right\}=0\quad \textrm{für}\quad \alpha \neq \beta.
\label{Parabose_Statistik_Algebra}
\end{eqnarray}
$p$ beschreibt die sogenannte Parabose"=Ordnung. Wenn man diese Algebra ($\ref{Parabose_Statistik_Algebra}$)
zu Grunde legt, so beschreiben die $b_r^{\alpha\dagger}$"=Operatoren beziehungsweise $b_r^{\alpha}$"=Operatoren
für ein fest definiertes $\alpha$ die Erzeugungs- beziehungsweise Vernichtungsoperatoren für Ur"=Alternativen,
die zum Zustand eines bestimmen Quantenobjektes gehören. Die Parabose-Ordnung $p$ beschreibt dann die
Zahl der Quantenobjekte, die in einem Zustand enthalten sind. Der Gesamtzustand bleibt gleich,
wenn man zwei Ur"=Alternativen vertauscht, die zum gleichen Quantenobjekt gehören, für die
also gilt: $\alpha=\beta$, und er kehrt das Vorzeichen um, wenn man zwei Ur"=Alternativen vertauscht,
die nicht zum gleichen Quantenobjekt gehören, für die also gilt: $\alpha \neq \beta$.
Natürlich kann man dann die zu festem $\alpha$ gehörigen $b_r^{\alpha}$"=Operatoren
gemäß ($\ref{a1234ABCD}$) mit den jeweiligen Erzeugungs- beziehungsweise Vernichtungsoperatoren
für die vier Basiszustände im Tensorraum dieses $\alpha$-ten Objektes identifizieren:

\begin{eqnarray}
&&b_1^{\alpha}=A^\alpha,\quad b_2^{\alpha}=B^{\alpha},\quad b_3^{\alpha}=C^{\alpha},\quad b_4^{\alpha}=D^{\alpha},\nonumber\\
&&b_1^{\alpha \dagger}=A^{\alpha \dagger},\quad b_2^{\alpha \dagger}=B^{\alpha \dagger},
\quad b_3^{\alpha \dagger}=C^{\alpha \dagger},\quad b_4^{\alpha \dagger}=D^{\alpha \dagger}.
\end{eqnarray}
Es soll nun der Propagator eines Quantenobjektes definiert werden. $T\left\{\mathcal{O}\left(t_1\right)\mathcal{O}\left(t_2\right)\right\}$
beschreibt das zeitgeordnete Produkt von Operatoren:

\begin{equation}
T\left\{\mathcal{O}\left(t_1\right)\mathcal{O}\left(t_2\right)\right\}
=\theta\left(t_1-t_2\right)\mathcal{O}\left(t_1\right)\mathcal{O}\left(t_2\right)
+\theta\left(t_2-t_1\right)\mathcal{O}\left(t_2\right)\mathcal{O}\left(t_1\right),
\end{equation}
wobei gilt:

\begin{equation}
\theta\left(t_1-t_2\right)=\begin{cases}1\quad \text{wenn}\quad t_1 \ge t_2 \\0\quad \text{wenn}\quad t_1 < t_2 \end{cases}.
\end{equation}
Wenn $|0\rangle$ den Vakuumzustand beschreibt, dann kann damit nun der Propagator zwischen zwei Orten und zwei Zeitpunkten
wie folgt bestimmt werden:

\begin{eqnarray}
&&\Delta\left(\textbf{x}^{'},\textbf{x},t^{'},t\right)=\langle 0|T\left\{\hat \Psi\left(\textbf{x}^{'},t^{'}\right)
\hat \Psi^{\dagger}\left(\textbf{x},t\right)\right\}|0\rangle\nonumber\\
&&=\langle 0|T\left\{\left[\sum_{N_{xyzn}^{'}}\hat \psi\left(N_{xyzn}^{'}\right) e^{-iE^{'}t^{'}}
f_{N_{xyzn}^{'}}\left(\textbf{x}^{'}\right)\right]
\left[\sum_{N_{xyzn}}\hat \psi^{\dagger}\left(N_{xyzn}\right) e^{iEt}
f_{N_{xyzn}}\left(\textbf{x}\right)\right]\right\}|0\rangle\nonumber\\
&&=\sum_{N_{xyzn}^{'}}\sum_{N_{xyzn}}\left[\theta\left(t^{'}-t\right)f_{N_{xyzn}^{'}}\left(\textbf{x}^{'}\right)
f_{N_{xyzn}}\left(\textbf{x}\right)\langle N_{xyzn}^{'}|e^{-iE^{'}t^{'}}
e^{iEt}|N_{xyzn}\rangle\right]\nonumber\\
&&=\sum_{N_{xyzn}^{'}}\sum_{N_{xyzn}}\left[\theta\left(t^{'}-t\right)f_{N_{xyzn}^{'}}\left(\textbf{x}^{'}\right)f_{N_{xyzn}}\left(\textbf{x}\right)
\sum_{\bar N_{xyzn}^{'}}\sum_{\bar N_{xyzn}} \langle \bar N_{xyzn}^{'}|\psi_{\bar N_{xyzn}^{'},N_{xyzn}^{'}}\left(t^{'}\right)
\psi_{\bar N_{xyzn},N_{xyzn}}\left(t\right)|\bar N_{xyzn}\rangle\right]\nonumber\\
&&=\sum_{N_{xyzn}^{'}}\sum_{N_{xyzn}}\left[\theta\left(t^{'}-t\right)f_{N_{xyzn}^{'}}\left(\textbf{x}^{'}\right)f_{N_{xyzn}}\left(\textbf{x}\right)
\sum_{\bar N_{xyzn}^{'}}\sum_{\bar N_{xyzn}} \psi_{\bar N_{xyzn}^{'},N_{xyzn}^{'}}\left(t^{'}\right)\psi_{\bar N_{xyzn},N_{xyzn}}\left(t\right)
\langle \bar N_{xyzn}^{'}|\bar N_{xyzn}\rangle\right]\nonumber\\
&&=\sum_{N_{xyzn}^{'}}\sum_{N_{xyzn}}\left[\theta\left(t^{'}-t\right)f_{N_{xyzn}^{'}}\left(\textbf{x}^{'}\right)f_{N_{xyzn}}\left(\textbf{x}\right)
\sum_{\bar N_{xyzn}^{'}}\sum_{\bar N_{xyzn}} \psi_{\bar N_{xyzn}^{'},N_{xyzn}^{'}}\left(t^{'}\right)\psi_{\bar N_{xyzn},N_{xyzn}}\left(t\right)
\delta_{\bar N_{xyzn}^{'},\bar N_{xyzn}}\right]\nonumber\\
&&=\sum_{N_{xyzn}^{'}}\sum_{N_{xyzn}}\left[\theta\left(t^{'}-t\right)f_{N_{xyzn}^{'}}\left(\textbf{x}^{'}\right)f_{N_{xyzn}}\left(\textbf{x}\right)
\sum_{\bar N_{xyzn}} \psi_{\bar N_{xyzn},N_{xyzn}^{'}}\left(t^{'}\right)\psi_{\bar N_{xyzn},N_{xyzn}}\left(t\right)\right]\nonumber\\
&&=\sum_{N_{xyzn}^{'}}\sum_{N_{xyzn}}\sum_{\bar N_{xyzn}}\left[\theta\left(t^{'}-t\right)
f_{N_{xyzn}^{'}}\left(\textbf{x}^{'}\right)f_{N_{xyzn}}\left(\textbf{x}\right)
\psi_{\bar N_{xyzn},N_{xyzn}^{'}}\left(t^{'}\right)\psi_{\bar N_{xyzn},N_{xyzn}}\left(t\right)\right].
\end{eqnarray}
Hierbei wurde natürlich einerseits die Definition ($\ref{Feldoperatoren_zeitabhaengig}$) und andererseits die
Tatsache verwendet, dass gilt: \mbox{$\hat \Psi\left(\textbf{x},t\right)|0\rangle=0$} und \mbox{$\langle 0|\hat \Psi^{\dagger}\left(\textbf{x},t\right)=0$}, weshalb beim zeitgeordneten Produkt der Term für den Fall $t^{'}<t$
natürlich einfach verschwindet. Die Komponenten $\psi_{\bar N_{xyzn},N_{xyzn}}\left(t\right)$ beschreiben die
Komponenten des sich aus der Zeitentwicklung gemäß ($\ref{ZeitentwicklungA}$) beziehungsweise ($\ref{ZeitentwicklungB}$)
ergebenden Zustandes. Hierbei wird durch die Zeitentwicklung ein Übergang des Basiszustandes $|N_{xyzn}\rangle$ eines
einzelnen symmetrischen Zustandes von Ur"=Alternativen, der durch den Erzeugungsoperator im Vielteilchen"=Hilbert"=Raum
erzeugt wurde, zu einer Linearkombination solcher Basiszustände beschrieben.

\subsection{Das Phänomen der Wechselwirkung als Verschränkung abstrakter Zustände}

Es ist nun die Aufgabe zu bewältigen, das Phänomen der Wechselwirkung zwischen verschiedenen Objekten
im Rahmen der Quantentheorie der Ur"=Alternativen zu beschreiben. Wechselwirkung bedeutet, dass sich
die Dynamik und damit die zeitliche Entwicklung zweier oder mehrerer Objekte nicht in einer voneinander
unabhängigen Weise vollzieht. Dies wiederum bedeutet, dass eine Abhängigkeitsbeziehung zwischen den
Zuständen der Objekte besteht, zwischen denen eine Wechselwirkung herrscht. Da im Rahmen der Quantentheorie
der Ur"=Alternativen keinerlei physikalische und insbesondere feldtheoretische Begriffe vorausgesetzt
werden dürfen, kann die Wechselwirkung nicht wie etwa in der gewöhnlichen Formulierung relativistischer
Quantenfeldtheorien durch ein punktweises Produkt von Feldern oder durch eine ähnliche Konzeption
begründet werden. Vielmehr muss sie sich wie alles andere auch alleine in einer nur auf die abstrakte
Realität der Alternativen oder speziell der Ur"=Alternativen bezogenen Weise ergeben und damit zugleich
in einer von vorneherein rein quantentheoretischen Weise definiert sein. In einer gewöhnlichen
Beschreibungsweise führt eine Wechselwirkung zwischen zwei Objekten zu einer anschließenden
Verschränkung der Zustände. Da eine beliebige Verschränkung die allgemeinste und abstrakteste
Weise ist, um eine Abhängigkeitsbeziehung zweier quantentheoretischer Zustände auszudrücken,
liegt es nahe, den Wechselwirkungsbegriff auf der fundamentalen Ebene überhaupt über den Begriff
der Verschränkung zwischen quantentheoretischen Zuständen zu definieren. Dies bedeutet im Rahmen
der Quantentheorie der Ur"=Alternativen, dass die Beziehung zweier sich in einer Wechselwirkungsbeziehung
befindlicher Objekte durch einen Zustand beschrieben werden kann, welcher eine Verschränkung der diese
beiden Objekte beschreibenden Zustände im Tensorraum der Ur"=Alternativen enthält. Wenn die Besetzungszahlen
zweier Objekte $1$ und $2$ mit $N^1_{ABCD}$ und $N^2_{ABCD}$ bezeichnet seien, so kann ein allgemeiner
die Wechselwirkungsbeziehung der beiden Objekte enthaltender Gesamtzustand in der folgenden Weise
zum Ausdruck gebracht werden:

\begin{equation}
|\Psi\rangle_{N^1 N^2}=\sum_{N^1_{ABCD},N^2_{ABCD}}\psi\left(N^1_{ABCD},N^2_{ABCD}\right)
|N^1_{ABCD}\rangle \otimes |N^2_{ABCD}\rangle,
\end{equation}
wobei die Koeffizienten der Basiszustände des Tensorproduktes der zu den beiden Objekten gehörigen
Tensorräume vieler Ur"=Alternativen, zu denen gemäß der Parabosedarstellung $b_r^{\alpha}$"=Operatoren
mit unterschiedlichem $\alpha$ gehören, nicht einfach ein direktes Produkt von auf die Einzelräume
der beiden Objekte bezogenen Koeffizienten darstellen darf, was bedeutet:

\begin{equation}
\psi\left(N^1_{ABCD},N^2_{ABCD}\right)\neq \psi_1\left(N^1_{ABCD}\right)*\psi_2\left(N^2_{ABCD}\right).
\end{equation}
Wenn man dies auf eine beliebige Zahl $M$ von verschiedenen Objekten überträgt, welcher gemäß
der Parabose"=Darstellung die Parabose"=Ordnung $p=M$ entspricht, so besitzt der entsprechende
Zustand die folgende Gestalt:

\begin{equation}
|\Psi\rangle_{N^1,...,N^M}=\sum_{N^1_{ABCD},...,N^M_{ABCD}}\psi\left(N^1_{ABCD},...,N^M_{ABCD}\right)
\left(|N^1_{ABCD}\rangle \otimes ... \otimes |N^M_{ABCD}\rangle\right),
\label{Zustand_N_verschraenkt}
\end{equation}
wobei natürlich auch hier in entsprechend analoger Weise gilt:

\begin{equation}
\psi\left(N^1_{ABCD},...,N^M_{ABCD}\right)\neq \psi_1\left(N^1_{ABCD}\right)*...*\psi_M\left(N^M_{ABCD}\right).
\end{equation}
Genaugenommen hat man es eigentlich gar nicht mehr mit voneinander getrennten Objekten zu tun, sondern
mit einem Gesamtobjekt, dass nur in einer gewissen Näherung als aus unterschiedlichen Objekten bestehend
angesehen werden darf. Dies ist eine unmittelbare Manifestation dessen, was unter dem Postulat der
Trennbarkeit der Alternativen gesagt wurde, nämlich dass die Wechselwirkung eine Korrektur der Näherung
der Trennbarkeit der Alternativen darstellt, die in Wirklichkeit das Ergebnis einer künstlichen
Separierung aus einer physikalischen oder kosmischen Gesamtrealität darstellen, die allerdings
notwendig ist, um eine Beschreibung der Natur im Rahmen einer Erfahrungswissenschaft wie der
theoretischen Physik überhaupt zu ermöglichen.

Wenn nun $\mathcal{O}_m$ einen zum Tensorraum des $m$"=ten Quantenobjektes gehörigen
Operator bezeichnet, so können die entsprechenden auf diesen Tensorraum bezogenen Operatoren
der Energie und der Impulse mit $E_m$, $P_{xm}$, $P_{ym}$ und $P_{zm}$ bezeichnet werden.
Da in jedem der $M$ symmetrischen Tensorräume vollkommen unabhängig von der Verschränkung
die abstrakte Gleichung ($\ref{Klein-Gordon-Gleichung}$) für sich gültig ist, so gilt:

\begin{equation}
\sum_{m=1}^M \left(E_m^2-P_{xm}^2-P_{ym}^2-P_{zm}^2\right)|\Psi\rangle_{N^1,...,N^M}=0.
\label{Energie-Impuls-Relation_Vielteilchen}
\end{equation}
Hierin ist natürlich noch nicht die Dynamik der Wechselwirkung zwischen den einzelnen Objekten enthalten. Hierzu muss
ein zusätzlicher Term zum gewöhnlichen Hamilton"=Operator hinzukommen. Wenn man den Hamilton"=Operator jedes einzelnen
Quantenobjektes gemäß der Definition des Energieoperators ($\ref{Energie-Impuls-Relation}$) zu Grunde legt, die auch
($\ref{Klein-Gordon-Gleichung}$) und damit ($\ref{Energie-Impuls-Relation_Vielteilchen}$) zu Grunde liegt, so lautet
der Gesamt"=Hamilton"=Operator eines Gesamt"=Zustandes $M$ freier Objekte $H_F^M$:

\begin{align}
&H_F^M=\frac{1}{2\sqrt{2}}\sum_{m=1}^M\left[-3\sum_{r=1}^4 \left(b_r^M b_r^M+b_r^{M\dagger}b_r^{M\dagger}\right)
+2b_1^M b_2^M+2b_1^M b_3^M+2b_1^M b_4^M+2b_2^M b_3^M
\right.\nonumber\\ &\left.
+2b_2^M b_4^M+2 b_3^M b_4^M+2b_1^{M\dagger}b_2^{M\dagger}+2b_1^{M\dagger} b_3^{M\dagger}+2 b_1^{M\dagger} b_4^{M\dagger}
+2b_2^{M\dagger}b_3^{M\dagger}+2b_2^{M\dagger} b_4^{M\dagger}+2 b_3^{M\dagger} b_4^{M\dagger}
\right.\nonumber\\ &\left.
+2\left(3b_1^{M\dagger} b_1^M+3b_2^{M\dagger} b_2^M+3b_3^{M\dagger} b_3^M+3b_4^{M\dagger} b_4^M
-b_1^{M\dagger} b_2^M-b_1^{M\dagger} b_3^M-b_1^{M\dagger} b_4^M-b_2^{M\dagger} b_1^M
\right.\right.\nonumber\\ &\left.\left.
-b_2^{M\dagger} b_3^M-b_2^{M\dagger} b_4^M-b_3^{M\dagger} b_1^M-b_3^{M\dagger} b_2^M-b_3^{M\dagger} b_4^M
-b_4^{M\dagger} b_1^M-b_4^{M\dagger} b_2^M-b_4^{M\dagger} b_3^M\right)+3\right]^{\frac{1}{2}}.
\end{align}
Der Term des Hamilton"=Operators, der die Wechselwirkung konstituiert und die Verschränkung erzeugt, muss ein Produkt
zwischen den in den verschiedenen Teilräumen wirkenden $b_r^m$"=Operatoren enthalten. Dies führt auf die folgende
allgemeine Bedingung für die Form eines solchen Wechselwirkungs"=Hamilton"=Operators von $M$ Objekten $H_W^M$:

\begin{equation}
H_W^M\left(b_r^m,b_r^{m\dagger}\right)\neq H^1\left(b_r^1,b_r^{1\dagger}\right)+...+H^M\left(b_r^M,b_r^{M\dagger}\right),
\end{equation}
wobei bei den $b_r^m$ der Index $r$ die Werte $1$ bis $4$ und der Index $m$ die Werte $1$ bis $M$ annehmen kann.
Damit lautet der Gesamt"=Hamilton"=Operator $H_G^M$ mehrerer Quantenobjekte in seiner allgemeinen Gestalt:

\begin{equation}
H_G^M=H_F^M\left(b_r^m,b_r^{m\dagger}\right)+H_W^M\left(b_r^m,b_r^{m\dagger}\right),
\end{equation}
Die entsprechende dynamische Entwicklung der Zustände ist gegeben durch die
entsprechende Schrödinger"=Gleichung:

\begin{equation}
i\partial_t|\Psi(t)\rangle_{N^1,...,N^M}=H_G^M |\Psi(t)\rangle_{N^1,...,N^M},
\end{equation}
was bedeutet:

\begin{equation}
|\Psi(t)\rangle_{N^1,...,N^M}=e^{-iH_G^M t}|\Psi(t_0)\rangle_{N^1,...,N^M}.
\end{equation}
Wenn sich der Gesamt"=Zustand der Objekte zu einem bestimmten Zeitpunkt $t_0$ als ein
Produkt der einzelnen Zustände der Objekte ohne Verschränkung darstellen lässt:

\begin{equation}
|\Psi(t_0)\rangle_{N^1,...,N^M}=\sum_{N_{ABCD}^1}\psi_1\left(N_{ABCD}^1,t_0\right)|N_{ABCD}^1\rangle \otimes ... \otimes \sum_{N_{ABCD}^M}\psi_M\left(N_{ABCD}^M,t_0\right)|N_{ABCD}^M\rangle,
\end{equation}
dann ergibt sich für die dynamische Entwicklung:

\begin{align}
&|\Psi(t)\rangle_{N^1,...,N^M}=e^{-iH_G^M t}|\Psi(t_0)\rangle_{N^1,...,N^M}\nonumber\\
&=e^{-i\left[H_F^M\left(b_r^m,b_r^{m\dagger}\right)+H_W^M\left(b_r^m,b_r^{m\dagger}\right)\right]t}\left[\sum_{N_{ABCD}^1}\psi_1\left(N_{ABCD}^1,t_0\right)|N_{ABCD}^1\rangle \otimes ... \otimes \sum_{N_{ABCD}^M}\psi_M\left(N_{ABCD}^M,t_0\right)|N_{ABCD}^M\rangle\right]\nonumber\\
&=e^{-i H_W^M\left(b_r^m,b_r^{m\dagger}\right)t}\left[e^{-iE^1 t}\sum_{N_{ABCD}^1}\psi_1\left(N_{ABCD}^1,t_0\right)|N_{ABCD}^1\rangle \otimes ...
\otimes e^{-iE^M t}\sum_{N_{ABCD}^M}\psi_M\left(N_{ABCD}^M,t_0\right)|N_{ABCD}^M\rangle\right]\nonumber\\
&=\sum_{N_{ABCD}^1, ... ,N_{ABCD}^M} f_W^M\left(N_{ABCD}^1, ... , N_{ABCD}^M, t\right)\nonumber\\
&\times \left[\psi_1\left(N_{ABCD}^1,t_0\right)e^{-iE^1t}|N_{ABCD}^1\rangle \otimes ... \otimes
\psi_M\left(N_{ABCD}^M,t_0\right)e^{-iE^M t}|N_{ABCD}^M\rangle\right].
\end{align}
Die zeitabhängige Funktion $f_W^M\left(N_{ABCD}^1, ..., N_{ABCD}^M, t\right)$ beschreibt die durch den Wechselwirkungsterm
des Hamilton"=Operators $H_W^M\left(b_r^m,b_r^{m\dagger}\right)$ induzierte Verschränkung der Zustände der einzelnen Objekte.
Natürlich kann man dies auch auf Zustände mit inneren Freiheitsgraden übertragen und die zu ($\ref{Dirac-Schroedinger-Gleichung}$)
analoge entsprechende Dirac"=Gleichung für ein System mit vielen Quantenobjekten lautet dann geschrieben als
Schrödinger"=Gleichung wie folgt:

\begin{equation}
i\partial_t|\Psi_\Gamma(t)\rangle_{N^1,...,N^M}=\left[H_D^M\left(b_r^m,b_r^{m\dagger}\right)
+H_W^M\left(b_r^m,b_r^{m\dagger}\right)\textbf{1}\right]|\Psi_\Gamma(t)\rangle_{N^1,...,N^M},
\end{equation}
wobei $|\Psi_\Gamma(t)\rangle_{N^1,...,N^M}$ ein verschränkter Zustand gemäß ($\ref{Zustand_N_verschraenkt}$) ist,
der zudem die inneren Freiheitsgrade gemäß ($\ref{Zustand_Tensorraum_Quantenzahlen}$) enthält, und $H_D^M$ die
Summe der Hamilton"=Operatoren gemäß ($\ref{Dirac-Hamilton-Operator}$) in den Hilbert"=Räumen der einzelnen
Quantenobjekte beschreibt:

\begin{equation}
H_D^M=-\gamma^0\sum_{m=1}^M \left[\gamma^1 P_{xm}\left(b_r^m,b_r^{m\dagger}\right)+\gamma^2 P_{ym}\left(b_r^m,b_r^{m\dagger}\right)
+\gamma^3 P_{zm}\left(b_r^m,b_r^{m\dagger}\right)\right].
\end{equation}
Beim Übergang in die Darstellung im Ortsraum nimmt diese Dirac"=Gleichung mit Wechselwirkungsterm
in der Gestalt der Schrödinger"=Gleichung dann folgende Form an:

\begin{equation}
i\partial_t\Psi_\Gamma\left(\textbf{x}_1,...,\textbf{x}_M\right)=
\left[i\gamma^0\sum_{m=1}^M\left(\gamma^1 \partial_{xm}+\gamma^2 \partial_{ym}+\gamma^3 \partial_{zm}\right)
+H_W^M\textbf{1}\right]\Psi_\Gamma\left(\textbf{x}_1,...,\textbf{x}_M\right).
\end{equation}
Hierin sind natürlich die inneren Symmetrien überhaupt noch nicht in die Wechselwirkung miteinbezogen. Wenn man eine
Wechselwirkung konstituieren möchte, die beim Übergang zur raum"=zeitlichen Darstellung näherungsweise in ein punktweises
Produkt zwischen den Zuständen übergeht, wie es den gewöhnlichen Wechselwirkungen der Elementarteilchenphysik entspricht,
so muss man einen Wechselwirkungsoperator der folgenden allgemeinen Form definieren:

\begin{equation}
H_W^M=h_W^M\left(b_r^m,b_r^{m\dagger}\right)\delta_{N^1,...,N^M}.
\label{Hamilton-Operator_Wechselwirkung}
\end{equation}
Wenn man diesen Operator auf den Produktzustand $|\Psi\rangle_{N^1,...,N^M}$ anwendet, so ergibt sich:

\begin{eqnarray}
&&H_W^M|\Psi\rangle_{N^1,...,N^M}=h_W^M\left(b_r^m,b_r^{m\dagger}\right)\delta_{N^1,...,N^M}|\Psi\rangle_{N^1,...,N^M}\nonumber\\
&&=\delta_{N^1,...,N^M} h_W^M\left(b_r^m,b_r^{m\dagger}\right)\left[\sum_{N_{ABCD}^1}\psi_1(N_{ABCD}^1)|N_{ABCD}^1\rangle \otimes ... \otimes \sum_{N_{ABCD}^M}\psi_M(N_{ABCD}^M)|N_{ABCD}^M\rangle \right]\nonumber\\
&&=\delta_{N^1,...,N^M} \sum_{N_{ABCD}^1...N_{ABCD}^M} f_h\left(N_{ABCD}^1,...,N_{ABCD}^M\right)\left[\psi_1(N_{ABCD}^1)|N_{ABCD}^1\rangle \otimes ...
\otimes \psi_M(N_{ABCD}^M)|N_{ABCD}^M\rangle \right]\nonumber\\
&&=\sum_{N_{ABCD}}f_h\left(N_{ABCD}\right)\left[\psi_1(N_{ABCD})|N_{ABCD}\rangle \otimes ... \otimes \psi_M(N_{ABCD})|N_{ABCD}\rangle \right].
\label{Produkt_Wechselwirkung_Tensorraum}
\end{eqnarray}
Dieser Ausdruck enthält nur Produkte, bei dem die Besetzungszahlen $N_{ABCD}$ der verschiedenen in Wechselwirkung stehenden
Objekte übereinstimmen. Hierbei tragen analog zum punktweisen Produkt der Wellenfunktionen, bei denen nur das Produkt der
Komponenten zu den gleichen Ortseigenzuständen beiträgt, nur die Komponenten zu gleichen Tensorraumbasiszuständen bei.
Dies wird sich beim Übergang in die Raum"=Zeit"=Darstellung aber auch in ein Produkt von Wellenfunktionen umwandeln:

\begin{eqnarray}
&&\sum_{N_{ABCD}}f_h\left(N_{ABCD}\right)\left[\psi_1(N_{ABCD})|N_{ABCD}\rangle \otimes ... \otimes \psi_M(N_{ABCD})|N_{ABCD}\rangle \right]\nonumber\\
&&\sum_{N_{xyzn}}f_h\left(N_{xyzn}\right)\left[\psi_1(N_{xyzn})|N_{xyzn}\rangle \otimes ... \otimes \psi_M(N_{xyzn})|N_{xyzn}\rangle \right]\nonumber\\
&&\longleftrightarrow \sum_{N_{xyzn}}f_h\left(N_{xyzn}\right)\left[\psi_1(N_{xyzn})f_{N_{xyzn}}\left(\textbf{x}\right)*
...*\psi_M(N_{xyzn})f_{N_{xyzn}}\left(\textbf{x}\right)\right].
\label{Produkt_Wechselwirkung_Tensorraum_Ortsraum}
\end{eqnarray}
Das Entscheidende bei diesem Produkt von Wellenfunktionen ist aber dennoch, dass es in Wirklichkeit ein Produkt ist,
dass die Komponenten bezüglich gleicher Basiszustände im diskreten Raum der Ur"=Alternativen miteinander multipliziert,
das sich dann nur indirekt räumlich darstellt, während gewöhnlich, also bei Feldtheorien, ein kontinuierliches punktweises
Produkt zu Grunde gelegt wird. Und eben ein solcher diskreter rein quantentheoretischer Wechselwirkungsbegriff könnte die
Unendlichkeiten umgehen, die in gewöhnlichen Quantenfeldtheorien auftreten.

\subsection{Konstruktion der realen Wechselwirkungen über das Korrespondenzprinzip}

Im letzten Unterabschnitt wurde ein allgemeiner Wechselwirkungsbegriff in der Quantentheorie der Ur"=Alternativen konstituiert,
der das Phänomen der Wechselwirkung in diesem abstrakten begrifflichen Rahmen allgemein charakterisiert. Aber letztendlich
geht es natürlich darum, basierend darauf die konkreten wirklichen Wechselwirkungen in der Natur zu erhalten und zu beschreiben.
Diesbezüglich erscheint es wohl als sinnvoll, sich zunächst einmal an die Art und Weise zu erinnern, in der die
Wechselwirkungen gewöhnlich in die Physik eingeführt werden. Dies geschieht durch lokale Eichsymmetrien, was bedeutet,
dass die Forderung der Invarianz unter lokalen Symmetrietransformationen an jedem Raum-Zeit Punkt zur Einführung der
Wechselwirkungsfelder und ihrer Kopplung an die sogenannten Materiefelder führt. Hierbei ist entscheidend, dass die
Wechselwirkungen des Standardmodells der Elementarteilchenphysik mit den inneren Symmetrien verbunden sind, die starke
Wechselwirkung mit der $SU(3)$"=Gruppe im Farbraum, die schwache Wechselwirkung mit der $SU(2)$"=Gruppe des Isospin und
der Elektromagnetismus mit der $U(1)$"=Gruppe der komplexen Phase in Bezug auf die elektrische Ladung. Die Gravitation
hingegen kann als lokale Eichtheorie in Bezug auf Raum"=Zeit"=Translationen beschrieben werden, also in Bezug auf eine
auf die Raum-Zeit bezogene Symmetrie. Natürlich kann eine eichtheoretische Beschreibungsweise in der Quantentheorie
der Ur"=Alternativen nicht verwendet werden, da hier die Raum-Zeit nur ein Darstellungsmedium ist, und die Wechselwirkung
ganz gemäß dem letzten Abschnitt begrifflich auf Verschränkungen zwischen den Zuständen der diskreten Ur"=Alternativen
gegründet ist. Aber vielleicht können hier die Eichtheorien als klassische Näherung einen Hinweis in Bezug auf die
rein quantentheoretische Formulierung im begrifflichen Rahmen der Ur"=Alternativen geben.

Der eigentliche Anspruch der von Weizsäckerschen Rekonstruktion der Physik besteht eigentlich darin, die Gestalt der
Naturgesetze bis in alle Einzelheiten zu begründen, also wirklich die Existenz jedes Objektes, jeder Wechselwirkung
und jedes darauf basierenden Phänomens exakt zu begründen. Dieser unglaubliche Anspruch kann im Rahmen dieser Arbeit
in Bezug auf die real existierenden Wechselwirkungen einstweilen noch nicht eingelöst werden. Es muss eigentlich nicht
erwähnt werden, dass auch keine andere Theorie in der Geschichte der theoretischen Physik bis auf den heutigen Tage
auch nur in die Nähe dessen gekommen wäre. Im Gegenteil, die Idee, dass dies basierend auf den Bedingungen der Möglichkeit
von Erfahrung möglich sein könnte, ist seitens von Weizsäcker in dieser Weise überhaupt erst entwickelt worden, wenngleich
Kant hier die gedankliche Vorarbeit geleistet hatte, indem er doch immerhin die gundlegenden Strukturen der Natur,
nicht hingegen die genauen Naturgesetze, als Bedingungen der Möglichkeit von Erfahrung überhaupt postulierte.
Was aber immerhin in dieser Arbeit erreicht werden kann, das ist die empirisch gefundenen Wechselwirkungen
in einen durch Ur"=Alternativen ausgedrückten Rahmen zu überführen. Diese Art der Überführung oder
Quantisierung in einem radikalen Sinne geht zunächst von den Feldgleichungen aus, interpretiert sie
als quantentheoretische Wellenfunktionen und überführt die Wellenfunktionen in Zustände im Tensorraum
der Ur"=Alternativen und die punktweisen Produkte zwischen ihnen in Beziehungen im Tensorraum
der Ur"=Alternativen. Dies bedeutet, dass man von der klassischen Theorie in die entsprechende
reine Quantentheorie der Ur"=Alternativen gelangt, indem man den zur bisherigen Betrachtung
umgekehrten Prozess durchläuft, also die Zustände im Tensorraum der Ur"=Alternativen nicht
auf eine raum"=zeitliche Darstellung abbildet, sondern den bereits bekannten raum"=zeitlichen
Ausdruck der Feldgleichungen als klassischen Grenzfall einer Darstellung von Zuständen im Tensorraum
der Ur"=Alternativen interpretiert und ihn entsprechend ersetzt. Diese Art der Überführung kann man
durchaus als eine Art der Quantisierung bezeichnen, nur als eine radikalere Quantisierung, die keine
Feldquantisierung mehr ist, sondern alles in eine reine quantentheoretische Beziehungsstruktur umwandelt
und nicht mehr auf Feldgrößen bezogen ist, da die raum"=zeitlichen Felder beziehungsweise Wellenfunktionen
eben wirklich nur noch als Darstellungen zu interpretieren sind. Die Überführung müsste in diesem
Ansatz mit den folgenden Quantisierungsregeln geschehen:

\begin{eqnarray}
\Psi\left(\textbf{x}\right)\quad&\longrightarrow&\quad |\Psi\rangle=\sum_{N_{ABCD}}\psi\left(N_{ABCD}\right)|N_{ABCD}\rangle,\nonumber\\
\partial^\mu \quad&\longrightarrow&\quad iP^\mu_{ABCD},\nonumber\\
\partial^\mu \Psi\left(\textbf{x}\right) \quad&\longrightarrow&\quad iP^\mu_{ABCD}|\Psi\rangle=i\sum_{N_{ABCD}}\mathcal{P}^\mu_\psi\left(N_{ABCD}\right)|N_{ABCD}\rangle=i|\mathcal{P}^\mu \Psi\rangle,
\label{Uebergang_Zustaende_Operatoren}
\end{eqnarray}
wobei hier natürlich die Komponenten $P^\mu_{ABCD}$ gemäß ($\ref{Ort_Impuls_Operatoren_ABCD}$), ($\ref{Energie_Operator}$)
und ($\ref{Viererimpuls_Tensorraum}$) definiert sind und ($\ref{Anwendung_Viererimpuls_Zustand}$) verwendet wurde.
Die punktweisen Produkte, welche in einer gewöhnlichen feldtheoretischen Beschreibungsweise die Wechselwirkung der
Felder definieren, müssen bei diesem über das Korrespondenzprinzip vollzogenen Übergang zu einer rein
quantentheoretischen Beschreibungsweise im Tensorraum der Ur"=Alternativen in folgender Weise unter
Verwendung von ($\ref{Produkt_Wechselwirkung_Tensorraum_Ortsraum}$) überführt werden:

\begin{equation}
\underbrace{\Psi\left(\textbf{x}\right)*...*\Psi\left(\textbf{x}\right)}_{M mal} \quad\longrightarrow\quad
\sum_{N_{ABCD}}\left[\psi_1\left(N_{ABCD}\right)|N_{ABCD}\rangle \otimes ... \otimes \psi_M\left(N_{ABCD}\right)|N_{ABCD}\rangle\right],
\label{Uebergang_Wechselwirkung_Produkt}
\end{equation}
was einem Wechselwirkungs"=Hamilton"=Operator der Form ($\ref{Hamilton-Operator_Wechselwirkung}$) mit
$h\left(b_r^m,b_r^{m\dagger}\right)=c$ entspricht, wobei $c$ eine Konstante ist. Wenn man nun
($\ref{Uebergang_Zustaende_Operatoren}$) und ($\ref{Uebergang_Wechselwirkung_Produkt}$)
zu Grunde legt, so kann man desweiteren folgenden Übergang bestimmen:

\begin{eqnarray}
\underbrace{\partial^\mu \Psi\left(\mathbf{x}\right)* ... *\partial^\nu \Psi\left(\mathbf{x}\right)}_{M mal}
&\longrightarrow& \sum_{N_{ABCD}}\left[iP_{ABCD}^\mu \psi_1\left(N_{ABCD}\right)|N_{ABCD}\rangle\otimes ...
\otimes iP_{ABCD}^\nu \psi_N\left(N_{ABCD}\right)|N_{ABCD}\rangle\right]\nonumber\\
&&=\sum_{N_{ABCD}}\left[i\mathcal{P}^\mu_{\psi 1}\left(N_{ABCD}\right)|N_{ABCD}\rangle\otimes ...
\otimes i\mathcal{P}^\nu_{\psi N}\left(N_{ABCD}\right)|N_{ABCD}\rangle\right],\nonumber\\
\label{Uebergang_Wechselwirkung_Produkt_Ableitungen}
\end{eqnarray}
wobei hier erneut ($\ref{Anwendung_Viererimpuls_Zustand}$) verwendet wurde. Mit diesen Regeln, die man als
in einem prinzipielleren Sinne verstandene Quantisierungsregeln bezeichnen könnte, kann man jede Feldtheorie
einschließlich aller ihrer Wechselwirkungen in eine rein quantentheoretische Beschreibungsweise im
Tensorraum der Ur"=Alternativen überführen. Natürlich muss man an das entsprechende Feld im
jeweiligen Fall noch die entsprechenden inneren Freiheitsgrade tensorieren.

\subsection{Elektromagnetismus}

Wenn nun die Methoden der Überführung einer feldtheoretischen Beschreibungsweise in eine reine Beziehungsstruktur abstrakter
Quanteninformation, die man auch als eine prinzipiellere Art der Quantisierung ansehen könnte, die im letzten Unterabschnitt
vorgeschlagen wurde, in Bezug auf den Elektromagnetismus ganz konkret anwenden will, so muss man zunächst einmal ein freies
elektromagnetisches Feld beziehungsweise den Zustand eines freien Photons konstruieren. Einen solchen Zustand erhält man,
indem man an einen allgemeinen Zustand im Tensorraum, der ein einzelnes freies Teilchen oder allgemeiner gesprochen
Quantenobjekt beschreibt, einen vektoriellen Freiheitsgrad tensoriert. Dieser muss natürlich wie im Falle eines Fermions
mit den zusätzlichen Quantenzahlen aus einzelnen Ur"=Alternativen konstruiert werden. Wenn man also zwei Ur"=Alternativen
$u$ und $v$ zu Grunde legt, so kann man zunächst einen Dirac"=Spinor konstruieren:

\begin{equation}
\chi=\left(\begin{matrix} u\\i\sigma^{2} v^{*}\end{matrix}\right),
\end{equation}
und diesen kann man auf einen Vektor abbilden:

\begin{equation}
A^\mu_\chi=\bar \chi \gamma^\mu \chi,
\label{Vektor_Photon}
\end{equation}
wobei bezüglich eines beliebigen Dirac"=Spinors $\psi_D$ wie gewöhnlich die folgende
Definition der Adjungierung gilt:

\begin{equation}
\bar \psi_D=\psi_D^{\dagger}\gamma^0.
\label{Adjungierung_Dirac}
\end{equation}
Den Vektor ($\ref{Vektor_Photon}$) kann man als den Spin"=Freiheitsgrad eines Photons mit einem Zustand im Tensorraum
über ein weiteres Tensorprodukt verbinden, wodurch man dann den Zustand eines vektoriellen Teilchens erhält,
welcher die folgende Gestalt aufweist:

\begin{equation}
|\Psi_A\rangle=A^\mu_{\chi N}=\sum_{N_{ABCD}}\psi\left(N_{ABCD}\right)|N_{ABCD}\rangle \otimes A^\mu_\chi.
\end{equation}
Dieser Zustand soll also ein Photon beschreiben. Wenn man die Energie"=Impuls"=Relation, die gegeben
ist in ($\ref{Energie-Impuls-Relation}$) und die man auch gemäß ($\ref{Viererimpuls_Relation}$)
darstellen kann, auf den Zustand eines Photons anwendet, so ergibt sich die folgende Gleichung:

\begin{equation}
\left(P_{ABCD}\right)^\mu \left(P_{ABCD}\right)_\mu|\Psi_A\rangle=0.
\end{equation}
Diese Gleichung wird natürlich in der Ortsdarstellung gemäß ($\ref{Klein-Gordon-Gleichung}$) zur einer gewöhnlichen Wellengleichung:

\begin{equation}
\left(\partial_t^2-\partial_x^2-\partial_y^2-\partial_z^2\right)A^\mu_{\chi N}\left(\textbf{x},t\right)=0.
\end{equation}
Bezüglich der Wechselwirkung muss man sich nun der Lagrangedichte als klassischem Grenzfall der Quantenelektrodynamik zuwenden:

\begin{equation}
\mathcal{L}_{QED}=\bar \Psi_D\left(\textbf{x},t\right)\left[i\gamma^\mu \left(\partial_\mu+iA_\mu\left(\textbf{x},t\right)\right)\right]
\Psi_D\left(\textbf{x},t\right)-\frac{1}{4}F_{\mu\nu}\left(\textbf{x},t\right)F^{\mu\nu}\left(\textbf{x},t\right),
\end{equation}
wobei $F_{\mu\nu}\left(\mathbf{x},t\right)=\partial_\mu A_\nu\left(\mathbf{x},t\right)-\partial_\nu A_\mu\left(\mathbf{x},t\right)$.
Das Dirac"=Spinorfeld $\Psi_D\left(\mathbf{x},t\right)$, welches nach der Quantisierung Elektronen beziehungsweise generell elektrisch
geladene Fermionen beschreibt, und das elektromagnetische Feld beschrieben durch das Potential $A_\mu\left(\mathbf{x},t\right)$,
welches nach der Quantisierung Photonen beschreibt, weisen keine Selbstwechselwirkung auf. Die Dynamik eines fermionischen
Teilchens, das mit einem Photon wechselwirkt, kann dann, wenn man einmal davon ausgeht, dass die internen Freiheitsgrade des
fermionischen Teilchens durch $\Gamma$ beschrieben werden wie es in ($\ref{Interne_Freiheitsgrade}$) definiert wurde, durch jene
Gleichung beschrieben werden, die sich aus dem Übergang der zum Lagrangian der Quantenelektrodynamik gehörigen Wellengleichung
in eine reine auf Ur"=Alternativen basierende Beschreibung im Sinne der Regeln aus dem letzten Unterabschnitt
($\ref{Uebergang_Zustaende_Operatoren}$), ($\ref{Uebergang_Wechselwirkung_Produkt}$) und
($\ref{Uebergang_Wechselwirkung_Produkt_Ableitungen}$) ergibt:

\begin{eqnarray}
&&i\gamma^\mu\left[\partial_\mu+iA_\mu\left(\textbf{x},t\right)\right]\Psi_D\left(\textbf{x},t\right)=0\\
&&\longrightarrow i\gamma^\mu\left\{i\left(P_{ABCD}\right)_\mu\Gamma \psi_D(N_{ABCD},t)|N_{ABCD}\rangle\right.\nonumber\\
&&\left.\quad\quad\quad\quad+i\sum_{N_{ABCD}}\left[\left(A_{\chi N}\right)_\mu\psi_A(N_{ABCD},t)|N_{ABCD}\rangle\otimes
\Gamma \psi_D(N_{ABCD},t)|N_{ABCD}\rangle\right]\right\}=0\nonumber\\
&&\quad\quad=-\gamma^\mu\left\{\Gamma \left[\mathcal{P}_{\psi_D}(N_{ABCD},t)\right]_\mu|N_{ABCD}\rangle\right.\nonumber\\
&&\left.\quad\quad\quad\quad\quad+\sum_{N_{ABCD}}\left[\left(A_{\chi N}\right)_\mu\psi_A(N_{ABCD},t)|N_{ABCD}\rangle\otimes
\Gamma \psi_D(N_{ABCD},t)|N_{ABCD}\rangle\right]\right\}=0,\nonumber
\end{eqnarray}
wobei ($\ref{Anwendung_Viererimpuls_Zustand}$) verwendet wurde. Die Entstehung von Massen durch Wechselwirkung wird in dieser
Arbeit nicht thematisiert.

\section{Die Gravitation in der Quantentheorie der Ur-Alternativen}

\subsection{Konstruktion des freien Gravitationsfeldes und metrische Struktur}

Grundsätzlich wird in dieser Arbeit davon ausgegangen, dass die gewöhnliche Beschreibung der Gravitation im Rahmen der
allgemeinen Relativitätstheorie in klassischer Näherung vollkommen korrekt ist, also auf der klassischen Ebene vor der
Quantisierung keine Verallgemeinerung notwendig ist. Die Tatsache, dass die Gravitation klassisch am sinnvollsten als
lokale Eichtheorie der Translationen aufgefasst werden kann \cite{Lyre:2004}, was zur Torsion als
Feldgröße führt, kann hier außer acht gelassen werden, da diese Formulierung mit der gewöhnlichen Formulierung der
allgemeinen Relativitätstheorie äquivalent ist. In \cite{Lyre:1996} wird die Konstruktion von Gravitonen im Rahmen der
Quantentheorie der Ur"=Alternativen in anderer Weise bereits thematisiert, aber die Frage ihrer Wechselwirkung in keiner
Weise behandelt. Natürlich müssen in der Quantentheorie der Ur"=Alternativen konsequent alle existierenden Realitäten und
Objekte aus Ur"=Alternativen konstruiert werden. Demnach müssen auch die gravitative Wechselwirkung und die metrische Struktur
der Raum"=Zeit aus Ur"=Alternativen begründet werden. Dazu ist es zunächst einmal wichtig, ein freies Gravitationsfeld zu
konstruieren. Dieses kann natürlich wie alle anderen Objekte auch nicht als ein gewöhnliches Feld angesehen werden,
das dann anschließend einer Quantisierung unterworfen wird. Vielmehr muss es sich auch aus abstrakten Quantenobjekten
konstituieren, die dann anschließend in die Raum"=Zeit abgebildet werden. Diese Objekte müssen in der Quantentheorie
der Ur"=Alternativen als Gravitonen auf Zuständen vieler Ur"=Alternativen im Tensorraum basieren, nur dass sie anstatt
der gewöhnlichen Quantenzahlen, also dem Spin, dem Isospin und dem Farbfreiheitsgrad, die zusätzliche Struktur eines
metrischen Tensors aufweisen.

Man könnte, wenn der Terminus des Gravitons in die Beschreibung hineingebracht wird, zunächst einwenden,
dass bei einer Beschreibung der Gravitation im Rahmen relativistischer Quantenfeldtheorien, in deren
Zusammenhang der Begriff des Gravitons in der Regel gebraucht wird, die Hintergrundunabhängigkeit der
allgemeinen Relativitätstheorie nicht gewahrt bleibt. Aber natürlich kann dieser Einwand im Rahmen der
Quantentheorie der Ur"=Alternativen in Wirklichkeit in keiner Weise sinnvoll erhoben werden, denn hier
herrscht prinzipiell eine viel radikalere Realisierung der Hintergrundunabhängigkeit als in der
allgemeinen Relativitätstheorie, da hier nicht nur eine relationalistische Raumauffassung zu Grunde liegt,
sondern Räumlichkeit und feldtheoretische Bezüge in keiner Weise mehr vorausgesetzt werden. Dies wurde ja
weiter oben bereits in aller Ausführlichkeit diskutiert und verleiht dem ganzen Ansatz der Quantentheorie
der Ur"=Alternativen gerade seine besondere Überzeugungskraft. Die hier konstruierten Gravitonen sind also
wie alle anderen Objekte auch keine Objekte im Raum, sondern stellen abstrakte Zustände im Tensorraum
der Ur"=Alternativen dar, die erst auf indirektem Wege eine raum-zeitliche Darstellung erhalten.
Zunächst muss der metrische Tensor aus Ur"=Alternativen konstruiert werden, um dann in Analogie
zu den Quantenzahlen der Objekte, welche Elementarteilchen repräsentieren sollen, das Tensorprodukt
mit einem symmetrischen Zustand im Tensorraum zu bilden, welches dann den Gesamtzustand des
Gravitons darstellt. Zur Konstruktion des metrischen Tensors werden vier Ur"=Alternativen
gebraucht, die als $u_{g1}$, $u_{g2}$, $v_{g1}$ und $v_{g2}$ bezeichnet seien:

\begin{equation}
u_{g1}=\left(\begin{matrix} a_{ug1}+ib_{ug1}\\c_{ug1}+id_{ug1} \end{matrix}\right),\quad
u_{g2}=\left(\begin{matrix} a_{ug2}+ib_{ug2}\\c_{ug2}+id_{ug2} \end{matrix}\right),\quad
v_{g1}=\left(\begin{matrix} a_{vg1}+ib_{vg1}\\c_{vg1}+id_{vg1} \end{matrix}\right),\quad
v_{g2}=\left(\begin{matrix} a_{vg2}+ib_{vg2}\\c_{vg2}+id_{vg2} \end{matrix}\right).
\end{equation}
Zunächst wird aus den beiden Ur"=Alternativen $u_{g1}$ und $u_{g2}$ ein Dirac"=Spinor konstruiert
und aus den Ur"=Alternativen $v_{g1}$ und $v_{g2}$ ein weiterer Dirac"=Spinor:

\begin{equation}
\chi_u=\left(\begin{matrix} u_{g1}\\i\sigma^2 u_{g2}^{*} \end{matrix}\right),
\quad \chi_v=\left(\begin{matrix} v_{g1}\\ i\sigma^2 v_{g2}^{*}\end{matrix}\right).
\end{equation}
Mit Hilfe der Relation, welche die Dirac"=Matrizen in eine Beziehung zur Minkowski"=Metrik
stellt ($\ref{Gamma-Matrizen-Relation}$), kann nun aus den beiden Dirac"=Spinoren $\chi_u$
und $\chi_v$ ein metrischer Tensor konstruiert werden, wobei die Definiton
($\ref{Adjungierung_Dirac}$) zu Grunde gelegt wird:

\begin{equation}
g^{\mu\nu}_{\chi}=\frac{1}{2}\left(\bar \chi_u \gamma^\mu \chi_u \bar \chi_v \gamma^\nu \chi_v
+\bar \chi_u \gamma^\nu \chi_u \bar \chi_v \gamma^\mu \chi_v\right).
\end{equation}
Wenn man nun diesen Ausdruck für den metrischen Tensor konkret durch die Komponenten der vier Ur"=Alternativen
ausdrücken will, so muss man zunächst die Komponenten der Vektoren bestimmen, aus denen er gebildet ist und
die folgende Gestalt haben:

\begin{eqnarray}
\bar \chi_u \gamma^0 \chi_u&=&a_{ug1}^2+b_{ug1}^2+c_{ug1}^2+d_{ug1}^2+a_{ug2}^2+b_{ug2}^2+c_{ug2}^2+d_{ug2}^2,\nonumber\\
\bar \chi_u \gamma^1 \chi_u&=&2a_{ug1}c_{ug1}+2b_{ug1}d_{ug1}+2a_{ug2}c_{ug2}+2b_{ug2}d_{ug2},\nonumber\\
\bar \chi_u \gamma^2 \chi_u&=&2a_{ug1}d_{ug1}-2b_{ug1}c_{ug1}+2a_{ug2}d_{ug2}-2b_{ug2}c_{ug2},\nonumber\\
\bar \chi_u \gamma^3 \chi_u&=&a_{ug1}^2+b_{ug1}^2-c_{ug1}^2-d_{ug1}^2+a_{ug2}^2+b_{ug2}^2-c_{ug2}^2-d_{ug2}^2,\nonumber\\
\bar \chi_v \gamma^0 \chi_v&=&a_{vg1}^2+b_{vg1}^2+c_{vg1}^2+d_{vg1}^2+a_{vg2}^2+b_{vg2}^2+c_{vg2}^2+d_{vg2}^2,\nonumber\\
\bar \chi_v \gamma^1 \chi_v&=&2a_{vg1}c_{vg1}+2b_{vg1}d_{vg1}+2a_{vg2}c_{vg2}+2b_{vg2}d_{vg2},\nonumber\\
\bar \chi_v \gamma^2 \chi_v&=&2a_{vg1}d_{vg1}-2b_{vg1}c_{vg1}+2a_{vg2}d_{vg2}-2b_{vg2}c_{vg2},\nonumber\\
\bar \chi_v \gamma^3 \chi_v&=&a_{vg1}^2+b_{vg1}^2-c_{vg1}^2-d_{vg1}^2+a_{vg2}^2+b_{vg2}^2-c_{vg2}^2-d_{vg2}^2.
\end{eqnarray}
Damit ergibt sich für die Komponenten des metrischen Tensors:

\begin{eqnarray}
g^{00}_\chi&=&a_{ug1}^2 a_{vg1}^2
+a_{ug1}^2 b_{vg1}^2
+a_{ug1}^2 c_{vg1}^2
+a_{ug1}^2 d_{vg1}^2
+a_{ug1}^2 a_{vg2}^2
+a_{ug1}^2 b_{vg2}^2
+a_{ug1}^2 c_{vg2}^2
+a_{ug1}^2 d_{vg2}^2\nonumber\\&&
+b_{ug1}^2 a_{vg1}^2
+b_{ug1}^2 b_{vg1}^2
+b_{ug1}^2 c_{vg1}^2
+b_{ug1}^2 d_{vg1}^2
+b_{ug1}^2 a_{vg2}^2
+b_{ug1}^2 b_{vg2}^2
+b_{ug1}^2 c_{vg2}^2
+b_{ug1}^2 d_{vg2}^2\nonumber\\&&
+c_{ug1}^2 a_{vg1}^2
+c_{ug1}^2 b_{vg1}^2
+c_{ug1}^2 c_{vg1}^2
+c_{ug1}^2 d_{vg1}^2
+c_{ug1}^2 a_{vg2}^2
+c_{ug1}^2 b_{vg2}^2
+c_{ug1}^2 c_{vg2}^2
+c_{ug1}^2 d_{vg2}^2\nonumber\\&&
+d_{ug1}^2 a_{vg1}^2
+d_{ug1}^2 b_{vg1}^2
+d_{ug1}^2 c_{vg1}^2
+d_{ug1}^2 d_{vg1}^2
+d_{ug1}^2 a_{vg2}^2
+d_{ug1}^2 b_{vg2}^2
+d_{ug1}^2 c_{vg2}^2
+d_{ug1}^2 d_{vg2}^2\nonumber\\&&
+a_{ug2}^2 a_{vg1}^2
+a_{ug2}^2 b_{vg1}^2
+a_{ug2}^2 c_{vg1}^2
+a_{ug2}^2 d_{vg1}^2
+a_{ug2}^2 a_{vg2}^2
+a_{ug2}^2 b_{vg2}^2
+a_{ug2}^2 c_{vg2}^2
+a_{ug2}^2 d_{vg2}^2\nonumber\\&&
+b_{ug2}^2 a_{vg1}^2
+b_{ug2}^2 b_{vg1}^2
+b_{ug2}^2 c_{vg1}^2
+b_{ug2}^2 d_{vg1}^2
+b_{ug2}^2 a_{vg2}^2
+b_{ug2}^2 b_{vg2}^2
+b_{ug2}^2 c_{vg2}^2
+b_{ug2}^2 d_{vg2}^2\nonumber\\&&
+c_{ug2}^2 a_{vg1}^2
+c_{ug2}^2 b_{vg1}^2
+c_{ug2}^2 c_{vg1}^2
+c_{ug2}^2 d_{vg1}^2
+c_{ug2}^2 a_{vg2}^2
+c_{ug2}^2 b_{vg2}^2
+c_{ug2}^2 c_{vg2}^2
+c_{ug2}^2 d_{vg2}^2\nonumber\\&&
+d_{ug2}^2 a_{vg1}^2
+d_{ug2}^2 b_{vg1}^2
+d_{ug2}^2 c_{vg1}^2
+d_{ug2}^2 d_{vg1}^2
+d_{ug2}^2 a_{vg2}^2
+d_{ug2}^2 b_{vg2}^2
+d_{ug2}^2 c_{vg2}^2
+d_{ug2}^2 d_{vg2}^2,\nonumber\\
\end{eqnarray}

\begin{eqnarray}
g^{11}_\chi&=&2a_{ug1}c_{ug1} a_{vg1}c_{vg1}
+2a_{ug1}c_{ug1} b_{vg1}d_{vg1}
+2a_{ug1}c_{ug1} a_{vg2}c_{vg2}
+2a_{ug1}c_{ug1} b_{vg2}d_{vg2}\nonumber\\
&&+2b_{ug1}d_{ug1} a_{vg1}c_{vg1}
+2b_{ug1}d_{ug1} b_{vg1}d_{vg1}
+2b_{ug1}d_{ug1} a_{vg2}c_{vg2}
+2b_{ug1}d_{ug1} b_{vg2}d_{vg2}\nonumber\\
&&+2a_{ug2}c_{ug2} a_{vg1}c_{vg1}
+2a_{ug2}c_{ug2} b_{vg1}d_{vg1}
+2a_{ug2}c_{ug2} a_{vg2}c_{vg2}
+2a_{ug2}c_{ug2} b_{vg2}d_{vg2}\nonumber\\
&&+2b_{ug2}d_{ug2} a_{vg1}c_{vg1}
+2b_{ug2}d_{ug2} b_{vg1}d_{vg1}
+2b_{ug2}d_{ug2} a_{vg2}c_{vg2}
+2b_{ug2}d_{ug2} b_{vg2}d_{vg2},\nonumber\\
\end{eqnarray}

\begin{eqnarray}
g^{22}_\chi&=&2a_{ug1}d_{ug1} a_{vg1}d_{vg1}
-2a_{ug1}d_{ug1} b_{vg1}c_{vg1}
+2a_{ug1}d_{ug1} a_{vg2}d_{vg2}
-2a_{ug1}d_{ug1} b_{vg2}c_{vg2}\nonumber\\
&&-2b_{ug1}c_{ug1} a_{vg1}d_{vg1}
+2b_{ug1}c_{ug1} b_{vg1}c_{vg1}
-2b_{ug1}c_{ug1} a_{vg2}d_{vg2}
+2b_{ug1}c_{ug1} b_{vg2}c_{vg2}\nonumber\\
&&+2a_{ug2}d_{ug2} a_{vg1}d_{vg1}
-2a_{ug2}d_{ug2} b_{vg1}c_{vg1}
+2a_{ug2}d_{ug2} a_{vg2}d_{vg2}
-2a_{ug2}d_{ug2} b_{vg2}c_{vg2}\nonumber\\
&&-2b_{ug2}c_{ug2} a_{vg1}d_{vg1}
+2b_{ug2}c_{ug2} b_{vg1}c_{vg1}
-2b_{ug2}c_{ug2} a_{vg2}d_{vg2}
+2b_{ug2}c_{ug2} b_{vg2}c_{vg2},\nonumber\\
\end{eqnarray}

\begin{eqnarray}
g^{33}_\chi&=&a_{ug1}^2 a_{vg1}^2
+a_{ug1}^2 b_{vg1}^2
-a_{ug1}^2 c_{vg1}^2
-a_{ug1}^2 d_{vg1}^2
+a_{ug1}^2 a_{vg2}^2
+a_{ug1}^2 b_{vg2}^2
-a_{ug1}^2 c_{vg2}^2
-a_{ug1}^2 d_{vg2}^2\nonumber\\&&
+b_{ug1}^2 a_{vg1}^2
+b_{ug1}^2 b_{vg1}^2
-b_{ug1}^2 c_{vg1}^2
-b_{ug1}^2 d_{vg1}^2
+b_{ug1}^2 a_{vg2}^2
+b_{ug1}^2 b_{vg2}^2
-b_{ug1}^2 c_{vg2}^2
-b_{ug1}^2 d_{vg2}^2\nonumber\\&&
-c_{ug1}^2 a_{vg1}^2
-c_{ug1}^2 b_{vg1}^2
+c_{ug1}^2 c_{vg1}^2
+c_{ug1}^2 d_{vg1}^2
-c_{ug1}^2 a_{vg2}^2
-c_{ug1}^2 b_{vg2}^2
+c_{ug1}^2 c_{vg2}^2
+c_{ug1}^2 d_{vg2}^2\nonumber\\&&
-d_{ug1}^2 a_{vg1}^2
-d_{ug1}^2 b_{vg1}^2
+d_{ug1}^2 c_{vg1}^2
+d_{ug1}^2 d_{vg1}^2
-d_{ug1}^2 a_{vg2}^2
-d_{ug1}^2 b_{vg2}^2
+d_{ug1}^2 c_{vg2}^2
+d_{ug1}^2 d_{vg2}^2\nonumber\\&&
+a_{ug2}^2 a_{vg1}^2
+a_{ug2}^2 b_{vg1}^2
-a_{ug2}^2 c_{vg1}^2
-a_{ug2}^2 d_{vg1}^2
+a_{ug2}^2 a_{vg2}^2
+a_{ug2}^2 b_{vg2}^2
-a_{ug2}^2 c_{vg2}^2
-a_{ug2}^2 d_{vg2}^2\nonumber\\&&
+b_{ug2}^2 a_{vg1}^2
+b_{ug2}^2 b_{vg1}^2
-b_{ug2}^2 c_{vg1}^2
-b_{ug2}^2 d_{vg1}^2
+b_{ug2}^2 a_{vg2}^2
+b_{ug2}^2 b_{vg2}^2
-b_{ug2}^2 c_{vg2}^2
-b_{ug2}^2 d_{vg2}^2\nonumber\\&&
-c_{ug2}^2 a_{vg1}^2
-c_{ug2}^2 b_{vg1}^2
+c_{ug2}^2 c_{vg1}^2
+c_{ug2}^2 d_{vg1}^2
-c_{ug2}^2 a_{vg2}^2
-c_{ug2}^2 b_{vg2}^2
+c_{ug2}^2 c_{vg2}^2
+c_{ug2}^2 d_{vg2}^2\nonumber\\&&
-d_{ug2}^2 a_{vg1}^2
-d_{ug2}^2 b_{vg1}^2
+d_{ug2}^2 c_{vg1}^2
+d_{ug2}^2 d_{vg1}^2
-d_{ug2}^2 a_{vg2}^2
-d_{ug2}^2 b_{vg2}^2
+d_{ug2}^2 c_{vg2}^2
+d_{ug2}^2 d_{vg2}^2,\nonumber\\
\end{eqnarray}

\begin{eqnarray}
g^{01}_\chi=g^{10}_\chi&=&a_{ug1}^2 a_{vg1}c_{vg1}
+a_{ug1}^2 b_{vg1}d_{vg1}
+a_{ug1}^2 a_{vg2}c_{vg2}
+a_{ug1}^2 b_{vg2}d_{vg2}\nonumber\\
&&+b_{ug1}^2 a_{vg1}c_{vg1}
+b_{ug1}^2 b_{vg1}d_{vg1}
+b_{ug1}^2 a_{vg2}c_{vg2}
+b_{ug1}^2 b_{vg2}d_{vg2}\nonumber\\
&&+c_{ug1}^2 a_{vg1}c_{vg1}
+c_{ug1}^2 b_{vg1}d_{vg1}
+c_{ug1}^2 a_{vg2}c_{vg2}
+c_{ug1}^2 b_{vg2}d_{vg2}\nonumber\\
&&+d_{ug1}^2 a_{vg1}c_{vg1}
+d_{ug1}^2 b_{vg1}d_{vg1}
+d_{ug1}^2 a_{vg2}c_{vg2}
+d_{ug1}^2 b_{vg2}d_{vg2}\nonumber\\
&&+a_{ug2}^2 a_{vg1}c_{vg1}
+a_{ug2}^2 b_{vg1}d_{vg1}
+a_{ug2}^2 a_{vg2}c_{vg2}
+a_{ug2}^2 b_{vg2}d_{vg2}\nonumber\\
&&+b_{ug2}^2 a_{vg1}c_{vg1}
+b_{ug2}^2 b_{vg1}d_{vg1}
+b_{ug2}^2 a_{vg2}c_{vg2}
+b_{ug2}^2 b_{vg2}d_{vg2}\nonumber\\
&&+c_{ug2}^2 a_{vg1}c_{vg1}
+c_{ug2}^2 b_{vg1}d_{vg1}
+c_{ug2}^2 a_{vg2}c_{vg2}
+c_{ug2}^2 b_{vg2}d_{vg2}\nonumber\\
&&+d_{ug2}^2 a_{vg1}c_{vg1}
+d_{ug2}^2 b_{vg1}d_{vg1}
+d_{ug2}^2 a_{vg2}c_{vg2}
+d_{ug2}^2 b_{vg2}d_{vg2}\nonumber\\
&&+a_{vg1}^2 a_{ug1}c_{ug1}
+a_{vg1}^2 b_{ug1}d_{ug1}
+a_{vg1}^2 a_{ug2}c_{ug2}
+a_{vg1}^2 b_{ug2}d_{ug2}\nonumber\\
&&+b_{vg1}^2 a_{ug1}c_{ug1}
+b_{vg1}^2 b_{ug1}d_{ug1}
+b_{vg1}^2 a_{ug2}c_{ug2}
+b_{vg1}^2 b_{ug2}d_{ug2}\nonumber\\
&&+c_{vg1}^2 a_{ug1}c_{ug1}
+c_{vg1}^2 b_{ug1}d_{ug1}
+c_{vg1}^2 a_{ug2}c_{ug2}
+c_{vg1}^2 b_{ug2}d_{ug2}\nonumber\\
&&+d_{vg1}^2 a_{ug1}c_{ug1}
+d_{vg1}^2 b_{ug1}d_{ug1}
+d_{vg1}^2 a_{ug2}c_{ug2}
+d_{vg1}^2 b_{ug2}d_{ug2}\nonumber\\
&&+a_{vg2}^2 a_{ug1}c_{ug1}
+a_{vg2}^2 b_{ug1}d_{ug1}
+a_{vg2}^2 a_{ug2}c_{ug2}
+a_{vg2}^2 b_{ug2}d_{ug2}\nonumber\\
&&+b_{vg2}^2 a_{ug1}c_{ug1}
+b_{vg2}^2 b_{ug1}d_{ug1}
+b_{vg2}^2 a_{ug2}c_{ug2}
+b_{vg2}^2 b_{ug2}d_{ug2}\nonumber\\
&&+c_{vg2}^2 a_{ug1}c_{ug1}
+c_{vg2}^2 b_{ug1}d_{ug1}
+c_{vg2}^2 a_{ug2}c_{ug2}
+c_{vg2}^2 b_{ug2}d_{ug2}\nonumber\\
&&+d_{vg2}^2 a_{ug1}c_{ug1}
+d_{vg2}^2 b_{ug1}d_{ug1}
+d_{vg2}^2 a_{ug2}c_{ug2}
+d_{vg2}^2 b_{ug2}d_{ug2},
\end{eqnarray}

\begin{eqnarray}
g^{02}_\chi&=&g^{20}_\chi=a_{ug1}^2 a_{vg1}d_{vg1}
-a_{ug1}^2 b_{vg1}c_{vg1}
+a_{ug1}^2 a_{vg2}d_{vg2}
-a_{ug1}^2 b_{vg2}c_{vg2}\nonumber\\
&&+b_{ug1}^2 a_{vg1}d_{vg1}
-b_{ug1}^2 b_{vg1}c_{vg1}
+b_{ug1}^2 a_{vg2}d_{vg2}
-b_{ug1}^2 b_{vg2}c_{vg2}\nonumber\\
&&+c_{ug1}^2 a_{vg1}d_{vg1}
-c_{ug1}^2 b_{vg1}c_{vg1}
+c_{ug1}^2 a_{vg2}d_{vg2}
-c_{ug1}^2 b_{vg2}c_{vg2}\nonumber\\
&&+d_{ug1}^2 a_{vg1}d_{vg1}
-d_{ug1}^2 b_{vg1}c_{vg1}
+d_{ug1}^2 a_{vg2}d_{vg2}
-d_{ug1}^2 b_{vg2}c_{vg2}\nonumber\\
&&+a_{ug2}^2 a_{vg1}d_{vg1}
-a_{ug2}^2 b_{vg1}c_{vg1}
+a_{ug2}^2 a_{vg2}d_{vg2}
-a_{ug2}^2 b_{vg2}c_{vg2}\nonumber\\
&&+b_{ug2}^2 a_{vg1}d_{vg1}
-b_{ug2}^2 b_{vg1}c_{vg1}
+b_{ug2}^2 a_{vg2}d_{vg2}
-b_{ug2}^2 b_{vg2}c_{vg2}\nonumber\\
&&+c_{ug2}^2 a_{vg1}d_{vg1}
-c_{ug2}^2 b_{vg1}c_{vg1}
+c_{ug2}^2 a_{vg2}d_{vg2}
-c_{ug2}^2 b_{vg2}c_{vg2}\nonumber\\
&&+d_{ug2}^2 a_{vg1}d_{vg1}
-d_{ug2}^2 b_{vg1}c_{vg1}
+d_{ug2}^2 a_{vg2}d_{vg2}
-d_{ug2}^2 b_{vg2}c_{vg2}\nonumber\\
&&+a_{vg1}^2 a_{ug1}d_{ug1}
-a_{vg1}^2 b_{ug1}c_{ug1}
+a_{vg1}^2 a_{ug2}d_{ug2}
-a_{vg1}^2 b_{ug2}c_{ug2}\nonumber\\
&&+b_{vg1}^2 a_{ug1}d_{ug1}
-b_{vg1}^2 b_{ug1}c_{ug1}
+b_{vg1}^2 a_{ug2}d_{ug2}
-b_{vg1}^2 b_{ug2}c_{ug2}\nonumber\\
&&+c_{vg1}^2 a_{ug1}d_{ug1}
-c_{vg1}^2 b_{ug1}c_{ug1}
+c_{vg1}^2 a_{ug2}d_{ug2}
-c_{vg1}^2 b_{ug2}c_{ug2}\nonumber\\
&&+d_{vg1}^2 a_{ug1}d_{ug1}
-d_{vg1}^2 b_{ug1}c_{ug1}
+d_{vg1}^2 a_{ug2}d_{ug2}
-d_{vg1}^2 b_{ug2}c_{ug2}\nonumber\\
&&+a_{vg2}^2 a_{ug1}d_{ug1}
-a_{vg2}^2 b_{ug1}c_{ug1}
+a_{vg2}^2 a_{ug2}d_{ug2}
-a_{vg2}^2 b_{ug2}c_{ug2}\nonumber\\
&&+b_{vg2}^2 a_{ug1}d_{ug1}
-b_{vg2}^2 b_{ug1}c_{ug1}
+b_{vg2}^2 a_{ug2}d_{ug2}
-b_{vg2}^2 b_{ug2}c_{ug2}\nonumber\\
&&+c_{vg2}^2 a_{ug1}d_{ug1}
-c_{vg2}^2 b_{ug1}c_{ug1}
+c_{vg2}^2 a_{ug2}d_{ug2}
-c_{vg2}^2 b_{ug2}c_{ug2}\nonumber\\
&&+d_{vg2}^2 a_{ug1}d_{ug1}
-d_{vg2}^2 b_{ug1}c_{ug1}
+d_{vg2}^2 a_{ug2}d_{ug2}
-d_{vg2}^2 b_{ug2}c_{ug2},
\end{eqnarray}

\begin{eqnarray}
g^{03}_\chi=g^{30}_\chi&=&a_{ug1}^2 a_{vg1}^2
+a_{ug1}^2 b_{vg1}^2
+b_{ug1}^2 a_{vg1}^2
+b_{ug1}^2 b_{vg1}^2
-c_{ug1}^2 c_{vg1}^2
-c_{ug1}^2 d_{vg1}^2
-d_{ug1}^2 c_{vg1}^2
-d_{ug1}^2 d_{vg1}^2\nonumber\\&&
+a_{ug2}^2 a_{vg1}^2
+a_{ug2}^2 b_{vg1}^2
+b_{ug2}^2 a_{vg1}^2
+b_{ug2}^2 b_{vg1}^2
-c_{ug2}^2 c_{vg1}^2
-c_{ug2}^2 d_{vg1}^2
-d_{ug2}^2 c_{vg1}^2
-d_{ug2}^2 d_{vg1}^2\nonumber\\&&
+a_{ug1}^2 a_{vg2}^2
+a_{ug1}^2 b_{vg2}^2
+b_{ug1}^2 a_{vg2}^2
+b_{ug1}^2 b_{vg2}^2
-c_{ug1}^2 c_{vg2}^2
-c_{ug1}^2 d_{vg2}^2
-d_{ug1}^2 c_{vg2}^2
-d_{ug1}^2 d_{vg2}^2\nonumber\\&&
+a_{ug2}^2 a_{vg2}^2
+a_{ug2}^2 b_{vg2}^2
+b_{ug2}^2 a_{vg2}^2
+b_{ug2}^2 b_{vg2}^2
-c_{ug2}^2 c_{vg2}^2
-c_{ug2}^2 d_{vg2}^2
-d_{ug2}^2 c_{vg2}^2
-d_{ug2}^2 d_{vg2}^2,\nonumber\\
\end{eqnarray}

\begin{eqnarray}
g^{12}_\chi=g^{21}_\chi&=&
2a_{ug1}c_{ug1} a_{vg1}d_{vg1}
-2a_{ug1}c_{ug1} b_{vg1}c_{vg1}
+2a_{ug1}c_{ug1} a_{vg2}d_{vg2}
-2a_{ug1}c_{ug1} b_{vg2}c_{vg2}\nonumber\\
&&+2b_{ug1}d_{ug1} a_{vg1}d_{vg1}
-2b_{ug1}d_{ug1} b_{vg1}c_{vg1}
+2b_{ug1}d_{ug1} a_{vg2}d_{vg2}
-2b_{ug1}d_{ug1} b_{vg2}c_{vg2}\nonumber\\
&&+2a_{ug2}c_{ug2} a_{vg1}d_{vg1}
-2a_{ug2}c_{ug2} b_{vg1}c_{vg1}
+2a_{ug2}c_{ug2} a_{vg2}d_{vg2}
-2a_{ug2}c_{ug2} b_{vg2}c_{vg2}\nonumber\\
&&+2b_{ug2}d_{ug2} a_{vg1}d_{vg1}
-2b_{ug2}d_{ug2} b_{vg1}c_{vg1}
+2b_{ug2}d_{ug2} a_{vg2}d_{vg2}
-2b_{ug2}d_{ug2} b_{vg2}c_{vg2}\nonumber\\
&&+2a_{vg1}c_{vg1} a_{ug1}d_{ug1}
-2a_{vg1}c_{vg1} b_{ug1}c_{ug1}
+2a_{vg1}c_{vg1} a_{ug2}d_{ug2}
-2a_{vg1}c_{vg1} b_{ug2}c_{ug2}\nonumber\\
&&+2b_{vg1}d_{vg1} a_{ug1}d_{ug1}
-2b_{vg1}d_{vg1} b_{ug1}c_{ug1}
+2b_{vg1}d_{vg1} a_{ug2}d_{ug2}
-2b_{vg1}d_{vg1} b_{ug2}c_{ug2}\nonumber\\
&&+2a_{vg2}c_{vg2} a_{ug1}d_{ug1}
-2a_{vg2}c_{vg2} b_{ug1}c_{ug1}
+2a_{vg2}c_{vg2} a_{ug2}d_{ug2}
-2a_{vg2}c_{vg2} b_{ug2}c_{ug2}\nonumber\\
&&+2b_{vg2}d_{vg2} a_{ug1}d_{ug1}
-2b_{vg2}d_{vg2} b_{ug1}c_{ug1}
+2b_{vg2}d_{vg2} a_{ug2}d_{ug2}
-2b_{vg2}d_{vg2} b_{ug2}c_{ug2},\nonumber\\
\end{eqnarray}

\begin{eqnarray}
g^{13}_\chi=g^{31}_\chi&=&a_{ug1}^2 a_{vg1}c_{vg1}
+a_{ug1}^2 b_{vg1}d_{vg1}
+a_{ug1}^2 a_{vg2}c_{vg2}
+a_{ug1}^2 b_{vg2}d_{vg2}\nonumber\\
&&+b_{ug1}^2 a_{vg1}c_{vg1}
+b_{ug1}^2 b_{vg1}d_{vg1}
+b_{ug1}^2 a_{vg2}c_{vg2}
+b_{ug1}^2 b_{vg2}d_{vg2}\nonumber\\
&&-c_{ug1}^2 a_{vg1}c_{vg1}
-c_{ug1}^2 b_{vg1}d_{vg1}
-c_{ug1}^2 a_{vg2}c_{vg2}
-c_{ug1}^2 b_{vg2}d_{vg2}\nonumber\\
&&-d_{ug1}^2 a_{vg1}c_{vg1}
-d_{ug1}^2 b_{vg1}d_{vg1}
-d_{ug1}^2 a_{vg2}c_{vg2}
-d_{ug1}^2 b_{vg2}d_{vg2}\nonumber\\
&&+a_{ug2}^2 a_{vg1}c_{vg1}
+a_{ug2}^2 b_{vg1}d_{vg1}
+a_{ug2}^2 a_{vg2}c_{vg2}
+a_{ug2}^2 b_{vg2}d_{vg2}\nonumber\\
&&+b_{ug2}^2 a_{vg1}c_{vg1}
+b_{ug2}^2 b_{vg1}d_{vg1}
+b_{ug2}^2 a_{vg2}c_{vg2}
+b_{ug2}^2 b_{vg2}d_{vg2}\nonumber\\
&&-c_{ug2}^2 a_{vg1}c_{vg1}
-c_{ug2}^2 b_{vg1}d_{vg1}
-c_{ug2}^2 a_{vg2}c_{vg2}
-c_{ug2}^2 b_{vg2}d_{vg2}\nonumber\\
&&-d_{ug2}^2 a_{vg1}c_{vg1}
-d_{ug2}^2 b_{vg1}d_{vg1}
-d_{ug2}^2 a_{vg2}c_{vg2}
-d_{ug2}^2 b_{vg2}d_{vg2}\nonumber\\
&&+a_{vg1}^2 a_{ug1}c_{ug1}
+a_{vg1}^2 b_{ug1}d_{ug1}
+a_{vg1}^2 a_{ug2}c_{ug2}
+a_{vg1}^2 b_{ug2}d_{ug2}\nonumber\\
&&+b_{vg1}^2 a_{ug1}c_{ug1}
+b_{vg1}^2 b_{ug1}d_{ug1}
+b_{vg1}^2 a_{ug2}c_{ug2}
+b_{vg1}^2 b_{ug2}d_{ug2}\nonumber\\
&&-c_{vg1}^2 a_{ug1}c_{ug1}
-c_{vg1}^2 b_{ug1}d_{ug1}
-c_{vg1}^2 a_{ug2}c_{ug2}
-c_{vg1}^2 b_{ug2}d_{ug2}\nonumber\\
&&-d_{vg1}^2 a_{ug1}c_{ug1}
-d_{vg1}^2 b_{ug1}d_{ug1}
-d_{vg1}^2 a_{ug2}c_{ug2}
-d_{vg1}^2 b_{ug2}d_{ug2}\nonumber\\
&&+a_{vg2}^2 a_{ug1}c_{ug1}
+a_{vg2}^2 b_{ug1}d_{ug1}
+a_{vg2}^2 a_{ug2}c_{ug2}
+a_{vg2}^2 b_{ug2}d_{ug2}\nonumber\\
&&+b_{vg2}^2 a_{ug1}c_{ug1}
+b_{vg2}^2 b_{ug1}d_{ug1}
+b_{vg2}^2 a_{ug2}c_{ug2}
+b_{vg2}^2 b_{ug2}d_{ug2}\nonumber\\
&&-c_{vg2}^2 a_{ug1}c_{ug1}
-c_{vg2}^2 b_{ug1}d_{ug1}
-c_{vg2}^2 a_{ug2}c_{ug2}
-c_{vg2}^2 b_{ug2}d_{ug2}\nonumber\\
&&-d_{vg2}^2 a_{ug1}c_{ug1}
-d_{vg2}^2 b_{ug1}d_{ug1}
-d_{vg2}^2 a_{ug2}c_{ug2}
-d_{vg2}^2 b_{ug2}d_{ug2},
\end{eqnarray}

\begin{eqnarray}
g^{23}_\chi=g^{32}_\chi&=&a_{ug1}^2 a_{vg1}d_{vg1}
-a_{ug1}^2 b_{vg1}c_{vg1}
+a_{ug1}^2 a_{vg2}d_{vg2}
-a_{ug1}^2 b_{vg2}c_{vg2}\nonumber\\
&&+b_{ug1}^2 a_{vg1}d_{vg1}
-b_{ug1}^2 b_{vg1}c_{vg1}
+b_{ug1}^2 a_{vg2}d_{vg2}
-b_{ug1}^2 b_{vg2}c_{vg2}\nonumber\\
&&-c_{ug1}^2 a_{vg1}d_{vg1}
+c_{ug1}^2 b_{vg1}c_{vg1}
-c_{ug1}^2 a_{vg2}d_{vg2}
+c_{ug1}^2 b_{vg2}c_{vg2}\nonumber\\
&&-d_{ug1}^2 a_{vg1}d_{vg1}
+d_{ug1}^2 b_{vg1}c_{vg1}
-d_{ug1}^2 a_{vg2}d_{vg2}
+d_{ug1}^2 b_{vg2}c_{vg2}\nonumber\\
&&+a_{ug2}^2 a_{vg1}d_{vg1}
-a_{ug2}^2 b_{vg1}c_{vg1}
+a_{ug2}^2 a_{vg2}d_{vg2}
-a_{ug2}^2 b_{vg2}c_{vg2}\nonumber\\
&&+b_{ug2}^2 a_{vg1}d_{vg1}
-b_{ug2}^2 b_{vg1}c_{vg1}
+b_{ug2}^2 a_{vg2}d_{vg2}
-b_{ug2}^2 b_{vg2}c_{vg2}\nonumber\\
&&-c_{ug2}^2 a_{vg1}d_{vg1}
+c_{ug2}^2 b_{vg1}c_{vg1}
-c_{ug2}^2 a_{vg2}d_{vg2}
+c_{ug2}^2 b_{vg2}c_{vg2}\nonumber\\
&&-d_{ug2}^2 a_{vg1}d_{vg1}
+d_{ug2}^2 b_{vg1}c_{vg1}
-d_{ug2}^2 a_{vg2}d_{vg2}
+d_{ug2}^2 b_{vg2}c_{vg2}\nonumber\\
&&+a_{vg1}^2 a_{ug1}d_{ug1}
-a_{vg1}^2 b_{ug1}c_{ug1}
+a_{vg1}^2 a_{ug2}d_{ug2}
-a_{vg1}^2 b_{ug2}c_{ug2}\nonumber\\
&&+b_{vg1}^2 a_{ug1}d_{ug1}
-b_{vg1}^2 b_{ug1}c_{ug1}
+b_{vg1}^2 a_{ug2}d_{ug2}
-b_{vg1}^2 b_{ug2}c_{ug2}\nonumber\\
&&-c_{vg1}^2 a_{ug1}d_{ug1}
+c_{vg1}^2 b_{ug1}c_{ug1}
-c_{vg1}^2 a_{ug2}d_{ug2}
+c_{vg1}^2 b_{ug2}c_{ug2}\nonumber\\
&&-d_{vg1}^2 a_{ug1}d_{ug1}
+d_{vg1}^2 b_{ug1}c_{ug1}
-d_{vg1}^2 a_{ug2}d_{ug2}
+d_{vg1}^2 b_{ug2}c_{ug2}\nonumber\\
&&+a_{vg2}^2 a_{ug1}d_{ug1}
-a_{vg2}^2 b_{ug1}c_{ug1}
+a_{vg2}^2 a_{ug2}d_{ug2}
-a_{vg2}^2 b_{ug2}c_{ug2}\nonumber\\
&&+b_{vg2}^2 a_{ug1}d_{ug1}
-b_{vg2}^2 b_{ug1}c_{ug1}
+b_{vg2}^2 a_{ug2}d_{ug2}
-b_{vg2}^2 b_{ug2}c_{ug2}\nonumber\\
&&-c_{vg2}^2 a_{ug1}d_{ug1}
+c_{vg2}^2 b_{ug1}c_{ug1}
-c_{vg2}^2 a_{ug2}d_{ug2}
+c_{vg2}^2 b_{ug2}c_{ug2}\nonumber\\
&&-d_{vg2}^2 a_{ug1}d_{ug1}
+d_{vg2}^2 b_{ug1}c_{ug1}
-d_{vg2}^2 a_{ug2}d_{ug2}
+d_{vg2}^2 b_{ug2}c_{ug2}.
\end{eqnarray}
Wenn man nun das Tensorprodukt mit einem allgemeinen Zustand vieler Ur"=Alternativen ($\ref{Zustand_Tensorraum}$) bildet,
so ergibt sich der Zustand für ein aus Ur"=Alternativen konstruiertes Graviton:

\begin{equation}
|\Psi_g\rangle=g_{N\chi}^{\mu\nu}=\sum_{N_{ABCD}}\psi(N_{ABCD})|N_{ABCD}\rangle \otimes g^{\mu\nu}_{\chi}.
\label{Gravitonzustand}
\end{equation}
Unter Verwendung von ($\ref{Zustand_Darstellung_Ortsraum}$) kann dieser Zustand in der Raum-Zeit dargestellt werden als:

\begin{equation}
|\Psi_g\rangle=g^{\mu\nu}_{N\chi} \quad\longrightarrow\quad g^{\mu\nu}_{N\chi}\left(\mathbf{x}\right)
=\sum_{N_{xyzn}}\psi\left(N_{xyzn}\right)f_{N_{xyzn}}\left(\mathbf{x}\right)g^{\mu\nu}_{\chi}.
\label{Gravitonzustand_Raum-Zeit-Darstellung}
\end{equation}
Ein Zustand, welcher viele Gravitonen enthält, entspricht dann der Konstruktion des rein quantentheoretischen Analogons
zum quantisierten Gravitationsfeld im begrifflichen Rahmen der Ur"=Alternativen. Die Basiszustände eines allgemeinen
solchen Zustandes sind durch die Zahl der Gravitonen in den Basiszuständen des Tensorraumes der Ur"=Alternativen
kombiniert mit den Basiszuständen der vier Ur"=Alternativen der Metrik definiert:

\begin{equation}
|\Psi_g\rangle_{\mathcal{N}}=\sum_{N_{ABCD}\otimes g^{\mu\nu}_\chi}\sum_{\mathcal{N}\left(N_{ABCD}\otimes g^{\mu\nu}_\chi\right)}\psi_{\mathcal{N}}
\left[\mathcal{N}\left(N_{ABCD}\otimes g^{\mu\nu}_\chi\right)\right]|\mathcal{N}\left(N_{ABCD}\otimes g^{\mu\nu}_\chi\right)\rangle.
\label{Gravitonenfeld}
\end{equation}
Natürlich gehorcht der Zustand des freien Gravitationsfeldes der Gleichung ($\ref{Klein-Gordon-Gleichung}$),
welche eine freie Wellengleichung ist:

\begin{equation}
\left(E^2-P_x^2-P_y^2-P_z^2\right)|\Psi_g(t)\rangle=0 \quad\longleftrightarrow\quad
\left(\partial_t^2-\partial_x^2-\partial_y^2-\partial_z^2\right)g^{\mu\nu}_{N\chi}\left(\mathbf{x},t\right)=0.
\end{equation}
Die Einsteinsche Feldgleichung geht in linearer Näherung in eine solche freie Wellengleichung über und deshalb
wundert es überhaupt nicht, dass auch der Zustand eines aus Ur"=Alternativen konstruierten Gravitons, solange noch
keine Wechselwirkung definiert ist, im Rahmen der Quantentheorie der Ur"=Alternativen ebenfalls zunächst dieser
freien Wellengleichung genügt. Um im Rahmen der Quantentheorie der Ur-Alternativen zum quantentheoretischen
Analogon der vollständigen Dynamik der allgemeinen Relativitätstheorie zu gelangen, welche ja eine
Selbstwechselwirkung des Gravitationsfeldes enthält, muss der allgemeine rein quantentheoretische
Wechselwirkungsbegriff zu Grunde gelegt werden, der im letzten Abschnitt konstituiert wurde.
Auf der Basis dieses Wechselwirkungsprozesses muss natürlich auch die Wechselwirkung der Gravitonen,
die in diesem Unterabschnitt zunächst nur isoliert betrachtet werden konnten, mit anderen Objekten
beschrieben werden, was aber hier nicht behandelt wird, obwohl dies im Prinzip dem gleichen Schema folgt.

\subsection{Die Dynamik der selbstwechselwirkenden Gravitation}

In diesem Unterabschnitt soll nun die Wechselwirkung der im letzten Unterabschnitt aus Ur"=Alternativen
konstruierten Gravitonenzustände ($\ref{Gravitonzustand}$) betrachtet werden. Um zur Dynamik der
selbstwechselwirkenden Gravitation in der Quantentheorie der Ur"=Alternativen zu gelangen, kann man
geleitet durch das Korrespondenzprinzip von der klassischen Einsteinschen Feldgleichung ausgehen,
die dann basierend auf den Quantisierungsregeln ($\ref{Uebergang_Zustaende_Operatoren}$),
($\ref{Uebergang_Wechselwirkung_Produkt}$) und ($\ref{Uebergang_Wechselwirkung_Produkt_Ableitungen}$)
in eine rein quantentheoretische auf Ur"=Alternativen basierende Beschreibung umgewandelt wird.
Die Einsteinsche Feldgleichung hat bekanntlich folgende Gestalt:

\begin{equation}
R_{\mu\nu}-\frac{1}{2}R g_{\mu\nu}=-\kappa T_{\mu\nu},
\end{equation}
wobei $\kappa$ die Gravitationskonstante $G$ enthält und $T_{\mu\nu}$ den Energie"=Impuls"=Tensor beschreibt.
Da die Wechselwirkung des Gravitationsfeldes mit anderen Quantenobjekten hier nicht betrachtet werden soll,
gilt $T_{\mu\nu}=0$ und dies bedeutet:

\begin{equation}
R_{\mu\nu}-\frac{1}{2} R g_{\mu\nu}=0 \quad\longleftrightarrow\quad R_{\mu\nu}=0,
\label{Einsteingleichung_freie}
\end{equation}
wobei der Ricci"=Tensor $R_{\mu\nu}$ über den Riemann"=Tensor:

\begin{equation}
R_{\mu\nu\rho}^{\quad\ \ \sigma}=\partial_\mu \Gamma_{\nu\rho}^{\sigma}-\partial_\nu \Gamma_{\mu\rho}^\sigma
+\Gamma_{\mu\rho}^{\lambda}\Gamma_{\nu\lambda}^{\sigma}-\Gamma_{\nu\rho}^{\lambda}\Gamma_{\mu\lambda}^{\sigma}
\end{equation}
in der folgenden Weise definiert ist:

\begin{equation}
R_{\mu\nu}=R_{\mu\rho\nu}^{\quad\ \ \rho}=\partial_\mu \Gamma_{\rho\nu}^{\rho}-\partial_\rho \Gamma_{\mu\nu}^\rho
+\Gamma_{\mu\nu}^{\lambda}\Gamma_{\rho\lambda}^{\rho}-\Gamma_{\rho\nu}^{\lambda}\Gamma_{\mu\lambda}^{\rho},
\label{Ricci-Tensor}
\end{equation}
der Ricci"=Skalar $R$ über den Ricci"=Tensor $R_{\mu\nu}$ in der folgenden Weise definiert ist:

\begin{equation}
R=g^{\mu\nu}R_{\mu\nu},
\end{equation}
und die Christoffelsymbole die folgende Gestalt haben:

\begin{equation}
\Gamma_{\mu\nu}^{\rho}=\frac{1}{2}g^{\rho\lambda}\left(\partial_\mu g_{\nu\lambda}+\partial_\nu g_{\mu\lambda}-\partial_\lambda g_{\mu\nu}\right).
\label{Christoffelsymbole}
\end{equation}
Wenn man ($\ref{Christoffelsymbole}$) in ($\ref{Ricci-Tensor}$) einsetzt, so ergibt sich für die Einsteinsche Feldgleichung ($\ref{Einsteingleichung_freie}$) ausgedrückt direkt durch den metrischen Tensor $g_{\mu\nu}$ die folgende Gestalt:

\begin{eqnarray}
&&R_{\mu\nu}=\frac{1}{2}\partial_\mu\left[g^{\rho\lambda}
\left(\partial_\rho g_{\nu\lambda}+\partial_\nu g_{\rho\lambda}-\partial_\lambda g_{\rho\nu}\right)\right]
-\frac{1}{2}\partial_\rho\left[g^{\rho\lambda}\left(\partial_\mu g_{\nu\lambda}+\partial_\nu g_{\mu\lambda}
-\partial_\lambda g_{\mu\nu}\right)\right]\nonumber\\
&&+\frac{1}{4}g^{\kappa\lambda}\left(\partial_\mu g_{\nu\lambda}+\partial_\nu g_{\mu\lambda}-\partial_\lambda g_{\mu\nu}\right)
g^{\rho\sigma}\left(\partial_\rho g_{\kappa\sigma}+\partial_\kappa g_{\rho\sigma}-\partial_\sigma g_{\rho\kappa}\right)\nonumber\\
&&-\frac{1}{4}g^{\kappa\lambda}\left(\partial_\rho g_{\nu\lambda}+\partial_\nu g_{\rho\lambda}-\partial_\lambda g_{\rho\nu}\right)
g^{\rho\sigma}\left(\partial_\mu g_{\kappa\sigma}+\partial_\kappa g_{\mu\sigma}-\partial_\sigma g_{\mu\kappa}\right)\nonumber\\
&&=\frac{1}{2}\left[\partial_\mu g^{\rho\lambda}\partial_\rho g_{\nu\lambda}+\partial_\mu g^{\rho\lambda}\partial_\nu g_{\rho\lambda}
-\partial_\mu g^{\rho\lambda}\partial_\lambda g_{\rho\nu}
+g^{\rho\lambda}\partial_\mu\partial_\rho g_{\nu\lambda}
+g^{\rho\lambda}\partial_\mu\partial_\nu g_{\rho\lambda}-g^{\rho\lambda}\partial_\mu\partial_\lambda g_{\rho\nu}\right.\nonumber\\
&&\left.-\partial_\rho g^{\rho\lambda}\partial_\mu g_{\nu\lambda}-\partial_\rho g^{\rho\lambda}\partial_\nu g_{\mu\lambda}
+\partial_\rho g^{\rho\lambda}\partial_\lambda g_{\mu\nu}
-g^{\rho\lambda}\partial_\rho \partial_\mu g_{\nu\lambda}-g^{\rho\lambda}\partial_\rho \partial_\nu g_{\mu\lambda}
+g^{\rho\lambda}\partial_\rho \partial_\lambda g_{\mu\nu}\right.\nonumber\\
&&\left.+\frac{1}{2}\left(g^{\kappa\lambda}g^{\rho\sigma}\partial_\mu g_{\nu\lambda}\partial_\rho g_{\kappa\sigma}
+g^{\kappa\lambda}g^{\rho\sigma}\partial_\nu g_{\mu\lambda}\partial_\rho g_{\kappa\sigma}
-g^{\kappa\lambda}g^{\rho\sigma}\partial_\lambda g_{\mu\nu}\partial_\rho g_{\kappa\sigma}
+g^{\kappa\lambda}g^{\rho\sigma}\partial_\mu g_{\nu\lambda}\partial_\kappa g_{\rho\sigma}
\right.\right.\nonumber\\&&\left.\left.
+g^{\kappa\lambda}g^{\rho\sigma}\partial_\nu g_{\mu\lambda}\partial_\kappa g_{\rho\sigma}
-g^{\kappa\lambda}g^{\rho\sigma}\partial_\lambda g_{\mu\nu}\partial_\kappa g_{\rho\sigma}
-g^{\kappa\lambda}g^{\rho\sigma}\partial_\mu g_{\nu\lambda}\partial_\sigma g_{\rho\kappa}
-g^{\kappa\lambda}g^{\rho\sigma}\partial_\nu g_{\mu\lambda}\partial_\sigma g_{\rho\kappa}
\right.\right.\nonumber\\&&\left.\left.
+g^{\kappa\lambda}g^{\rho\sigma}\partial_\lambda g_{\mu\nu}\partial_\sigma g_{\rho\kappa}
-g^{\kappa\lambda}g^{\rho\sigma}\partial_\rho g_{\nu\lambda}\partial_\mu g_{\kappa\sigma}
+g^{\kappa\lambda}g^{\rho\sigma}\partial_\nu g_{\rho\lambda}\partial_\mu g_{\kappa\sigma}
-g^{\kappa\lambda}g^{\rho\sigma}\partial_\lambda g_{\rho\nu}\partial_\mu g_{\kappa\sigma}
\right.\right.\nonumber\\&&\left.\left.
+g^{\kappa\lambda}g^{\rho\sigma}\partial_\rho g_{\nu\lambda}\partial_\kappa g_{\mu\sigma}
+g^{\kappa\lambda}g^{\rho\sigma}\partial_\nu g_{\rho\lambda}\partial_\kappa g_{\mu\sigma}
-g^{\kappa\lambda}g^{\rho\sigma}\partial_\lambda g_{\rho\nu}\partial_\kappa g_{\mu\sigma}
-g^{\kappa\lambda}g^{\rho\sigma}\partial_\rho g_{\nu\lambda}\partial_\sigma g_{\mu\kappa}
\right.\right.\nonumber\\&&\left.\left.
-g^{\kappa\lambda}g^{\rho\sigma}\partial_\nu g_{\rho\lambda}\partial_\sigma g_{\mu\kappa}
+g^{\kappa\lambda}g^{\rho\sigma}\partial_\lambda g_{\rho\nu}\partial_\sigma g_{\mu\kappa}\right)\right]\nonumber\\
&&=\frac{1}{2}\left[
\partial_\mu g^{\rho\lambda}\partial_\rho g_{\nu\lambda}
+\partial_\mu g^{\rho\lambda}\partial_\nu g_{\rho\lambda}
-\partial_\mu g^{\rho\lambda}\partial_\lambda g_{\rho\nu}
+g^{\rho\lambda}\partial_\mu\partial_\nu g_{\rho\lambda}
-g^{\rho\lambda}\partial_\mu\partial_\lambda g_{\rho\nu}
\right.\nonumber\\&&\left.
-\partial_\rho g^{\rho\lambda}\partial_\mu g_{\nu\lambda}
-\partial_\rho g^{\rho\lambda}\partial_\nu g_{\mu\lambda}
+\partial_\rho g^{\rho\lambda}\partial_\lambda g_{\mu\nu}
-g^{\rho\lambda}\partial_\rho \partial_\nu g_{\mu\lambda}
+g^{\rho\lambda}\partial_\rho \partial_\lambda g_{\mu\nu}
\right.\nonumber\\&&
+g^{\kappa\lambda}g^{\rho\sigma}\partial_\rho g_{\nu\lambda}\partial_\kappa g_{\mu\sigma}
-g^{\kappa\lambda}g^{\rho\sigma}\partial_\rho g_{\nu\lambda}\partial_\mu g_{\kappa\sigma}
-g^{\kappa\lambda}g^{\rho\sigma}\partial_\lambda g_{\rho\nu}\partial_\kappa g_{\mu\sigma}
\nonumber\\&&\left.
+\frac{1}{2}\left(g^{\kappa\lambda}g^{\rho\sigma}\partial_\mu g_{\nu\lambda}\partial_\kappa g_{\rho\sigma}
+g^{\kappa\lambda}g^{\rho\sigma}\partial_\nu g_{\mu\lambda}\partial_\kappa g_{\rho\sigma}
\right.\right.\nonumber\\&&\left.\left.
-g^{\kappa\lambda}g^{\rho\sigma}\partial_\lambda g_{\mu\nu}\partial_\kappa g_{\rho\sigma}
+g^{\kappa\lambda}g^{\rho\sigma}\partial_\nu g_{\rho\lambda}\partial_\mu g_{\kappa\sigma}
\right)\right]=0.
\label{Einsteingleichung_freie_Metrik}
\end{eqnarray}
Diese klassische dynamische Grundgleichung für das Gravitationsfeld muss in einer Weise quantisiert werden,
dass sie in eine Beschreibungsweise im Sinne der Ur"=Alternativen überführt wird. Im Rahmen dieser Beschreibung ist
auch das Gravitationsfeld lediglich eine Darstellung eines dahinter stehenden Zustandes im Tensorraum der Ur"=Alternativen und
das Produkt des Gravitationsfeldes mit sich selbst entspricht einer Verschränkung solcher Zustände. Diese Art der
Quantisierung ist deutlich radikaler als die gewöhnliche Quantisierung, bei der Vertauschungsrelationen zwischen
dem Gravitationsfeld und der kanonisch konjugierten Feldgröße gefordert werden, welche also kontinuierliche
Feldgrößen zu Grunde legt. Auch im Rahmen der Schleifenquantengravitation mit den Spin"=Netzwerken bleibt der
Raum eine unabhängige Realität und werden mit Holonomien weiterhin Größen zu Grunde gelegt, die sich auf Beziehungen
in einer eigenständigen Raum-Zeit beziehen. Bei den Wechselwirkungen geht man gewöhnlich weiterhin im
klassischen Sinne von punktweisen Produkten in der Raum"=Zeit aus. Die Frage der Wechselwirkung ist
nämlich die entscheidende. In dem hier versuchten Modell der Ur"=Alternativen gibt es überhaupt kein
Gravitationsfeld mehr, sondern nur Kombinationen von Ur"=Alternativen als diskreten Einheiten und
Verschränkungen zwischen diesen, die sich lediglich raum-zeitlich darstellen und in dieser Darstellung
näherungsweise zu der Einsteinschen Feldgleichung führen. Es muss also zunächst in dem Sinne eine
Quantisierung der Gravitation vorgenommen werden, dass das Gravitationsfeld in einen Zustand von
Gravitonen überführt wird, die durch Zustände im Tensorraum der Ur"=Alternativen beschrieben werden,
beziehungsweise zunächst in den Zustand eines einzelnen Gravitons. Dies geschieht durch eine
Kombination von ($\ref{Uebergang_Zustaende_Operatoren}$) und ($\ref{Gravitonzustand}$):

\begin{eqnarray}
g^{\mu\nu}\left(\mathbf{x},t\right)\quad\longrightarrow &&
|\Psi_g(t)\rangle=g^{\mu\nu}_{\chi N}(t)=\sum_{N_{ABCD}}\psi\left(N_{ABCD},t\right)|N_{ABCD}\rangle \otimes g^{\mu\nu}_{\chi}\nonumber\\
&&\longleftrightarrow\quad g^{\mu\nu}_{\chi N}\left(\mathbf{x},t\right)=\sum_{N_{xyzn}}\psi\left(N_{xyzn},t\right)
f_{N_{xyzn}}(\mathbf{x})g^{\mu\nu}_\chi.
\label{Quantisierung_Metrik}
\end{eqnarray}
Die Ableitungen müssen dementsprechend zu Impulsoperatoren im Tensorraum werden, indem man die Definition
($\ref{Viererimpuls_Tensorraum}$) und den Übergang ($\ref{Uebergang_Zustaende_Operatoren}$) zu Grunde legt,
was bedeutet:

\begin{eqnarray}
\partial^\rho g^{\mu\nu}\left(\mathbf{x},t\right)\quad\longrightarrow &&
iP^{\rho}_{ABCD}|\Psi_g(t)\rangle=iP^\rho_{ABCD}\sum_{N_{ABCD}}\psi\left(N_{ABCD},t\right)|N_{ABCD}\rangle \otimes g^{\mu\nu}_\chi\nonumber\\
&&=i|\mathcal{P}^\rho\Psi\rangle=i\sum_{N_{ABCD}}\mathcal{P}^\rho_\psi\left(N_{ABCD},t\right)|N_{ABCD}\rangle \otimes g^{\mu\nu}_\chi
\equiv i\mathcal{P}^{\rho}g^{\mu\nu}_{N \chi}(t)\nonumber\\
&&\longleftrightarrow\quad i\sum_{N_{xyzn}}\mathcal{P}^\rho_\psi\left(N_{xyzn},t\right)f_{N_{xyzn}}(\mathbf{x}) g^{\mu\nu}_\chi
\equiv i\mathcal{P}^{\rho}g^{\mu\nu}_{\chi N}\left(\mathbf{x},t\right).
\label{Quantisierung_Metrik_Ableitung}
\end{eqnarray}
Um nun zu einer vollständigen quantentheoretischen Beschreibung der Gravitation zu gelangen, muss die Wechselwirkung
integriert werden. Es müssen also Zustände mehrerer Gravitonen gemäß ($\ref{Gravitonenfeld}$) betrachtet werden,
welche jedoch gemäß ($\ref{Zustand_N_verschraenkt}$) in Zustände mit Verschränkung überführt werden.
Und um diese zu definieren, müssen die punktweisen Produkte des Gravitationsfeldes in ein Produkt im Tensorraum
gemäß ($\ref{Uebergang_Wechselwirkung_Produkt}$) umgewandelt werden, das einen verschränkten Zustand gemäß
($\ref{Zustand_N_verschraenkt}$) definiert. Dies bedeutet konkret, dass ein Produkt des metrischen Tensors
mit sich selbst in der folgenden Weise umgewandelt wird:

\begin{eqnarray}
&&g^{\mu\nu}\left(\mathbf{x},t\right)g^{\rho\sigma}\left(\mathbf{x},t\right)\quad\longrightarrow\nonumber\\
&&\delta_{N^1,N^2}\left[\sum_{N_{ABCD}^1}\left(\psi_1\left(N_{ABCD}^1,t\right)|N_{ABCD}^1\rangle\otimes g_{\chi 1}^{\mu\nu}\right)
\otimes \sum_{N_{ABCD}^2}\left(\psi_2\left(N_{ABCD}^2,t\right)|N_{ABCD}^2\rangle \otimes g_{\chi 2}^{\rho\sigma}\right)\right]\nonumber\\
&&=\sum_{N_{ABCD}}\left[\left(\psi_1\left(N_{ABCD},t\right)|N_{ABCD}\rangle\otimes g_{\chi 1}^{\mu\nu}\right)
\otimes \left(\psi_2\left(N_{ABCD},t\right)|N_{ABCD}\rangle \otimes g_{\chi 2}^{\rho\sigma}\right)\right]
\equiv \sum_{N} g_{1 \chi N}^{\mu\nu}(t) g_{2 \chi N}^{\rho\sigma}(t)\nonumber\\
&&\longleftrightarrow\quad \sum_{N_{xyzn}}\left[\psi_1\left(N_{xyzn},t\right)f_{N_{xyzn}}\left(\mathbf{x}\right)g_{\chi 1}^{\mu\nu}
\psi_2\left(N_{xyzn},t\right)f_{N_{xyzn}}\left(\mathbf{x}\right)g_{\chi 2}^{\rho\sigma}\right]\nonumber\\
&&\quad\quad\quad\equiv\sum_{N}g_{1 \chi N}^{\mu\nu}\left(\mathbf{x},t\right)g_{2 \chi N}^{\rho\sigma}\left(\mathbf{x},t\right).
\label{Quantisierung_Produkt}
\end{eqnarray}
Und ein Produkt von Ableitungen des metrischen Tensors mit sich selbst wird gemäß ($\ref{Uebergang_Wechselwirkung_Produkt_Ableitungen}$)
in der folgenden Weise umgewandelt:

\begin{align}
&\partial^\kappa g^{\mu\nu}\left(\mathbf{x},t\right)\partial^\lambda g^{\rho\sigma}\left(\mathbf{x},t\right)\quad\longrightarrow\nonumber\\
&\delta_{N^1,N^2}\left[iP^{\kappa}_{ABCD}\sum_{N_{ABCD}^1}\left(\psi_1\left(N_{ABCD}^1,t\right)|N_{ABCD}^1\rangle\otimes g_{\chi 1}^{\mu\nu}\right)
\otimes iP^{\lambda}_{ABCD}\sum_{N_{ABCD}^2}\left(\psi_2\left(N_{ABCD}^2,t\right)
|N_{ABCD}^2\rangle \otimes g_{\chi 2}^{\rho\sigma}\right)\right]\nonumber\\
&=-\delta_{N^1,N^2}\left[\sum_{N_{ABCD}^1}\left(\mathcal{P}^{\kappa}_{\psi_1}\left(N_{ABCD}^1,t\right)|N_{ABCD}^1\rangle\otimes
g_{\chi 1}^{\mu\nu}\right) \otimes \sum_{N_{ABCD}^2}\left(\mathcal{P}^{\lambda}_{\psi_2}\left(N_{ABCD}^2,t\right)
|N_{ABCD}^2\rangle \otimes g_{\chi 2}^{\rho\sigma}\right)\right]\nonumber\\
&=-\sum_{N_{ABCD}}\left[\left(\mathcal{P}^{\kappa}_{\psi_1}\left(N_{ABCD},t\right)|N_{ABCD}\rangle\otimes g_{\chi 1}^{\mu\nu}\right)
\otimes \left(\mathcal{P}^{\lambda}_{\psi_2}\left(N_{ABCD},t\right)|N_{ABCD}\rangle \otimes g_{\chi 2}^{\rho\sigma}\right)\right]\nonumber\\
&\equiv-\sum_{N}\mathcal{P}^\kappa_1 g_{1 \chi N}^{\mu\nu}(t)\mathcal{P}^\lambda_2 g_{2 \chi N}^{\rho\sigma}(t)
\ \longleftrightarrow\ -\sum_{N_{xyzn}}\left[\mathcal{P}^{\kappa}_{\psi_1}\left(N_{xyzn},t\right)
f_{N_{xyzn}}\left(\mathbf{x}\right)g_{\chi 1}^{\mu\nu}
\mathcal{P}^{\lambda}_{\psi_2}\left(N_{xyzn},t\right)f_{N_{xyzn}}\left(\mathbf{x}\right)g_{\chi 2}^{\rho\sigma}\right]\nonumber\\
&\quad\quad\quad\quad\quad\quad\quad\quad\quad\quad\quad\quad\quad\quad\quad\quad\equiv-\sum_{N}\mathcal{P}^\kappa_1 g_{1 \chi N}^{\mu\nu}
\left(\mathbf{x},t\right)\mathcal{P}^\lambda_2 g_{2 \chi N}^{\rho\sigma}\left(\mathbf{x},t\right).
\label{Quantisierung_Produkt_Ableitungen}
\end{align}
Wenn man nun in der freien Einsteingleichung ($\ref{Einsteingleichung_freie_Metrik}$) den metrischen Tensor
gemäß ($\ref{Quantisierung_Metrik}$) in einen Zustand im Tensorraum der Ur"=Alternativen umwandelt und die
auftretenden Produkte gemäß ($\ref{Quantisierung_Produkt}$) und ($\ref{Quantisierung_Produkt_Ableitungen}$)
in das quantentheoretische Analogon überführt, so ergibt sich eine Gleichung, welche die dynamische
Grundgleichung der selbstwechselwirkenden Gravitation ohne Kopplung an andere Objekte im Rahmen der
Quantentheorie der Ur"=Alternativen darstellt. Diese Gleichung stellt eine reine Beziehungsstruktur
von Ur"=Alternativen dar ohne direkten Bezug zum Begriff eines Feldes. Lediglich durch einen indirekten
Übergang in die raum"=zeitliche Darstellung, wie sie durch die Darstellung der Tensorraumzustände als
Funktionen im physikalischen Raum in zum obigen Prozess der Quantisierung umgekehrter Richtung vollzogen
werden kann, erscheint sie näherungsweise wie eine quantentheoretische Wellengleichung, also eine
quantisierte Einsteingleichung. Aber die Gleichung selbst beschreibt die Natur rein quantentheoretisch,
also nur basierend auf einer Beziehungsstruktur abstrakter Informationseinheiten. Wenn die verschiedenen
in den Wechselwirkungsprozess einbezogenen aus Ur"=Alternativen konstruierten Gravitonen durch
eine von $1$ bis $4$ laufende Nummerierung bezeichnet werden, so erhält die dynamische Grundgleichung
der Gravitation unter Verwendung von ($\ref{Quantisierung_Metrik}$), ($\ref{Quantisierung_Metrik_Ableitung}$),
($\ref{Quantisierung_Produkt}$) und ($\ref{Quantisierung_Produkt_Ableitungen}$) die folgende Gestalt: 

\begin{align}
&R_{\mu\nu}^\psi(t)=\frac{1}{2}\left[
\sum_{N}\mathcal{P}_{1 \mu} g_{1 \chi N}^{\rho\lambda}(t)\mathcal{P}_{2 \rho} g^{2 \chi N}_{\nu\lambda}(t)
+\sum_{N}\mathcal{P}_{1 \mu} g_{1 \chi N}^{\rho\lambda}(t)\mathcal{P}_{2 \nu} g^{2 \chi N}_{\rho\lambda}(t)
-\sum_{N}\mathcal{P}_{1 \mu} g_{1 \chi N}^{\rho\lambda}(t)\mathcal{P}_{2 \lambda} g^{2 \chi N}_{\rho\nu}(t)
\right.\nonumber\\&\left.
+\sum_{N}g_{1 \chi N}^{\rho\lambda}(t)\mathcal{P}_{2 \mu}\mathcal{P}_{2 \nu} g^{2 \chi N}_{\rho\lambda}(t)
-\sum_{N}g_{1 \chi N}^{\rho\lambda}(t)\mathcal{P}_{2 \mu}\mathcal{P}_{2 \lambda} g^{2 \chi N}_{\rho\nu}(t)
-\sum_{N}\mathcal{P}_{1 \rho} g_{1 \chi N}^{\rho\lambda}(t)\mathcal{P}_{2 \mu} g^{2 \chi N}_{\nu\lambda}(t)
\right.\nonumber\\&\left.
-\sum_{N}\mathcal{P}_{1 \rho} g_{1 \chi N}^{\rho\lambda}(t)\mathcal{P}_{2 \nu} g^{2 \chi N}_{\mu\lambda}(t)
+\sum_{N}\mathcal{P}_{1 \rho} g_{1 \chi N}^{\rho\lambda}(t)\mathcal{P}_{2 \lambda} g^{2 \chi N}_{\mu\nu}(t)
-\sum_{N}g_{1 \chi N}^{\rho\lambda}(t)\mathcal{P}_{2 \rho}\mathcal{P}_{2 \nu} g^{2 \chi N}_{\mu\lambda}(t)
\right.\nonumber\\&\left.
+\sum_{N}g_{1 \chi N}^{\rho\lambda}(t)\mathcal{P}_{2 \rho}\mathcal{P}_{2 \lambda} g^{2 \chi N}_{\mu\nu}(t)
+\sum_{N}g_{1 \chi N}^{\kappa\lambda}(t)g_{2 \chi N}^{\rho\sigma}\mathcal{P}_{3 \rho}(t)
g^{3 \chi N}_{\nu\lambda}\mathcal{P}_{4 \kappa}g^{4 \chi N}_{\mu\sigma}
\right.\nonumber\\&\left.
-\sum_{N}g_{1 \chi N}^{\kappa\lambda}(t)g_{2 \chi N}^{\rho\sigma}\mathcal{P}_{3 \rho}(t)
g^{3 \chi N}_{\nu\lambda}(t)\mathcal{P}_{4 \mu}g^{4 \chi N}_{\kappa\sigma}(t)
-\sum_{N}g_{1 \chi N}^{\kappa\lambda}(t)g_{2 \chi N}^{\rho\sigma}\mathcal{P}_{3 \lambda}(t)
g^{3 \chi N}_{\rho\nu}(t)\mathcal{P}_{4 \kappa}g^{4 \chi N}_{\mu\sigma}(t)
\right.\nonumber\\&\left.
+\frac{1}{2}\left(
\sum_{N}g_{1 \chi N}^{\kappa\lambda}(t)g_{2 \chi N}^{\rho\sigma}\mathcal{P}_{3 \mu}(t)
g^{3 \chi N}_{\nu\lambda}(t)\mathcal{P}_{4 \kappa}g^{4 \chi N}_{\rho\sigma}(t)
+\sum_{N}g_{1 \chi N}^{\kappa\lambda}(t)g_{2 \chi N}^{\rho\sigma}\mathcal{P}_{3 \nu}(t)
g^{3 \chi N}_{\mu\lambda}(t)\mathcal{P}_{4 \kappa}g^{4 \chi N}_{\rho\sigma}(t)
\right.\right.\nonumber\\&\left.\left.
-\sum_{N}g_{1 \chi N}^{\kappa\lambda}(t)g_{2 \chi N}^{\rho\sigma}(t)\mathcal{P}_{3 \lambda}
g^{3 \chi N}_{\mu\nu}(t)\mathcal{P}_{4 \kappa}g^{4 \chi N}_{\rho\sigma}(t)
+\sum_{N}g_{1 \chi N}^{\kappa\lambda}(t)g_{2 \chi N}^{\rho\sigma}(t)\mathcal{P}_{3 \nu}
g^{3 \chi N}_{\rho\lambda}(t)\mathcal{P}_{4 \mu}g^{4 \chi N}_{\kappa\sigma}(t)\right)\right]=0.\nonumber\\
\end{align}

\section{Zusammenfassung und Diskussion}

In dieser Arbeit wurde zunächst deutlich gemacht, dass eine einheitliche Beschreibung der Natur eine Kopernikanische
Wende bezüglich der Interpretation der Natur des Raumes unumgänglich macht. Dies bedeutet, dass die physikalische
Realität gemäß der Quantentheorie nicht in der Weise zu verstehen ist, dass gegenständliche und geometrisch beschriebene
Objekte in einem vorgegebenen physikalischen Raum existieren, sondern rein logische abstrakte Objekte, die noch keinerlei
feldtheoretische Begriffe voraussetzen, umgekehrt die Existenz des physikalischen Raumes, der formal mit der Zeit zur
Raum"=Zeit verbunden werden kann, mit der bekannten Struktur begründen. Basierend auf dieser zentralen Erkenntnis
des inneren Wesens der Natur gemäß der Quantentheorie wurde dann die Quantentheorie der Ur"=Alternativen des
Carl Friedrich von Weizsäcker als die konsequente Realisierung eines solchen rein quantentheoretischen Realitätsbegriffes
dargestellt, der von feldtheoretischen Begriffen vollkommen unabhängig ist. Die Ur"=Alternativen sind elementare
quantentheoretische Informationseinheiten. Allerdings werden diese im Gegensatz zu ihrer gewöhnlichen Verwendung
in der Physik im Rahmen der Quantentheorie der Ur"=Alternativen in einem sehr viel grundsätzlicheren Sinne interpretiert. 
Die Ur"=Alternativen sind nicht Information, die sich im Raum befinden oder in einem Netzwerk ausgetauscht würde,
die einen Träger bräuchte oder sich auf bereits vorhandene physikalische Objekte beziehen würde. Vielmehr kommt
dieser Information absolute Bedeutung zu. Es gibt daher in der Quantentheorie der Ur"=Alternativen überhaupt nichts
anderes als die Information. Umgekehrt kann die Existenz aller anderen Realitäten wie der physikalischen Objekte,
deren Wechselwirkungen und der physikalische Raum überhaupt nur aus dieser im schlechthinnigen Sinne abstrakten
Information begründet werden. Insofern führt die Quantentheorie der Ur"=Alternativen die Veränderung des
Realitätsbegriffes in der Quantentheorie zu ihrer letzten Konsequenz. Nachdem die grundlegende begriffliche
und philosophische Basis der Quantentheorie der Ur"=Alternativen dargestellt wurde, wie sie seitens
Carl Friedrich von Weizsäcker entwickelt wurde, wurden eigene weiterführende Ansätze in Bezug auf
eine Beschreibung der realen Physik basierend auf diesem begrifflichen Grundrahmen entwickelt.
Hierzu gehörte ein bestimmter Ansatz zu einer Abbildung der Zustände im Tensorraum der Ur"=Alternativen
in die Raum"=Zeit, ein Ansatz zur Integration der inneren Symmetrien der Elementarteilchenphysik, ein rein
quantentheoretischer Begriff der Wechselwirkung und schließlich ein Versuch, basierend auf diesen neuen
Entwicklungen zu einer rein quantentheoretischen und im schlechthinnigen Sinne hintergrundunabhängigen
Beschreibungsweise der Gravitation zu gelangen. Hierbei wurde das Korrespondenzprinzip zu Grunde gelegt,
also eine solche Dynamik basierend auf Ur"=Alternativen konstruiert, sodass sich nach Abbildung der
Wechselwirkungsbeziehung der verschiedenen aus Ur"=Alternativen gebildeten Gravitonen in die Raum-Zeit
in einer klassischen Näherung die gewöhnliche Einsteinsche Feldgleichung ergibt.

Natürlich ist in dieser Arbeit nur ein begrifflicher und mathematischer Grundansatz zur Beschreibung der konkreten
Physik im begrifflichen Rahmen der Ur"=Alternativen entwickelt worden. Dass die Quantentheorie der Ur"=Alternativen
bezüglich ihrer begrifflichen Grundbasis mit ihrem rein quantentheoretischen Realitätsbegriff im Prinzip
von ihrer grundlegenden Idee her ganz sicher nicht nur vielversprechend ist, sondern dass hier die
richtige Richtung mit Sicherheit eingeschlagen wurde, daran kann angesichts der überwältigenden
argumentativen Substanz keinerlei Zweifel bestehen. Ob die konkrete mathematische Ausgestaltung
und die meinerseits entwickelten weiterführenden Konzepte in exakt dieser Weise zur Wahrheit führen können,
das kann zumindest nicht mit Sicherheit gesagt werden, aber auch dies scheint wenigstens vielversprechend zu sein.
Dies gilt auch dann, wenn die mathematischen Konzepte ganz sicher noch auf eine bessere formale Basis gestellt
werden müssen. Zudem wäre es von entscheidender Bedeutung zu zeigen, dass sich basierend auf der Auflösung der
Zustände quantentheoretischer Objekte in die rein logischen Objekte der Ur"=Alternativen und entsprechender diskreter
Zustandsräume und dem auf abstrakten Beziehungen von Ur"=Alternativen sich gründenden Wechselwirkungsbegriff
bei konkreten Berechnungen von physikalischen Vorgängen keine Divergenzen ergeben und das Verfahren der
Renormierung auf diese Weise umgangen werden kann. Die Begründung der Existenz aller Wechselwirkungen
sowie aller fundamentalen Objekte der Elementarteilchenphysik wäre das ideale Endziel. Zudem müsste noch eine Beziehung
der Abbildung der Zustände des Tensorraumes der Ur"=Alternativen zur globalen Topologie des Kosmos hinzukommen,
die in der Quantentheorie der Ur"=Alternativen gewöhnlich eigentlich über den Zustandsraum einer einzigen
Ur"=Alternative erfolgt, der aufgrund der Normierungsbedingung aus ($\ref{Ur-Alternative}$)
die Toplogie einer $\mathbb{S}^3$ aufweist, die man dann als globale räumliche Struktur
des Kosmos interpretiert. Auch müsste man natürlich irgendwann versuchen, genaue Vorhersagen
für konkrete Phänomenen zu machen, für welche die Quantentheorie der Ur"=Alternativen ganz spezifische
Ergebnisse liefert. Phänomene wie das EPR-Paradoxon zeigen nur, dass eine Beschreibung der Natur
jenseits feldtheoretischer Begriffe unumgänglich ist, aber noch nicht, dass die Realisierung
dessen in genau der Weise vollzogen werden muss wie dies im Rahmen der Quantentheorie der
Ur"=Alternativen geschieht. In jedem Falle aber wurde mit dieser Arbeit gezeigt, dass es
prinzipiell durchaus möglich sein könnte, die konktrete Physik im Rahmen der Quantentheorie der Ur"=Alternativen
zu beschreiben und daher eine einheitliche Beschreibung der Physik in einem rein quantentheoretischen Rahmen
zu erhalten, in dem der neue Realitätsbegriff, den die Quantentheorie eröffnet hat, in konsequenter Weise
realisiert ist. Und dieser Realitätsbegriff basiert nur auf abstrakter Information und damit letztendlich
auf reiner Logik in der Zeit, wobei die Logik ähnlich wie bei Hegel eine ontologische Bedeutung erhält.

\textbf{Danksagung:} Ich danke Bernd Henschenmacher für anregende Diskussionen und Gedanken
über Grundfragen der Quantentheorie und insbesondere bezüglich der Oktonionen.

\end{document}